\documentclass{article}

\newif\ifblind
\blindfalse

\PassOptionsToPackage{numbers, compress}{natbib}
\usepackage[main,final]{neurips_2026}

\usepackage{graphicx}
\usepackage[utf8]{inputenc}
\usepackage[T1]{fontenc}
\usepackage{hyperref}
\usepackage{url}
\usepackage{booktabs}
\usepackage{amsmath}
\usepackage{amssymb}
\usepackage{amsfonts}
\usepackage{bm}
\usepackage{nicefrac}
\usepackage{microtype}
\usepackage{xcolor}
\usepackage{enumitem}
\usepackage{makecell}
\usepackage{multirow}
\usepackage{colortbl}
\usepackage{float}
\usepackage{microtype}

\usepackage{xcolor}

\newcommand{\method}{STRAND}
\newcommand{\moduleDNA}{DNA Perturbation Module}
\newcommand{\moduleRNA}{Latent Cellular Transport Module}
\newcommand{\xhdr}[1]{\vspace{1mm}\noindent\textbf{#1.}}

\definecolor{bestcolor}{RGB}{220,235,255}
\definecolor{secondcolor}{RGB}{235,235,235}

\newcommand{\hb}{\mathbf{h}}

\newcommand{\mb}{\mathbf{m}}

\newcommand{\pb}{\mathbf{p}}
\newcommand{\qb}{\mathbf{q}}

\newcommand{\ub}{\mathbf{u}}

\newcommand{\xb}{\mathbf{x}}
\newcommand{\yb}{\mathbf{y}}
\newcommand{\zb}{\mathbf{z}}

\newcommand{\Eb}{\mathbf{E}}

\newcommand{\Ib}{\mathbf{I}}

\newcommand{\Wb}{\mathbf{W}}
\newcommand{\Xb}{\mathbf{X}}
\newcommand{\Yb}{\mathbf{Y}}

\title{STRAND: Sequence-Conditioned Transport for Single-Cell Perturbations}

\author{%
  Boyang Fu$^{1}$ \quad
  Sameer Gabbita$^{2}$ \quad
  George Dasoulas$^{3}$ \quad
  Xiang Lin$^{1}$ \quad
  Shanghua Gao$^{1}$ \\
  \bfseries
  Xiaorui Su$^{1}$ \quad
  Soumya Ghosh$^{3}$ \quad
  Marinka Zitnik$^{1,4,5}$\thanks{Corresponding author: \texttt{marinka@hms.harvard.edu}} \\[2mm]
  \normalfont
  \begin{tabular}{c}
    $^{1}$Department of Biomedical Informatics, Harvard Medical School, Boston, MA, USA \\
    $^{2}$Department of Biomedical Engineering, Johns Hopkins University, Baltimore, MD, USA \\
    $^{3}$Merck \& Co., Inc., Cambridge, MA, USA \\
    $^{4}$Kempner Institute for the Study of Natural and Artificial Intelligence, \\
    Harvard University, Allston, MA, USA \\
    $^{5}$Broad Institute of MIT and Harvard, Cambridge, MA, USA
  \end{tabular}
}

\begin{document}

\maketitle

\begin{abstract}
Prediction of the transcriptional response of a cell to a genetic perturbation is a central problem in the computational modeling of gene regulation. Perturbations of the same gene can produce different transcriptional responses depending on the targeted locus, such as an alternative transcription start site or a distal regulatory element, and gene-level perturbation models represent these distinct interventions with a single gene identifier. We introduce STRAND, a generative model that encodes the regulatory DNA sequence at the perturbation locus and uses this encoding to parameterize a conditional transport process from control to perturbed single-cell states. Sequence conditioning enables zero-shot prediction for perturbation targets absent from training and supports locus-resolved inputs at 128,bp model resolution within a 524,kb sequence window. On CRISPRi screens in K562, Jurkat, and RPE1 cells, STRAND achieves the best average rank on few-shot and unseen-gene benchmarks, with gains of up to 33\% in perturbation discrimination, and improves transfer to held-out cell lines by up to 0.12 in $\Delta$Pearson. Ablation studies attribute these gains to sequence conditioning and to conditional transport. At the locus level, STRAND distinguishes alternative transcription start sites and validated regulatory elements that gene-identifier models represent as a single perturbation. STRAND is available at \url{https://github.com/mims-harvard/STRAND}.
\end{abstract}

\section{Introduction} \label{introduction}

Prediction of the transcriptional response of a cell to a genetic perturbation is a central problem in the computational modeling of gene regulation~\cite{Roohani2023,Lotfollahi2023,ahlmann2025deep,Adduri2025,park2026causal,lorch2026latent,dong2026stack}. Specific DNA sequences, including promoters, enhancers, and alternative transcription start sites (TSSs), mediate regulatory effects, and a single gene can contain several such elements with distinct regulatory functions~\cite{nasser2021genome,avsec2021effective,linder2025predicting,pampari2025chrombpnet,avsec2025alphagenome}. Most single-cell perturbation models, including recent single-cell foundation models, represent perturbations at the gene level, as discrete identifiers or as static nodes in a graph~\cite{cui2024scgpt,wenkel2025txpert,passigan2025interpretable,zhu2025scouter,dong2026stack}. These models therefore assign the same representation to perturbations that target different genomic loci within one gene. Experimental technologies such as CRISPRi, CRISPRa, and base editing intervene at precise genomic locations, which creates a resolution gap between experimental interventions and the gene-level inputs of predictive models.

\begin{figure}[t]
    \centering
    \includegraphics[width=1.\columnwidth]{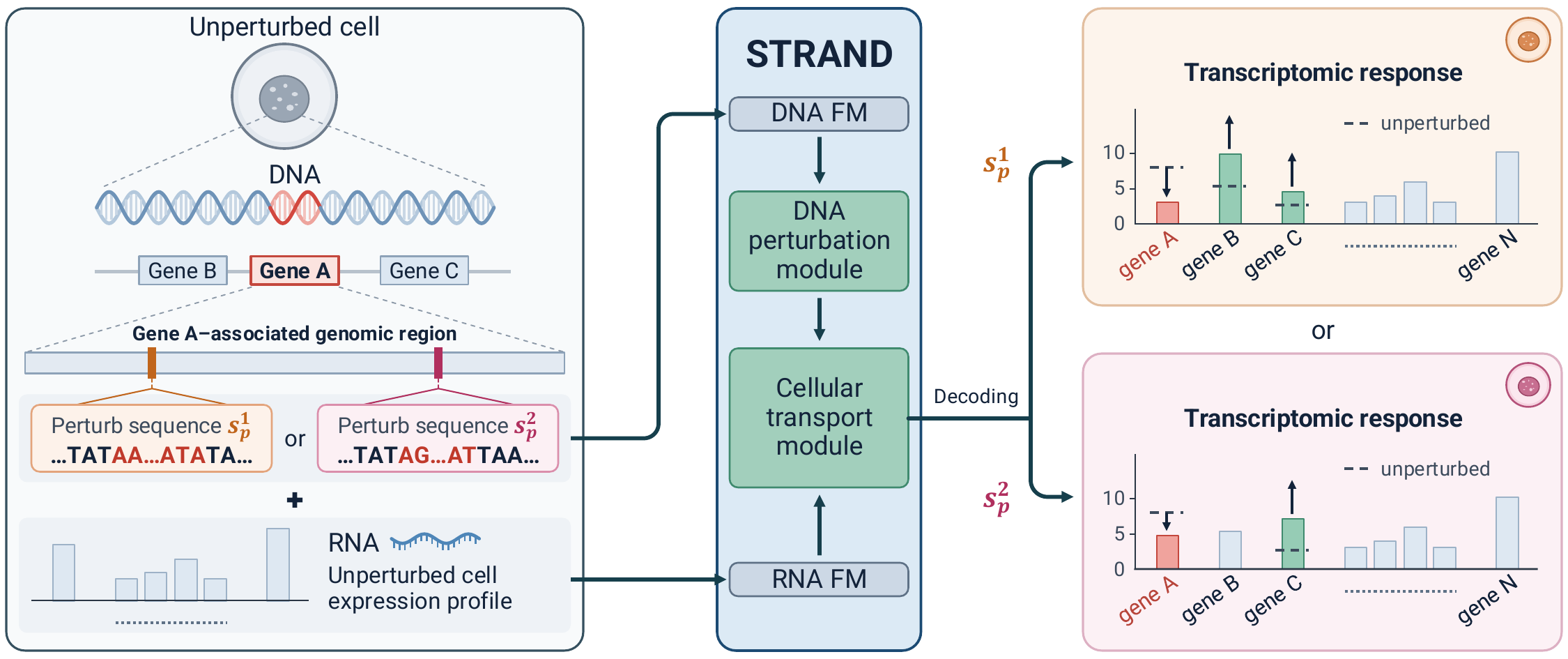}
    \vspace{-10pt}
    \caption{\textbf{Overview of \method{}.} \method{} conditions perturbation effects on regulatory DNA sequence to perform in-silico profiling of locus-specific perturbations at 128 bp model resolution within a 524 kb sequence context. Perturbations targeting the same gene at different genomic locations ($s_p^1$, $s_p^2$) can induce distinct transcriptional responses.}
    \label{fig:arch}
    \vspace{-10pt}
\end{figure}

The resolution gap is important because perturbations with the same gene label produce different cellular outcomes depending on the targeted regulatory sequence. Perturbation of the coding region of \textit{BCL11A} is lethal to cells, while perturbation of a specific enhancer of the same gene induces therapeutic effects and leaves the cell viable~\cite{zeng2025gene,frangoul2021crispr}. Within a single enhancer, a small subset of nucleotides controls gene regulation, and perturbations at nearby positions have no measurable effect~\cite{canver2015bcl11a}. These observations motivate the modeling of perturbations at sequence-level resolution. Sequence-level resolution also gives access to non-coding regulatory sequence, which constitutes the majority of the genome and contains many disease-associated genetic variants that gene-level perturbation models cannot represent~\cite{maurano2012systematic,nasser2021genome}.

Prediction of global transcriptomic responses from DNA sequence presents three difficulties. (1) {\em The input space is large.} Regulatory effects depend on hundreds of thousands of nucleotides, and the mapping from raw DNA sequence to transcriptomic change is highly non-linear~\cite{cheng2025dnalongbench,fu2023fast}. Recent DNA foundation models learn general regulatory grammar from sequence~\cite{linder2025predicting,avsec2025alphagenome,patel2024dart}, with training objectives defined over regulatory tracks instead of perturbation-induced state changes. Perturbation effects are therefore available only through post-hoc analyses within local sequence windows, and these analyses leave out the propagation of a perturbation through cell-type-specific regulatory programs~\cite{wei2025genome}. (2) {\em Regulatory effects are context-dependent.} The effect of a sequence motif depends on chromatin state, long-range interactions, and cellular context, all of which vary across cell types~\cite{song2025decoding}. A model of this dependence links sequence-level changes to transcriptional responses conditioned on cellular state. (3) {\em The space of sequence perturbations is combinatorial.} Available single-cell perturbation data are sparse across perturbations, samples, and contexts~\cite{Peidli2024,huang2025x}, which makes direct supervision from DNA sequence to perturbation response infeasible.

Existing perturbation predictors condition on proxy descriptors of the perturbed gene, including discrete covariates~\cite{bereket2023modelling,tu2024supervised,gaudelet2024season}, protein function~\cite{Adduri2025,dong2026stack}, gene regulatory structure~\cite{wu2022predicting,Roohani2023,wenkel2025txpert}, and text-derived embeddings~\cite{wu2025contextualizing,zhu2025scouter}. DNA foundation models learn sequence representations with objectives over regulatory tracks, which places global transcriptomic responses outside their prediction targets. Gene-level predictors interpolate within a fixed set of genes and define a closed perturbation vocabulary of gene identifiers. Generalization to unseen perturbations then depends on the transfer of gene-level correlations, which yields limited gains over simple baselines~\cite{wu2024perturbench}. Gene-identifier conditioning also leaves locus-defined perturbations outside the input space of these models (Table~\ref{tab:perturbation_resolution}). STRAND changes the perturbation variable itself. We condition prediction on regulatory DNA sequence and a perturbation mask at a specified genomic region, which replaces the closed vocabulary of gene identifiers with a sequence-defined input.

\xhdr{Present Work}
We introduce \method{}, a generative model that predicts single-cell
transcriptional responses from an unperturbed cell state and the regulatory DNA sequence at a target region (Figure~\ref{fig:arch}). A locus-anchored DNA perturbation module converts the targeted sequence context into a perturbation condition, and the model uses this condition to predict the distribution of perturbed cell states in the latent space of an RNA foundation model. Prior sequence-to-expression models predict a predefined set of regulatory tracks, whereas \method{} predicts global transcriptomic state changes from genomic sequence. Our contributions are as follows:
\begin{itemize}[leftmargin=5pt, labelsep=0.5em, topsep=0pt, partopsep=0pt, parsep=0pt, itemsep=2pt]

\item \textbf{Genome-sequence perturbation conditioning beyond gene identifiers.} \method{} expands architectural input support from protein-coding gene labels ($\sim$1.5\% of the genome) to 128~bp bins across $\sim$95\% of the genome (Table~\ref{tab:perturbation_resolution}), with the perturbation variable defined by regulatory DNA sequence and a perturbation mask at a user-specified locus instead of a closed vocabulary of annotated genes. This representation supports predictions outside the input space of gene-identifier models. \method{} ranks the functionally dominant \textit{FLVCR1b} alternative TSS above the canonical \textit{FLVCR1a} TSS in K562 (E-distance $0.683$ vs.\ $0.564$), resolves context-specific perturbation heterogeneity on a fine coordinate grid, and recovers a focal peak near the validated \textit{e-HDAC6} enhancer $\sim$13.65~kb downstream of the TSS, within $400$~bp of the CRISPRi-mapped center.

\item \textbf{Improved generalization from sequence-grounded perturbation representations.} Across CRISPRi benchmarks in K562, Jurkat, and RPE1, \method{} attains the best average rank on few-shot and unseen-gene benchmarks, with gains of up to $33\%$ in perturbation discrimination and up to $0.12$ $\Delta$Pearson in held-out-cell-line transfer over the next-best model. Matched ablations replace the learned perturbation module with fixed average-pooled DNA embeddings, which reduces top-DE-gene Pearson correlation by $0.07$--$0.10$ and direction match by $0.03$--$0.07$ and supports the contribution of learned sequence-grounded perturbation representations.
\end{itemize}

\section{Related Work} \label{sec:related_work}

\begin{table}[h]
\centering
\caption{\textbf{Perturbation-input scope.}
Gene-level perturbation predictors condition on gene or perturbation identifiers,
whereas \method{} conditions on regulatory DNA sequence at a specified locus.
Input scope denotes architectural input support, not empirical validation.}
\fontsize{8.5pt}{10pt}\selectfont
\vspace{-4pt}
\setlength{\tabcolsep}{5pt}
\begin{tabular}{llll}
\toprule
\textbf{Model class} & \textbf{Perturbation input} & \textbf{Response prediction} & \textbf{Input scope} \\
\midrule
Gene-level predictors
& Gene / label
& Yes
& Annotated genes ($\sim$1.5\%) \\
DNA FMs
& DNA sequence
& No$^\dagger$
& Regulatory tracks / local readouts \\
\textbf{\method{}}
& \textbf{Locus DNA sequence}
& \textbf{Yes}
& \textbf{128 bp bins; $\sim$95\% genome} \\
\bottomrule
\end{tabular}
\vspace{-4pt}
\begin{flushleft}
\fontsize{7.5pt}{8.5pt}\selectfont
$^\dagger$DNA foundation models predict regulatory signals but do not directly predict global single-cell perturbation-response.
\end{flushleft}
\vspace{-10pt}
\label{tab:perturbation_resolution}
\end{table}

\xhdr{Gene- and Metadata-Conditioned Perturbation Prediction}
Single-cell perturbation predictors differ in architecture, and most specify perturbations through gene identifiers, treatment labels, covariates, graph-derived metadata, or learned embeddings of these entities. Counterfactual models such as scGen and CPA learn latent spaces in which perturbations correspond to approximately additive or covariate-conditioned shifts, and related VAE formulations extend this view to sparse mechanism shifts~\cite{Lotfollahi2019,Lotfollahi2023,bereket2023modelling}. Graph- and relation-aware models such as GEARS, CellOracle, BioLord, and PertAdapt structure perturbation space with gene--gene, GO-derived, or regulatory-network metadata~\cite{Roohani2023,Kamimoto2023,Piran2024,bai2025pertadapt}. Recent virtual-cell models, including TxPert, STATE, and STACK, improve gene-level perturbation prediction across benchmarks~\cite{wenkel2025txpert,Adduri2025,dong2026stack}. Distributional formulations replace point prediction with optimal-transport, flow-matching, diffusion, or trajectory-based objectives, with conditioning on perturbation labels, treatments, genes, or covariates~\cite{Bunne2023,klein2025cellflow,He2025,Luo2024}.

Across these architectures, the perturbation vocabulary stays at the level of a gene, treatment, covariate, or related metadata label. Such labels support interpolation across observed or related entries, and they place unseen genomic loci and sub-gene regulatory contexts outside the input space of the model. Prior benchmarks report that added model complexity gives inconsistent improvements in the recovery of perturbation effects, with mode collapse and batch effects affecting the quality of the learned responses~\cite{wu2024perturbench,vinas2025systema,ahlmann2025deep,Luecken2021}. These results indicate a role for the perturbation input representation alongside the model architecture. \method{} changes this input representation from a gene-level identifier to regulatory DNA sequence at a user-specified genomic region, which gives the condition access to upstream regulatory context and removes the closed-vocabulary constraint.

Drug-response models condition on molecular graphs, dose--response assumptions, or molecular similarity instead of regulatory DNA context~\cite{szalata2024benchmark,liang2025scppdm,hetzel2022predicting,yu2025perturbnet,gonzalez2025combinatorial}. We focus our empirical comparisons on genetic perturbation methods and treat drug-response models as complementary approaches.

\xhdr{Single-Cell Foundation Models as Representation Backbones}
Single-cell foundation models such as Geneformer, scGPT, scFoundation, and SCimilarity~\cite{theodoris2023transfer,cui2024scgpt,hao2024large,heimberg2025cell}, together with text-based models such as PerturbQA, Scouter, and rbio1~\cite{wu2025contextualizing,zhu2025scouter,istrate2025rbio1}, learn transferable cell-state representations from large single-cell corpora. These models provide representation backbones for perturbation prediction, with the perturbation itself specified by a gene name, a perturbation identifier, or a textual description. \method{} uses an RNA foundation model for the representation of cell state and defines the perturbation condition by regulatory DNA sequence instead of a gene-level identifier.

\xhdr{DNA Sequence Models}
DNA foundation models trained on regulatory tracks learn sequence features predictive of chromatin accessibility, transcription factor binding, gene expression, and isoform usage~\cite{avsec2021effective,lal2024decoding,garau2024multi,linder2025predicting,hingerl2025flashzoi,avsec2025alphagenome}. These models are sequence-to-track predictors, and perturbation effects follow from post-hoc differencing of reference and edited sequence predictions across selected tracks~\cite{wei2025genome}, within fixed genomic windows and for the cellular contexts present during training. \method{} treats the DNA model as a regulatory knowledge base and learns an explicit interventional module that maps locus-specific sequence context and cellular state to genome-wide transcriptomic perturbation responses.

\begin{figure*}[t]
    \centering    \includegraphics[width=1.\linewidth]{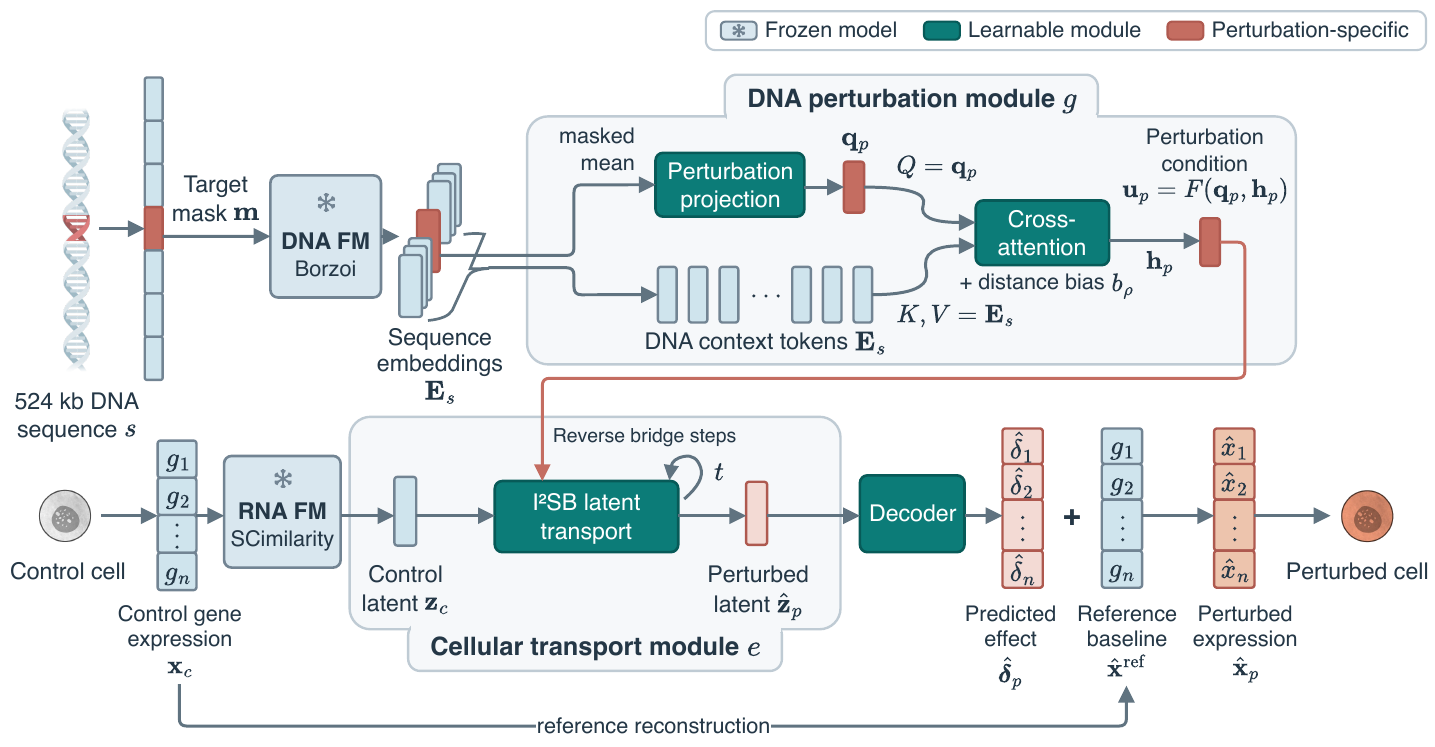}
    \vspace{-5pt}
    \caption{\textbf{\method{} architecture design.}
    The framework integrates genomic and transcriptomic modalities to predict perturbed cell expression.
The pipeline consists of three main learnable components:
(1) A \textbf{DNA Perturbation Module} $g$ that processes representations from a frozen Borzoi model and the perturbation mask $\mb$ to produce the perturbation condition $\ub_p$;
(2) A \textbf{Cellular Transport Module} $e$ utilizing a control-anchored I2SB bridge conditioned on $\ub_p$ to transport the control cell embedding $\zb_c$ to the perturbed cell embedding $\hat{\zb}_{p}$; and
(3) A \textbf{Gene Level Decoding Module} that maps $\hat{\zb}_{p}$ to a gene-level effect $\hat{\boldsymbol{\delta}}_p$, which is added to the reconstructed control reference $\hat{\xb}^{\mathrm{ref}}$ to give the predicted perturbed expression $\hat{\xb}_p$.
}
    \label{fig:arch_detail}
    \vspace{-10pt}
\end{figure*}

\section{\method{} Model} \label{sec:methods}
\label{sec:method}

\method{} predicts the distribution of perturbed single-cell expression profiles from an observed control population and the regulatory DNA sequence at a targeted locus. The design of \method{} follows from two modeling choices about the definition of a perturbation and the form of the perturbation response. We define a perturbation by its genomic coordinates and the surrounding regulatory sequence, so that perturbations at different loci of the same gene have distinct representations. We model the perturbation response as transport from control to perturbed cell states, which preserves cell-to-cell heterogeneity and makes the effect of each perturbation dependent on the basal state of the cell.

\xhdr{Architecture Overview}
\method{} consists of two frozen pretrained encoders and three learnable modules (Figure~\ref{fig:arch_detail}). The pretrained DNA encoder provides representations of the sequence over the targeted genomic context, and the pretrained RNA encoder provides a shared latent representation of cellular state. The DNA Perturbation Module uses the perturbation mask and the local sequence context to compute a perturbation condition. The Latent Cellular Transport Module learns a conditional map from an observed control state to a perturbed state given this condition, and a decoder predicts gene-level expression from the transported latent state. In this design, sequence information acts as a condition on control-anchored transport, so the same sequence-defined perturbation can have different effects across cellular states. We next define the problem setup and describe each module and the training objective.

\xhdr{Problem Setup}
Let $\xb \in \mathbb{R}^G$ denote the log-normalized gene expression profile of a single cell. We consider two empirical distributions over cell states: a control cell distribution $\xb_c \sim p_0(\xb)$ and a perturbed cell distribution $\xb_p \sim p_1(\xb \mid s, \mb)$, where $s \in \mathcal{A}^w$ denotes the local genomic sequence ($\mathcal{A}=\{A,C,G,T\}$, $w=524{,}288$\,bp) and $\mb \in \{0,1\}^L$ marks the perturbed positions (App.~\ref{sec:supp_tss_protocol}). The DNA encoder is Flashzoi~\cite{hingerl2025flashzoi}, a re-engineered Borzoi~\cite{linder2025predicting} backbone, which maps $s$ to $L = w/B = 4096$ token embeddings $\Eb_s \in \mathbb{R}^{L \times d_0}$ at $B=128$\,bp model resolution. The RNA encoder is SCimilarity~\cite{heimberg2025cell}, $e_\psi: \xb \mapsto \zb \in \mathbb{R}^d$ (so $\zb_c = e_\psi(\xb_c)$, $\zb_p = e_\psi(\xb_p)$). Both encoders are kept frozen and enter the rest of the pipeline only
through their output embeddings, so either can be replaced by an
alternative DNA or RNA foundation model without changes to the
downstream architecture.

Our goal is to learn a conditional transport model $\mathbf{z}_p \sim \mathcal{G}_\theta(\mathbf{z}_c, s, \mathbf{m})$ that transports a control latent $\mathbf{z}_c$ to a perturbed latent $\mathbf{z}_p$ conditioned on the perturbation parameters. Since control and perturbed cells are unpaired, we use a scalable Optimal Transport solver \citep{halmos2025hierarchical} returning a coupling $\pi(\mathbf{z}_c, \mathbf{z}_p)$ over (control, perturbed) pairs minimizing total transport cost; we sample $(\mathbf{z}_c, \mathbf{z}_p) \sim \pi$ for training. Robustness to solver choice and feature space is reported in App.~\ref{sec:ot_details}.

\xhdr{DNA Perturbation Module}
The DNA encoder provides token-level regulatory features not specialized to any particular perturbation
query. The DNA Perturbation Module $g_\phi$ converts the perturbation
mask $\mb$ into a sequence-derived condition $\ub_p = g_\phi(\Eb_s, \mb)$
for the latent transport. A useful condition should be
\emph{locus-anchored}, so that the representation reflects the targeted
positions, and \emph{context-aware}, because regulatory activity depends
on neighboring motifs and broader sequence context within the encoder
window.

Let $\mathcal{I}_p = \{i : m_i = 1\}$ denote the perturbed sequence-token
indices and $c_p = |\mathcal{I}_p|^{-1}\sum_{i \in \mathcal{I}_p} i$
their token-level center. We construct a locus-anchored query as:
\[
\qb_p
=
f_q\!\left(
\frac{1}{|\mathcal{I}_p|}\sum_{i \in \mathcal{I}_p} \Eb_{s,i}
\right),
\]
which then attends to the full sequence via single-query multi-head
cross-attention with a learned distance-based positional bias
$b_\rho(\cdot)$:
\[
\alpha_j
=
\mathrm{softmax}_j\!\left(
\frac{(\Wb_Q \qb_p)^\top (\Wb_K \Eb_{s,j})}{\sqrt{d_k}}
+ b_\rho(|j - c_p|)
\right),
\qquad
\hb_p
=
\sum_{j=1}^{L} \alpha_j\, \Wb_V \Eb_{s,j}.
\]
The final perturbation condition is
$\ub_p = F_{\mathrm{proj}}([\qb_p \,\|\, \hb_p])$, where $F_{\mathrm{proj}}$
is a two-layer MLP with LayerNorm and GELU. Thus $\qb_p$ anchors the
condition to the perturbed locus while $\hb_p$ retrieves the regulatory
context relevant to it. Multi-head attention and projection details are
deferred to App.~\ref{sec:supp_perturb_module}.

\xhdr{Latent Cellular Transport Module} We model perturbation response as distributional transport to preserve cell-state heterogeneity. We learn a stochastic transition mapping each control latent $\zb_c$ into an OT-paired perturbed latent $\zb_p$, anchored on observed control cells (rather than an unconditional noise prior); this is important in multi-cell-line settings where perturbation effects are defined relative to different basal states. We formulate this transport via an endpoint-conditioned bridge-matching objective \citep{liu_i2sb_2023}, with paired endpoints $(\mathbf{z}_c, \mathbf{z}_p)$ in the SCimilarity latent. The stochastic interpolant between the paired latents is:
\begin{equation}
\begin{aligned}
\zb_t & = w_0(t)\zb_p + w_1(t)\zb_c + \sigma(t)\pmb{\varepsilon}, \\
 \pmb{\varepsilon} & \sim \mathcal{N}(0, \Ib), \;t \in \{0,\dots,T\}.
\end{aligned}
\end{equation}
where $w_0(t), w_1(t)$ and $\sigma(t)$ are determined by the I2SB noise schedule
via $\sigma_t^2$ and $\bar\sigma_t^2$ (see App.~\ref{sec:supp_theory}).
We learn a conditional noise prediction network $e_\theta(\zb_t, t, \ub_p, \zb_c)$ trained via Bridge Matching:
\[
\mathcal L_{\text{bridge}}
=
\mathbb E_{\mathbf{z}_p,\mathbf{z}_c,t}
\Big[
\big\|
e_\theta(\mathbf{z}_t,t,\mathbf{u}_p,\mathbf{z}_c)
-
\boldsymbol{\varepsilon}^{\text{I2SB}}
\big\|_2^2
\Big],
\]
with \(
\boldsymbol{\varepsilon}^{\text{I2SB}}
\;\doteq\;
\frac{\mathbf{z}_t - \mathbf{z}_p}{\sigma_t}.
\)
This training objective follows I2SB~\cite{liu_i2sb_2023},
with full derivations in App.~\ref{sec:supp_theory}. This formulation enables
\method{} to generate perturbed states by transporting control states under
sequence-conditioned perturbation semantics. $e_\theta$ learns a guided diffusion process conditioned jointly on the control state $\mathbf{z}_c$ and the sequence embedding $\mathbf{u}_p$, such that the influence of a given DNA perturbation is modulated by the baseline cellular state, yielding cell-type-specific transition dynamics with detailed configurations at App.~\ref{sec:supp_default_config}.

\xhdr{RNA Expression Decoding}
We decode the transported latent state with a residual hurdle decoder. A lightweight
autoencoder first forms a denoised control baseline
$\hat{\xb}^{\mathrm{ref}}_i=d_\eta(e_\eta(\xb_{c,i}))$. Given the transported latent
$\hat{\zb}_{p,i}$, the decoder predicts, for each gene $j$, a nonzero probability
$p^{\mathrm{nz}}_{ij}$ and truncated-Gaussian residual parameters
$(\mu_{ij},\sigma_{ij})$ relative to $\hat{x}^{\mathrm{ref}}_{ij}$. The
reconstruction objective is the negative log-likelihood of this hurdle
truncated-Gaussian residual model. At inference, all reported point predictions
use the corresponding hurdle predictive mean value (App.~\ref{eq:guassian-loss}).

\xhdr{Sequence--Effect Alignment}
As an auxiliary objective, we encourage each perturbation's DNA embedding
to be predictive of its average transcriptional effect. For perturbation
$k$, let $\mathcal{B}_k$ index the OT-matched cell pairs in a minibatch. We compute the perturbation-level averages
\(
\bar{\mathbf{u}}_k = \frac{1}{|\mathcal{B}_k|}\sum_{i\in\mathcal{B}_k}\mathbf{u}_{p,i},
\)
\(
\Delta\bar{\mathbf{z}}_k = \frac{1}{|\mathcal{B}_k|}\sum_{i\in\mathcal{B}_k}
\bigl(\mathbf{z}_{p,i} - \mathbf{z}_{c,i}\bigr),
\)
where $\Delta\bar{\mathbf{z}}_k$ is the mean control-to-perturbed shift in
the SCimilarity latent. Both quantities are stabilized by an exponential
moving average across minibatches. A symmetric InfoNCE loss
\citep{radford2021learning} on
$\bigl(\mathrm{Proj}_{\text{DNA}}(\bar{\mathbf{u}}_k),\,
\Delta\bar{\mathbf{z}}_k\bigr)$ pulls matching DNA--RNA pairs together
and pushes non-matching ones apart, yielding a contrastive estimate of
compatibility between regulatory DNA context and average transcriptional
effect. Implementation and ablations: App.~\ref{sec:supp_align}.

\xhdr{Overall \method{} Objective}
\label{sec:overall-loss}
The model is trained end-to-end by minimizing a weighted sum of the bridge matching, reconstruction, and alignment losses as:
    $\mathcal{L} = \lambda_1 \mathcal{L}_{\text{bridge}} + \lambda_2 \mathcal{L}_{\text{CLIP}} + \lambda_3 \mathcal{L}_{\text{rec}}.$
We employ uncertainty weighting~\cite{kendall2018multi} to dynamically adjust $\lambda_i$ during training.

\section{Experimental Setup}
\label{sec:setup}
\xhdr{Data Source and Preprocessing}
We use single-cell CRISPRi perturbation screens curated in PerturbQA~\cite{wu2025contextualizing}, which aggregates screens from two independent source studies: the K562 and RPE1 screens of \citet{replogle2022mapping} (single-sgRNA CRISPRi, 10x Genomics) and the Jurkat and HepG2 screens of \citet{nadig2025transcriptome} (dual-sgRNA CRISPRi, 10x Genomics). We restrict our analysis to K562, Jurkat, and RPE1, excluding HepG2 due to low statistical power for differential expression evaluation arising from reduced cell counts per perturbation (median 45 cells in HepG2 versus 72 in RPE1, 83 in Jurkat, and 121 in K562~\cite{nadig2025transcriptome}). We further constructed a cross-cell-line dataset (termed \textit{Combined}) using the three cell lines, with details at App.~\ref{sec:combined-data-description}.
For \method{}, we couple control and perturbed cells using Optimal Transport (OT), with details in App.~\ref{sec:ot_details}.

\xhdr{Transfer Design}
Beyond within-cell-line evaluation, we assess generalization with a \emph{cross-study, leave-one-context-out} transfer design: in each fold, one cell line is held out entirely, and the model is trained only on the remaining two screens. Holding out K562 or RPE1 tests transfer across cell lines within the \citet{replogle2022mapping} study, while holding out Jurkat tests transfer to a screen from an independent study (\citet{nadig2025transcriptome}) with a different sgRNA design; in this fold, the target source study is fully unseen during training. For an external held-out context, we train a single model on all three screens and test it on HCT116 from X-Atlas/Orion~\cite{huang2025x}, a third study that uses dual-sgRNA CRISPRi with the FiCS fix--cryopreserve Perturb-seq protocol (App.~\ref{sec:supp_hct116}). In all held-out contexts, no perturbation profiles, labels, or target-context fine-tuning from the held-out screen are used.

\xhdr{Existing Models}
Performance varies across perturbation prediction benchmarks~\cite{vinas2025systema, ahlmann2025deep, wu2024perturbench}. We therefore compare \method{} with five representative gene-level baselines: \emph{GEARS}~\cite{Roohani2023}, \emph{BioLord}~\cite{Piran2024}, \emph{STATE}~\cite{Adduri2025}, \emph{Linear}~\cite{linder2025predicting}, and \emph{PerturbMean}~\cite{vinas2025systema}. For the hold-out cell line setting, STACK, whose in-context formulation
is directly aligned with the setting, is further included.  Implementation details are provided in App.~\ref{sec:supp_baselines}.

\xhdr{Evaluation Metrics}
We adopt evaluation protocols tailored to each experimental setting.
For models trained and evaluated on individual cell lines, we adhere to the benchmarking standards established by the Virtual Cell Challenge and the STATE framework~\cite{Adduri2025}. For each test perturbation, we sample $N=256$ random control cells from the corresponding cell line. Each model generates the predicted perturbed population $\hat{p}_{1}$,  with the same sample size, which is compared against the ground-truth distribution.
Following the CellEval framework~\cite{Adduri2025}, we evaluate performance along three complementary axes:
(i) \textbf{Global Fit} (MSE, $\Delta$Pearson),
(ii) \textbf{DE Signal} (Dir.\ Match, Prec.@N, Ov.@N), and
(iii) \textbf{Distributional Fidelity} (Discr.\ Cos~\cite{wu2024perturbench}).
We exclude metrics that treat Wilcoxon-based DE calls as hard ground-truth labels (e.g., DE classification or recall-based metrics), which have been shown to produce severely inflated false discoveries ~\cite{squair2021confronting,murphy2023avoiding}.

For joint-cell-line (\textit{Combined}), we additionally report metrics assessing the distributional fidelity and population-level accuracy following~\cite{klein2025cellflow} on settings not natively supported by CellEval.
We evaluate \textbf{Energy Distance} and \textbf{pseudo-bulk $R^2$} following a similar protocol as in~\cite{klein2025cellflow}.  Detailed metric definitions are provided in App.~\ref{sec:supp_metrics}.
\section{Results} \label{sec:results}

We evaluate \method{} at two levels of resolution, first on gene-level perturbation benchmarks and then on locus-level analyses of regulatory sequence. At the gene level, we compare \method{} with gene-level perturbation predictors on few-shot prediction, prediction for unseen perturbation targets, and transfer to held-out cell lines and an independent study (Section~\ref{sec:results_gene_level}). Ablations of the DNA Perturbation Module and the latent transport backbone then quantify the contributions of sequence conditioning and conditional transport to this performance (Section~\ref{sec:ablation}). At the locus level, we use in silico perturbation scans to assess the ability of \method{} to distinguish alternative transcription start sites, recover cell-type-specific regulatory activity at lineage-defining genes, and localize a validated distal enhancer in non-coding sequence (Section~\ref{sec:tss-analysis}). These analyses address predictions outside the input support of gene-identifier models, and we evaluate them against independent experimental evidence from prior CRISPRi and functional studies.

\subsection{Gene-Level Perturbation Prediction}
\label{sec:results_gene_level}

\begin{figure}[t]
    \centering
    \includegraphics[width=0.6\linewidth]{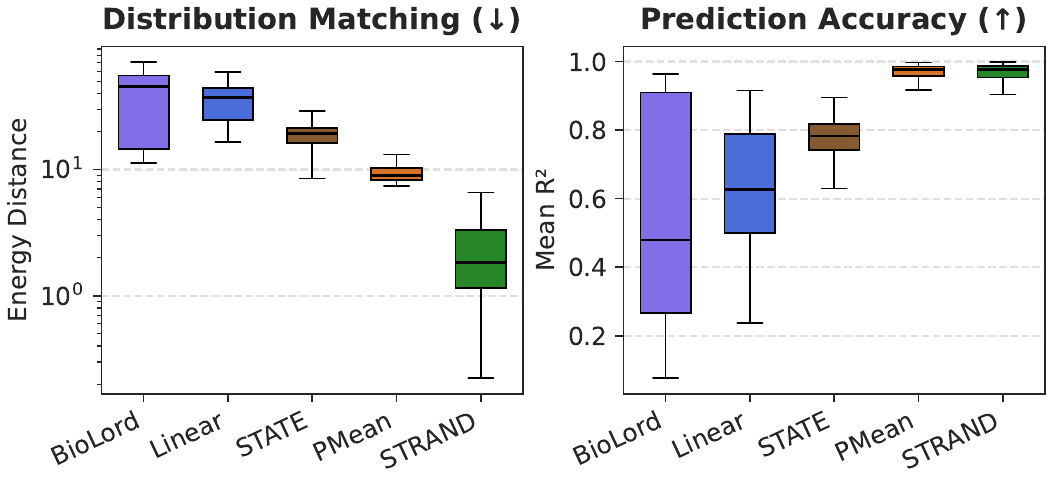}
    \vspace{-10pt}
\caption{
\textbf{Comparison to gene-level baselines}. \method{} is compared against {PMean} (PerturbMean), STATE, Linear, and Biolord on the \textit{Combined} data in few-shot setting. GEARS is excluded as it doesn't support cross-cell-line joint training.
}
\label{fig3:dist-bench}
    \vspace{-10pt}
\end{figure}

\xhdr{Few-Shot  Gene Perturbations}
To model realistic data scarcity, we move $k=5$ cells per test gene into training. Models are trained independently per cell line and evaluated on held-out test cells (App.~\ref{sec:supp_data_stats}). \method{} achieves the best average rank ($1.57$–$1.86$) across all metrics and cell lines (Table~\ref{tab:agg-all-tables}, ~\ref{tab:fewshot_results_scaled}). On \textit{Combined} (Fig.~\ref{fig3:dist-bench}), STRAND reduces E-distance by $\sim$ 4$\times$ versus the next-best method.

Simple baselines (Linear, PerturbMean) achieve low MSE but poor discrimination ($\approx 0.5$); STATE produces high-variance predictions (lower bias on individual cells but worse global fit, K562 MSE $5.36$ vs.\ \method{}'s $0.80$), reflecting a bias-variance tradeoff between distributional fidelity and pointwise reconstruction. \method{} matches simple baselines' MSE on K562 while improving Discr.\ Cos from STATE's $0.54$ to $0.71$ (a $\sim 32\%$ relative improvement; per-cell-line in App.~Table~\ref{tab:fewshot_results_scaled}).

\begin{table*}[t]
\centering
\scriptsize
\setlength{\tabcolsep}{3pt}
\caption{\textbf{Gene perturbation performance comparison.}
Single cell-line evaluation: models are trained and evaluated \emph{independently} on each cell line (K562, Jurkat, RPE1) and metrics averaged across cell lines under few-shot ($k{=}5$) and zero-shot settings. Per–cell-line results in Tables~\ref{stab:low-sample},~\ref{stab:zero-shot},~\ref{stab:zero-shot-subset}. Zero-shot with  GEARS-supported perturbation subset in Table~\ref{stab:zero-shot-subset}.
Dir.\ Match, Prec.@N, Ov.@N, and Discr.\ Cos are scaled by $\times 10$; MSE by $\times 100$. E-Dist and $\Delta$Pearson are unscaled.
}\resizebox{\textwidth}{!}{%
\begin{tabular}{lcccccccc|c}
\toprule
 &  &
 \multicolumn{2}{c}{\textbf{Distributional Fidelity}} &
 \multicolumn{3}{c}{\textbf{DE Signal}} &
 \multicolumn{2}{c}{\textbf{Global Fit}} &
 \textbf{Summary} \\
\cmidrule(lr){3-4}
\cmidrule(lr){5-7}
\cmidrule(lr){8-9}
\cmidrule(lr){10-10}
\textbf{Task} & \textbf{Method} &
\textbf{Discr. Cos $\uparrow$} &
\textbf{E-Dist $\downarrow$} &
\textbf{Dir. Match $\uparrow$} &
\textbf{Prec.@N $\uparrow$} &
\textbf{Ov.@N $\uparrow$} &
\textbf{$\Delta$Pearson $\uparrow$} &
\textbf{MSE $\downarrow$} &
\textbf{Avg. Rank $\downarrow$} \\
\midrule
\multirow{6}{*}{\begin{tabular}{@{}l@{}}Few-shot\\ \end{tabular}}
& BioLord
& 5.04 $\pm$ 0.09 & 9.41 $\pm$ 0.11 & 6.10 $\pm$ 0.04 & 0.72 $\pm$ 0.02 & 0.40 $\pm$ 0.02 & 0.16 $\pm$ 0.00 & 1.38 $\pm$ 0.04 & 4.71 \\
& GEARS
& 5.40 $\pm$ 0.10 & 1.64 $\pm$ 0.06 & 7.05 $\pm$ 0.05 & 0.71 $\pm$ 0.02 & 0.77 $\pm$ 0.03 & 0.31 $\pm$ 0.01 & 1.19 $\pm$ 0.03 & 3.14 \\
& Linear
& 5.04 $\pm$ 0.09 & 8.90 $\pm$ 0.08 & 3.91 $\pm$ 0.03 & 1.00 $\pm$ 0.04 & 1.03 $\pm$ 0.03 & 0.19 $\pm$ 0.01 & 1.11 $\pm$ 0.03 & 3.29 \\
& PerturbMean
& 5.02 $\pm$ 0.09 & 8.91 $\pm$ 0.08 & 8.43 $\pm$ 0.03 & 0.22 $\pm$ 0.01 & 0.58 $\pm$ 0.02 & 0.52 $\pm$ 0.01 & 1.03 $\pm$ 0.03 & 3.29 \\
& STATE
& 6.40 $\pm$ 0.09 & 12.46 $\pm$ 0.11 & 5.75 $\pm$ 0.06 & 0.74 $\pm$ 0.02 & 0.38 $\pm$ 0.02 & 0.13 $\pm$ 0.01 & 5.06 $\pm$ 0.09 & 4.71 \\
\rowcolor{gray!10}
& STRAND
& 7.10 $\pm$ 0.09 & 2.07 $\pm$ 0.05 & 7.69 $\pm$ 0.05 & 0.74 $\pm$ 0.02 & 1.21 $\pm$ 0.04 & 0.39 $\pm$ 0.01 & 1.04 $\pm$ 0.03 & \textbf{1.86} \\
\cmidrule(lr){1-10}

\multirow{5}{*}{Zero-shot}
& BioLord
& 5.05 $\pm$ 0.09 & 9.44 $\pm$ 0.11 & 6.15 $\pm$ 0.04 & 0.77 $\pm$ 0.02 & 0.45 $\pm$ 0.02 & 0.15 $\pm$ 0.00 & 1.30 $\pm$ 0.04 & 3.43 \\
& Linear
& 5.03 $\pm$ 0.09 & 8.91 $\pm$ 0.08 & 3.83 $\pm$ 0.03 & 1.00 $\pm$ 0.04 & 1.02 $\pm$ 0.03 & 0.20 $\pm$ 0.01 & 1.04 $\pm$ 0.03 & 2.43 \\
& PerturbMean
& 5.02 $\pm$ 0.09 & 8.92 $\pm$ 0.08 & 8.43 $\pm$ 0.03 & 0.22 $\pm$ 0.01 & 0.58 $\pm$ 0.02 & 0.52 $\pm$ 0.01 & 1.03 $\pm$ 0.03 & 2.43 \\
& STATE &
\multicolumn{8}{c}{Not Applicable} \\
\rowcolor{gray!10}
& STRAND
& 5.17 $\pm$ 0.09 & 2.38 $\pm$ 0.07 & 7.25 $\pm$ 0.05 & 0.78 $\pm$ 0.02 & 1.15 $\pm$ 0.04 & 0.34 $\pm$ 0.01 & 1.10 $\pm$ 0.03 & \textbf{1.71} \\
\bottomrule
\end{tabular}
}
\label{tab:agg-all-tables}
\end{table*}

\xhdr{Unseen Gene Perturbations} We next evaluate the generalization of \method{} to perturbations absent from training. We adopt the train-test splits of PerturbQA~\cite{wu2025contextualizing}, where the training set covers a designated set of perturbations and the test perturbations have no labels or expression profiles during training (App.~\ref{sec:supp_data_stats}). We compare \method{} with methods that support unseen-gene prediction in this setting~\cite{Adduri2025}. \method{} achieves the best overall rank among the evaluated methods (Tables~\ref{tab:agg-all-tables} and~\ref{tab:zeroshot_gene_results_scaled}), with the highest Discrimination Score in every cell line. This metric measures the similarity of each predicted perturbation effect to its own ground-truth effect relative to all other perturbations, and separates models that produce perturbation-specific responses from those that collapse to a shared mean response~\cite{wu2024perturbench}.

\xhdr{Unseen Cell-Line Perturbations}
We test cross-context transfer on the \textit{Combined} dataset with leave-one-cell-line-out evaluation, training on two cell lines and testing on the 300 most prevalent perturbations of the third. In this out-of-distribution setting, \method{} is the only method whose top-$K$ effect correlation and direction-match curves stay above Identity, STATE, and STACK across all three held-out cell lines (Fig.~\ref{fig:supp_cross_cellline}). At the top-100 DE genes, \method{} attains an average direction match of $0.587$, above the $0.50$ chance level and above STACK ($0.500$) and STATE ($0.481$). Pearson gains concentrate on the genes with the strongest responses and decrease as $K$ grows. On HCT116, from an independent study and protocol, the model trained on all three cell lines reaches a direction match of $0.568$ at the top-100 DE genes, above STATE ($0.522$) and STACK ($0.490$), and this result uses no retraining and no HCT116 perturbation data (Fig.~\ref{fig:supp_cross_cellline}, bottom row; App.~\ref{sec:supp_hct116}).

\subsection{Ablation Studies}
\label{sec:ablation}

We ablate the \moduleDNA{} and the \moduleRNA{} to quantify their contributions, and Table~\ref{stab:clip-ablation} reports an ablation of the sequence-expression alignment loss.

\begin{figure}[h]
    \centering
\begin{minipage}[t]{0.49\linewidth}
  \vspace{0pt}
  \centering
  \includegraphics[width=\linewidth]{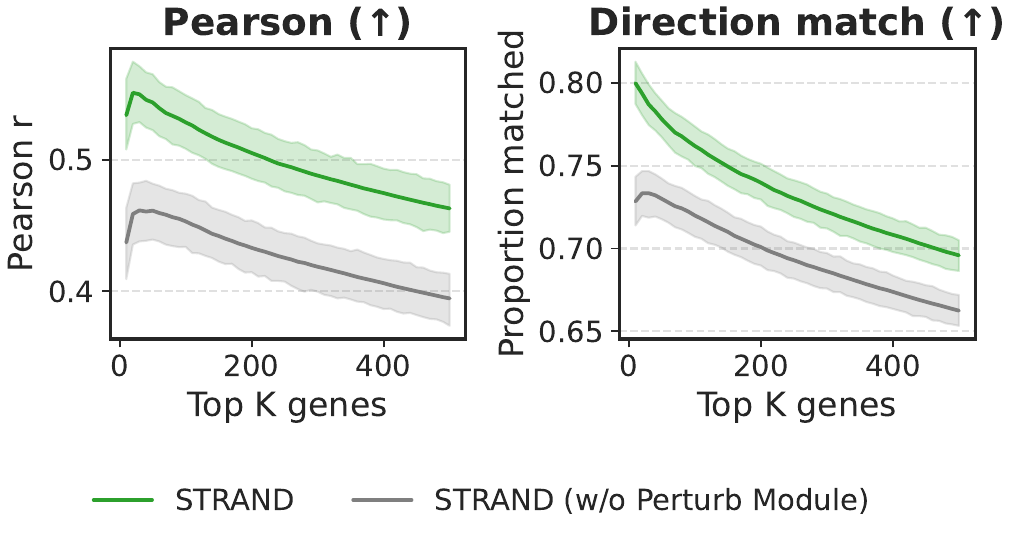}
\end{minipage}
\hfill
\begin{minipage}[t]{0.5\linewidth}
  \vspace{0pt}
  \centering
  \includegraphics[width=\linewidth]{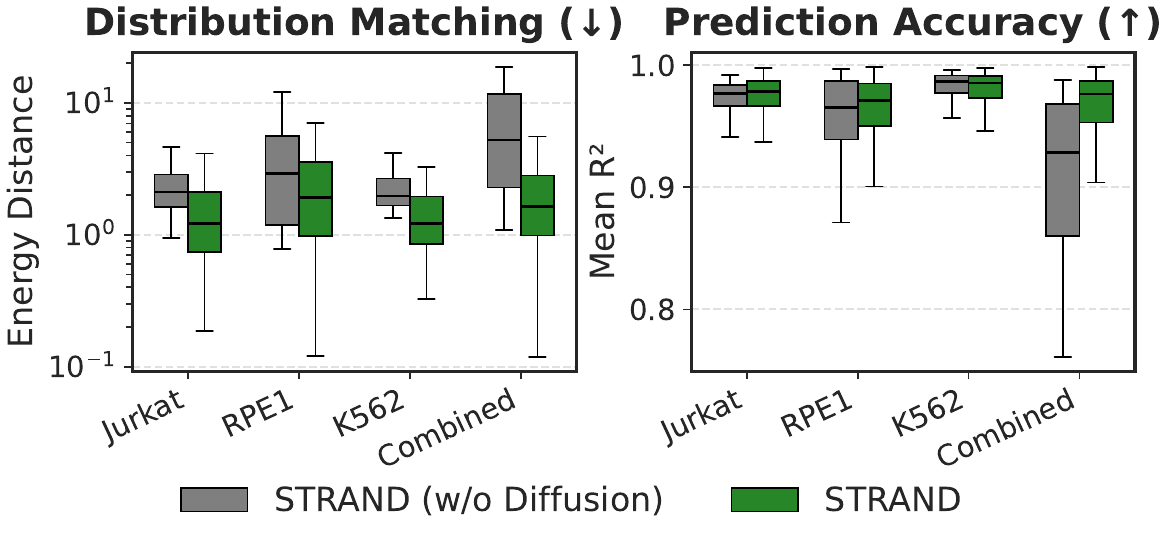}
\end{minipage}
    \vspace{-10pt}
    \caption{
    \textbf{Ablation of STRAND components.}
    \textbf{Left:} Ablation of the STRAND DNA Perturbation Module. Models are trained on the \textit{Combined} dataset. Plots show statistics for the top-$K$ genes with the largest absolute expression changes. Shaded regions denote 95\% CIs.
    \textbf{Right:} Ablation of the \method{} Cellular Transport Module. The generative model demonstrates better fidelity in modeling the perturbation distribution. More comprehensive comparison in Table~\ref{tab:perturbation_latent_diffusion_ablation_se}.
    }
    \label{fig:strand-ablation}
    \vspace{-10pt}
\end{figure}

\xhdr{\moduleDNA{}}
We hypothesize that co-training with the perturbation module improves the model's ability to retrieve the finer interventional relationship between regulatory activation and transcriptional response from the pretrained DNA embeddings. To verify this, we compared our full model against a baseline where the perturbation module is replaced by average-pooled, fixed DNA embeddings (termed \textbf{w/o Perturb Module}). Using the few-shot setting, we computed the Pearson correlation and proportion of directional match between the predicted and ground-truth expression changes $\Delta ({\xb}_p - {\xb}_c)$  for the top-$K$ most differentially expressed genes. Figure~\ref{fig:strand-ablation} illustrates the performance difference across $K$. At $K{=}100$, the module-based model shows an average of $16.6\%$ improvement in Pearson correlation and $5.8\%$ improvement in effect direction match over the fixed baseline (averaged across K562, Jurkat, and RPE1). This confirms that learning perturbation-specific embeddings is important for prioritizing significant biological effects.

\xhdr{Latent Transport Backbone}
We ablate STRAND's latent transport backbone in two steps. First, to test
whether distributional transport is needed beyond deterministic feature
fusion, we replace the stochastic transport module with a deterministic
MLP predictor on $[\mathbf{z}_c; \mathbf{z}^{\mathrm{DNA}}]$, holding all
other components and losses fixed. Relative to this predictive-fusion
baseline, STRAND reduces E-Distance by $38.9\%$ and increases $R^2$ by
$27.6\%$ on the combined dataset (Figure~\ref{fig:strand-ablation}).
Second, holding the OT coupling, DNA perturbation module, decoder, and
data splits fixed, we replace the $\mathrm{I}^2$SB bridge objective with
flow matching. Across Jurkat, K562, and RPE1, bridge matching improves $18$ of
$21$ metric--cell-line comparisons and reduces mean MSE from $0.0145$ to
$0.0099$; full results in App.~\ref{sec:supp_bridge_ablation}.

\subsection{Position-Resolved Validation Analysis}\label{sec:tss-analysis}
\method{} conditions directly on DNA sequence, enabling virtual perturbation profiling at user-specified genomic coordinates and distinguishing alternative regulatory elements within a gene. The validation qualitatively shows predictions that gene-level baselines structurally cannot make. We validate this by (i) prioritizing alternative TSSs with known functional differences, (ii) recovering cell-type-specific regulatory landscapes at lineage-defining genes, and (iii) extrapolating to non-coding regions of the genome.

For inference-time tractability, we slide a $\pm 200$\,bp perturbation mask in 100\,bp steps (50\,bp around e-HDAC6) and compute E-distance between predicted perturbed and control expression.  Finer-grid scanning is feasible with increased inference time. We use the same perturbation-coordinate protocol as in the benchmark; inference details are provided in App.~\ref{sec:supp_default_config}.

\xhdr{Alternative TSS Discovery}
Genes initiate transcription at transcription start sites (TSSs). When multiple TSSs exist for a gene, perturbation designs and gene-level models typically focus on the canonical site, even though alternative TSSs can be functionally dominant.

We examined \textit{FLVCR1} in K562 cells, which uses alternative TSSs to produce two distinct isoforms. \textit{FLVCR1a}, transcribed from the canonical TSS, localizes to the plasma membrane and functions as a heme exporter. In contrast, \textit{FLVCR1b} is transcribed from an alternative TSS within the first intron and produces a mitochondrial heme exporter required for erythroid differentiation~\cite{chiabrando2012mitochondrial}. Experimental studies show that \textit{FLVCR1b}, rather than \textit{FLVCR1a}, drives hemoglobinization and erythroid differentiation in K562 cells. Silencing \textit{FLVCR1b} blocks differentiation~\cite{chiabrando2012mitochondrial}.

Our \emph{in silico} perturbations ranked the \textit{FLVCR1b} TSS (FLVCR1-202, chr1:212,858,916) highest among all of the five \textit{FLVCR1} transcripts (E-dist $=0.683$), exceeding the canonical \textit{FLVCR1a} TSS (E-dist $=0.564$; Table~\ref{stab:TSS-screen}). The functional importance of the alternative \textit{FLVCR1b} promoter in K562 cells is a distinction that gene-level models (GEARS, BioLord, STATE) cannot capture.

\begin{figure}[h!t]
\vspace{-8pt}
    \centering    \includegraphics[width=\linewidth]{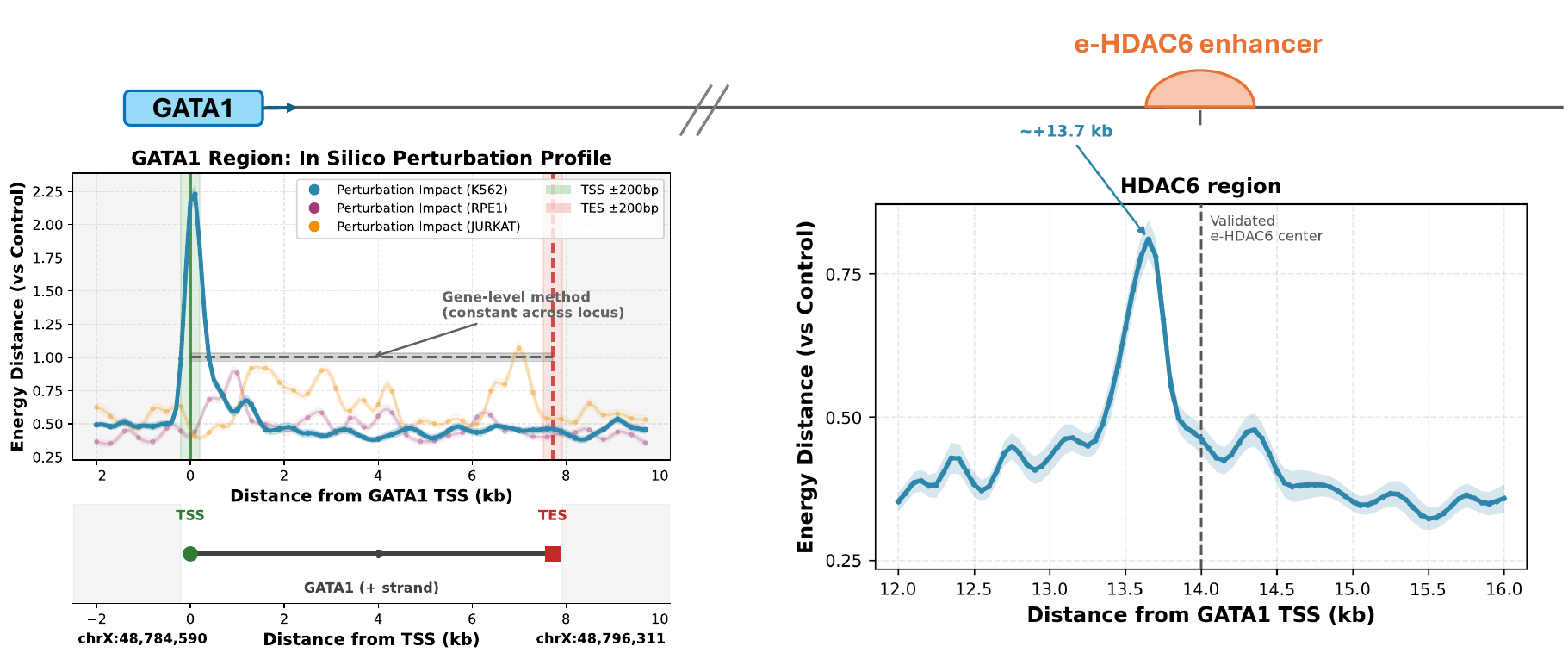}
    \vspace{-15pt}
\caption{\textbf{Virtual perturbation screening at the GATA1/HDAC6 regulatory region.} \method{} performs an in silico scan with a sliding $\pm 200$\,bp mask (100\,bp steps; 50\,bp in the right panel),
revealing cell-type-specific sensitivity at the \textbf{K562} promoter
(left) and localizing a sharp peak near the well-established e-HDAC6
enhancer~\cite{fulco2016systematic} $\sim$13.7\,kb downstream in non-coding
region (right). Gene-level methods (\textbf{gray}) yield a constant
signal across the gene body and cannot extrapolate to non-coding regions.}
\label{fig:GATA1-k562}
    \vspace{-10pt}
\end{figure}

\xhdr{In Silico Screening}
We examine three representative gene–cell-line pairs with well-characterized cell-type-specific functions.
\textit{GATA1 in K562 cells:} GATA1 is a master transcription factor essential for erythroid differentiation, and K562 is an erythroleukemia cell line where GATA1 plays a critical regulatory role \cite{moriguchi2014regulatory}. \method{} predicts a cell-type-specific response: K562 shows a sharp peak in perturbation sensitivity at the TSS, while RPE1 and Jurkat cells exhibit minimal response throughout the locus (Figure~\ref{fig:GATA1-k562}). This specificity aligns with GATA1's known lineage-restricted function.
Similarly, we demonstrate \method{}’s ability to identify two additional cell-line–specific gene activities, CD3D in Jurkat (Fig.~\ref{sfig:jurkat-cd3d}) and KRT18 in RPE1 (Fig.~\ref{sfig:rpe-krt18}), at the correct sequence locations. Detailed documentation is provided in App.~\ref{sec:supp_case_study}.

\xhdr{Distal Non-Coding Regulatory Evidence}
Sliding-window scans also recover a secondary peak $\sim$13.65\,kb downstream of the \textit{GATA1} TSS in K562, within $400$\,bp of the validated \textit{e-HDAC6} enhancer center reported in CRISPRi screens (Fig.~\ref{fig:GATA1-k562}; App.~\ref{sec:supp_hdac6}). The peak reaches an E-distance of $0.811$, approximately twice the background level across the scanned window.

\section{Conclusion} \label{sec:conclusion}

Gene regulation operates on DNA sequence, and experimental perturbation technologies intervene at defined genomic coordinates, so perturbations at different loci of the same gene produce different transcriptional responses. Gene-level perturbation models represent these interventions with a single identifier, which separates the resolution of the model from the resolution of the experiment. \method{} closes this separation by using regulatory DNA sequence as the conditioning variable for single-cell perturbation transport. We train a perturbation adaptor over pretrained DNA foundation-model representations jointly with a control-anchored bridge-matching objective, which converts sequence context at target loci into perturbation embeddings for generative transport from control to perturbed cell states. The resulting model extrapolates to unseen sequence-defined perturbations and performs position-resolved inference at 128~bp model resolution under available CRISPRi annotations, over an input space larger than the one available to gene-identifier conditioning (Table~\ref{tab:perturbation_resolution}). App.~\ref{sec:limitations} discusses limitations on sequence resolution and context generalization.

\ifblind\else
{\small \begin{ack}
 We thank Ruthie Johnson, Nanxiang Zhao, Yepeng Huang, and Intae Moon for their helpful feedback. We thank Peter Halmos for his feedback on experimental design.
B.F., X.L., and M.Z. gratefully acknowledge the support of the Merck Award. Sa.G. was supported by the Dr. Susanne E. Churchill Summer Institute in Biomedical Informatics (SIBMI) at Harvard Medical School.
B.F., Sa.G., X.L., Sh.G., X.S., and M.Z. gratefully acknowledge support, in part, by NSF CAREER 2339524, Harvard Data Science Initiative, Amazon Faculty Research, Google Research Scholar Program, AstraZeneca Research, Roche Alliance with Distinguished Scientists (ROADS) Program, Sanofi iDEA-iTECH Award, GlaxoSmithKline Award, Boehringer Ingelheim Award, Merck Award, Optum AI Research Collaboration Award, Pfizer Research, Gates Foundation (INV-079038), Chan Zuckerberg Initiative, John and Virginia Kaneb Fellowship at Harvard Medical School, Biswas Computational Biology Initiative in partnership with the Milken Institute, Harvard Medical School Dean’s Innovation Fund for the Use of Artificial Intelligence, and the Kempner Institute for the Study of Natural and Artificial Intelligence at Harvard University. This work was delivered as part of the Aurora project supported by the Cancer Grand Challenges partnership funded by Cancer Research UK (CGCAI1-Mar26/100001).
Any opinions, findings, conclusions or recommendations expressed in this material are those of the authors and do not necessarily reflect the views of the funders.
Figures~\ref{fig:arch} and~\ref{fig:arch_detail} were created with BioRender.com.
All authors declare no competing interests.
\end{ack}}
\fi
\clearpage

\bibliographystyle{plainnat}
\bibliography{refs}

@article{Peidli2024,
  title = {scPerturb: harmonized single-cell perturbation data},
  volume = {21},
  ISSN = {1548-7105},
  url = {http://dx.doi.org/10.1038/s41592-023-02144-y},
  DOI = {10.1038/s41592-023-02144-y},
  number = {3},
  journal = {Nature Methods},
  publisher = {Springer Science and Business Media LLC},
  author = {Peidli,  Stefan and Green,  Tessa D. and Shen,  Ciyue and Gross,  Torsten and Min,  Joseph and Garda,  Samuele and Yuan,  Bo and Schumacher,  Linus J. and Taylor-King,  Jake P. and Marks,  Debora S. and Luna,  Augustin and Bl\"{u}thgen,  Nils and Sander,  Chris},
  year = {2024},
  month = jan,
  pages = {531–540}
}

@article{Lotfollahi2023,
  title = {Predicting cellular responses to complex perturbations in high‐throughput screens},
  volume = {19},
  ISSN = {1744-4292},
  url = {http://dx.doi.org/10.15252/msb.202211517},
  DOI = {10.15252/msb.202211517},
  number = {6},
  journal = {Molecular Systems Biology},
  publisher = {Springer Science and Business Media LLC},
  author = {Lotfollahi,  Mohammad and Klimovskaia Susmelj,  Anna and De Donno,  Carlo and Hetzel,  Leon and Ji,  Yuge and Ibarra,  Ignacio L and Srivatsan,  Sanjay R and Naghipourfar,  Mohsen and Daza,  Riza M and Martin,  Beth and Shendure,  Jay and McFaline‐Figueroa,  Jose L and Boyeau,  Pierre and Wolf,  F Alexander and Yakubova,  Nafissa and G\"{u}nnemann,  Stephan and Trapnell,  Cole and Lopez‐Paz,  David and Theis,  Fabian J},
  year = {2023},
  month = may 
}

@article{Roohani2023,
  title = {Predicting transcriptional outcomes of novel multigene perturbations with GEARS},
  volume = {42},
  ISSN = {1546-1696},
  url = {http://dx.doi.org/10.1038/s41587-023-01905-6},
  DOI = {10.1038/s41587-023-01905-6},
  number = {6},
  journal = {Nature Biotechnology},
  publisher = {Springer Science and Business Media LLC},
  author = {Roohani,  Yusuf and Huang,  Kexin and Leskovec,  Jure},
  year = {2023},
  month = aug,
  pages = {927–935}
}

@article{Kamimoto2023,
  title = {Dissecting cell identity via network inference and in silico gene perturbation},
  volume = {614},
  ISSN = {1476-4687},
  url = {http://dx.doi.org/10.1038/s41586-022-05688-9},
  DOI = {10.1038/s41586-022-05688-9},
  number = {7949},
  journal = {Nature},
  publisher = {Springer Science and Business Media LLC},
  author = {Kamimoto,  Kenji and Stringa,  Blerta and Hoffmann,  Christy M. and Jindal,  Kunal and Solnica-Krezel,  Lilianna and Morris,  Samantha A.},
  year = {2023},
  month = feb,
  pages = {742–751}
}

@article{Lotfollahi2019,
  title = {scGen predicts single-cell perturbation responses},
  volume = {16},
  ISSN = {1548-7105},
  url = {http://dx.doi.org/10.1038/s41592-019-0494-8},
  DOI = {10.1038/s41592-019-0494-8},
  number = {8},
  journal = {Nature Methods},
  publisher = {Springer Science and Business Media LLC},
  author = {Lotfollahi,  Mohammad and Wolf,  F. Alexander and Theis,  Fabian J.},
  year = {2019},
  month = jul,
  pages = {715–721}
}

@article{Luecken2021,
  title = {Benchmarking atlas-level data integration in single-cell genomics},
  volume = {19},
  ISSN = {1548-7105},
  url = {http://dx.doi.org/10.1038/s41592-021-01336-8},
  DOI = {10.1038/s41592-021-01336-8},
  number = {1},
  journal = {Nature Methods},
  publisher = {Springer Science and Business Media LLC},
  author = {Luecken,  Malte D. and B\"{u}ttner,  M. and Chaichoompu,  K. and Danese,  A. and Interlandi,  M. and Mueller,  M. F. and Strobl,  D. C. and Zappia,  L. and Dugas,  M. and Colomé-Tatché,  M. and Theis,  Fabian J.},
  year = {2021},
  month = dec,
  pages = {41–50}
}

@article{Adduri2025,
  title={Predicting cellular responses to perturbation across diverse contexts with State},
  author={Adduri, Abhinav K and Gautam, Dhruv and Bevilacqua, Beatrice and Naghipourfar, Mohsen and Imran, Alishba and Shah, Rohan and Teyssier, Noam and Verma, Rishi and Carpenter, Christopher and Eraslan, Basak and others},
  journal={Cell},
  volume={189},
  number={19},
  pages={5914--5931},
  year={2026},
  publisher={Elsevier}
}

@article{Piran2024,
  title = {Disentanglement of single-cell data with biolord},
  volume = {42},
  ISSN = {1546-1696},
  url = {http://dx.doi.org/10.1038/s41587-023-02079-x},
  DOI = {10.1038/s41587-023-02079-x},
  number = {11},
  journal = {Nature Biotechnology},
  publisher = {Springer Science and Business Media LLC},
  author = {Piran,  Zoe and Cohen,  Niv and Hoshen,  Yedid and Nitzan,  Mor},
  year = {2024},
  month = jan,
  pages = {1678–1683}
}

@article{He2025,
  title={Squidiff: predicting cellular development and responses to perturbations using a diffusion model},
  author={He, Siyu and Zhu, Yuefei and Tavakol, Daniel Naveed and Ye, Haotian and Lao, Yeh-Hsing and Zhu, Zixian and Xu, Cong and Chauhan, Shradha and Garty, Guy and Tomer, Raju and others},
  journal={Nature methods},
  volume={23},
  number={1},
  pages={65--77},
  year={2026},
  publisher={Nature Publishing Group US New York}
}

@article{Luo2024,
  title = {scDiffusion: conditional generation of high-quality single-cell data using diffusion model},
  volume = {40},
  ISSN = {1367-4811},
  url = {http://dx.doi.org/10.1093/bioinformatics/btae518},
  DOI = {10.1093/bioinformatics/btae518},
  number = {9},
  journal = {Bioinformatics},
  publisher = {Oxford University Press (OUP)},
  author = {Luo,  Erpai and Hao,  Minsheng and Wei,  Lei and Zhang,  Xuegong},
  editor = {Mathelier,  Anthony},
  year = {2024},
  month = aug 
}

@article{Bunne2023,
  title = {Learning single-cell perturbation responses using neural optimal transport},
  volume = {20},
  ISSN = {1548-7105},
  url = {http://dx.doi.org/10.1038/s41592-023-01969-x},
  DOI = {10.1038/s41592-023-01969-x},
  number = {11},
  journal = {Nature Methods},
  publisher = {Springer Science and Business Media LLC},
  author = {Bunne,  Charlotte and Stark,  Stefan G. and Gut,  Gabriele and del Castillo,  Jacobo Sarabia and Levesque,  Mitch and Lehmann,  Kjong-Van and Pelkmans,  Lucas and Krause,  Andreas and R\"{a}tsch,  Gunnar},
  year = {2023},
  month = sep,
  pages = {1759–1768}
}

@article{wenkel2025txpert,
  title={TxPert: Leveraging Biochemical Relationships for Out-of-Distribution Transcriptomic Perturbation Prediction},
  author={Wenkel, Frederik and Tu, Wilson and Masschelein, Cassandra and Shirzad, Hamed and Eastwood, Cian and Whitfield, Shawn T and Bendidi, Ihab and Russell, Craig and Hodgson, Liam and Mesbahi, Yassir El and others},
  journal={arXiv preprint arXiv:2505.14919},
  year={2025}
}

@article{ahlmann2025deep,
  title={Deep-learning-based gene perturbation effect prediction does not yet outperform simple linear baselines},
  author={Ahlmann-Eltze, Constantin and Huber, Wolfgang and Anders, Simon},
  journal={Nature Methods},
  volume={22},
  number={8},
  pages={1657--1661},
  year={2025},
  publisher={Nature Publishing Group US New York}
}

@article{cui2024scgpt,
  title={scGPT: toward building a foundation model for single-cell multi-omics using generative AI},
  author={Cui, Haotian and Wang, Chloe and Maan, Hassaan and Pang, Kuan and Luo, Fengning and Duan, Nan and Wang, Bo},
  journal={Nature methods},
  volume={21},
  number={8},
  pages={1470--1480},
  year={2024},
  publisher={Nature Publishing Group US New York}
}

@article{hao2024large,
  title={Large-scale foundation model on single-cell transcriptomics},
  author={Hao, Minsheng and Gong, Jing and Zeng, Xin and Liu, Chiming and Guo, Yucheng and Cheng, Xingyi and Wang, Taifeng and Ma, Jianzhu and Zhang, Xuegong and Song, Le},
  journal={Nature methods},
  volume={21},
  number={8},
  pages={1481--1491},
  year={2024},
  publisher={Nature Publishing Group US New York}
}

@article{vinas2025systema,
  title={Systema: a framework for evaluating genetic perturbation response prediction beyond systematic variation},
  author={Vi{\~n}as Torn{\'e}, Ramon and Wiatrak, Maciej and Piran, Zoe and Fan, Shuyang and Jiang, Liangze and Teichmann, Sarah A and Nitzan, Mor and Brbi{\'c}, Maria},
  journal={Nature Biotechnology},
  pages={1--10},
  year={2025},
  publisher={Nature Publishing Group US New York}
}

@article{zhu2025scouter,
  title={Scouter predicts transcriptional responses to genetic perturbations with large language model embeddings},
  author={Zhu, Ouyang and Li, Jun},
  journal={Nature Computational Science},
  volume={6},
  number={1},
  pages={21--28},
  year={2026},
  publisher={Nature Publishing Group US New York}
}

@article{bai2025pertadapt,
  title={PertAdapt: Unlocking Single-Cell Foundation Models for Genetic Perturbation Prediction via Condition-Sensitive Adaptation},
  author={Bai, Ding and Song, Le and Xing, Eric},
  journal={bioRxiv},
  year={2025},
  publisher={Cold Spring Harbor Laboratory}
}

@article{halmos2025hierarchical,
  title={Hierarchical Refinement: Optimal Transport to Infinity and Beyond},
  author={Halmos, Peter and Gold, Julian and Liu, Xinhao and Raphael, Benjamin J},
  journal={arXiv preprint arXiv:2503.03025},
  year={2025}
}

@article{heimberg2025cell,
  title={A cell atlas foundation model for scalable search of similar human cells},
  author={Heimberg, Graham and Kuo, Tony and DePianto, Daryle J and Salem, Omar and Heigl, Tobias and Diamant, Nathaniel and Scalia, Gabriele and Biancalani, Tommaso and Turley, Shannon J and Rock, Jason R and others},
  journal={Nature},
  volume={638},
  number={8052},
  pages={1085--1094},
  year={2025},
  publisher={Nature Publishing Group UK London}
}

@article{song2025decoding,
  title={Decoding heterogeneous single-cell perturbation responses},
  author={Song, Bicna and Liu, Dingyu and Dai, Weiwei and McMyn, Natalie F and Wang, Qingyang and Yang, Dapeng and Krejci, Adam and Vasilyev, Anatoly and Untermoser, Nicole and Loregger, Anke and others},
  journal={Nature cell biology},
  pages={493--504},
  year={2025},
  publisher={Nature Publishing Group UK London}
}

@article{yu2025perturbnet,
  title={Perturbnet predicts single-cell responses to unseen chemical and genetic perturbations},
  author={Yu, Hengshi and Qian, Weizhou and Song, Yuxuan and Welch, Joshua D},
  journal={Molecular Systems Biology},
  volume={21},
  number={8},
  pages={960--982},
  year={2025}
}

@article{pampari2025chrombpnet,
  title={ChromBPNet: bias factorized, base-resolution deep learning models of chromatin accessibility reveal cis-regulatory sequence syntax, transcription factor footprints and regulatory variants},
  author={Pampari, Anusri and Shcherbina, Anna and Kvon, Evgeny Z and Kosicki, Michael and Nair, Surag and Kundu, Soumya and Kathiria, Arwa S and Risca, Viviana I and Kuningas, Kristiina and Alasoo, Kaur and others},
  journal={BioRxiv},
  year={2025}
}

@article{park2026causal,
  title={Causal Network Recovery in Perturb-seq Experiments Using Proxy and Instrumental Variables},
  author={Park, Kwangmoon and Li, Hongzhe},
  journal={arXiv preprint arXiv:2601.01830},
  year={2026}
}

@article{wu2022predicting,
  title={Predicting cellular responses with variational causal inference and refined relational information},
  author={Wu, Yulun and Barton, Robert A and Wang, Zichen and Ioannidis, Vassilis N and De Donno, Carlo and Price, Layne C and Voloch, Luis F and Karypis, George},
  journal={arXiv preprint arXiv:2210.00116},
  year={2022}
}

@article{linder2025predicting,
  title={Predicting cell type-specific coverage profiles from DNA sequence},
  author={Linder, Johannes and Yuan, Han and Kelley, David R},
  journal={bioRxiv},
  year={2025},
  publisher={Cold Spring Harbor Laboratory}
}

@article{gaudelet2024season,
  title={Season combinatorial intervention predictions with Salt \& Peper},
  author={Gaudelet, Thomas and Del Vecchio, Alice and Carrami, Eli M and Cudini, Juliana and Kapourani, Chantriolnt-Andreas and Uhler, Caroline and Edwards, Lindsay},
  journal={arXiv preprint arXiv:2404.16907},
  year={2024}
}

@article{patel2024dart,
  title={DART-Eval: A comprehensive DNA language model evaluation benchmark on regulatory DNA},
  author={Patel, Aman and Singhal, Arpita and Wang, Austin and Pampari, Anusri and Kasowski, Maya and Kundaje, Anshul},
  journal={Advances in Neural Information Processing Systems},
  volume={37},
  pages={62024--62061},
  year={2024}
}

@inproceedings{tu2024supervised,
  title={A supervised contrastive framework for learning disentangled representations of cellular perturbation data},
  author={Tu, Xinming and H{\"u}tter, Jan-Christian and Wang, Zitong Jerry and Kudo, Takamasa and Regev, Aviv and Lopez, Romain},
  booktitle={Machine Learning in Computational Biology},
  pages={90--100},
  year={2024},
  organization={PMLR}
}

@article{istrate2025rbio1,
  title={rbio1-training scientific reasoning LLMs with biological world models as soft verifiers},
  author={Istrate, Ana-Maria and Milletari, Fausto and Castrotorres, Fabrizio and Tomczak, Jakub M and Torkar, Michaela and Li, Donghui and Karaletsos, Theofanis},
  journal={bioRxiv},
  year={2025},
  publisher={Cold Spring Harbor Laboratory}
}

@article{passigan2025interpretable,
  title={Interpretable Perturbation Modeling Through Biomedical Knowledge Graphs},
  author={Passigan, Pascal and Ning, Angelina and others},
  journal={arXiv preprint arXiv:2512.22251},
  year={2025}
}

@article{bereket2023modelling,
  title={Modelling cellular perturbations with the sparse additive mechanism shift variational autoencoder},
  author={Bereket, Michael and Karaletsos, Theofanis},
  journal={Advances in Neural Information Processing Systems},
  volume={36},
  pages={1--12},
  year={2023}
}

@article{huang2025x,
  title={X-Atlas/Orion: Genome-wide Perturb-seq Datasets via a Scalable Fix-Cryopreserve Platform for Training Dose-Dependent Biological Foundation Models},
  author={Huang, Ann C and Hsieh, Tsung-Han S and Zhu, Jiang and Michuda, Jackson and Teng, Ashton and Kim, Soohong and Rumsey, Elizabeth M and Lam, Sharon K and Anigbogu, Ikenna and Wright, Philip and others},
  journal={bioRxiv},
  year={2025},
  publisher={Cold Spring Harbor Laboratory}
}

@inproceedings{hetzel2022predicting,
  title={Predicting Cellular Responses to Novel Drug Perturbations at a Single-Cell Resolution},
  author={Hetzel, Leon and Böhm, Simon and Kilbertus, Niki and Günnemann, Stephan and Lotfollahi, Mohammad and Theis, Fabian J},
  booktitle={NeurIPS 2022},
  year={2022}
}

@article{liang2025scppdm,
  title={scPPDM: A Diffusion Model for Single-Cell Drug-Response Prediction},
  author={Liang, Zhaokang and Zhuang, Shuyang and Jiao, Xiaoran and Mao, Weian and Chen, Hao and Shen, Chunhua},
  journal={arXiv preprint arXiv:2510.11726},
  year={2025}
}

@article{szalata2024benchmark,
  title={A benchmark for prediction of transcriptomic responses to chemical perturbations across cell types},
  author={Sza{\l}ata, Artur and Benz, Andrew and Cannoodt, Robrecht and Cortes, Mauricio and Fong, Jason and Kuppasani, Sunil and Lieberman, Richard and Liu, Tianyu and Mas-Rosario, Javier A and Meinl, Rico and others},
  journal={Advances in Neural Information Processing Systems},
  volume={37},
  pages={20566--20616},
  year={2024}
}

@article{wei2025genome,
  title={GenoME: a MoE-based generative model for individualized, multimodal prediction and perturbation of genomic profiles},
  author={Wei, Jiachen and Xue, Yue and Chai, Hao and Gao, Yi Qin},
  journal={bioRxiv},
  year={2025},
  publisher={Cold Spring Harbor Laboratory}
}

@misc{liu_i2sb_2023,
    title = {{I$^2$SB}: {Image}-to-{Image} {Schrödinger} {Bridge}},
    shorttitle = {{I$^2$SB}},
    url = {http://arxiv.org/abs/2302.05872},
    doi = {10.48550/arXiv.2302.05872},
    urldate = {2025-04-29},
    publisher = {arXiv},
    author = {Liu, Guan-Horng and Vahdat, Arash and Huang, De-An and Theodorou, Evangelos A. and Nie, Weili and Anandkumar, Anima},
    month = may,
    year = {2023},
    note = {arXiv:2302.05872 [cs]},
}

@article{wu2025contextualizing,
  title={Contextualizing biological perturbation experiments through language},
  author={Wu, Menghua and Littman, Russell and Levine, Jacob and Qiu, Lin and Biancalani, Tommaso and Richmond, David and Huetter, Jan-Christian},
  journal={arXiv preprint arXiv:2502.21290},
  year={2025}
}

@article{canver2015bcl11a,
  title={BCL11A enhancer dissection by Cas9-mediated in situ saturating mutagenesis},
  author={Canver, Matthew C and Smith, Elenoe C and Sher, Falak and Pinello, Luca and Sanjana, Neville E and Shalem, Ophir and Chen, Diane D and Schupp, Patrick G and Vinjamur, Divya S and Garcia, Sara P and others},
  journal={Nature},
  volume={527},
  number={7577},
  pages={192--197},
  year={2015},
  publisher={Nature Publishing Group UK London}
}

@article{maurano2012systematic,
  title={Systematic localization of common disease-associated variation in regulatory DNA},
  author={Maurano, Matthew T and Humbert, Richard and Rynes, Eric and Thurman, Robert E and Haugen, Eric and Wang, Hao and Reynolds, Alex P and Sandstrom, Richard and Qu, Hongzhu and Brody, Jennifer and others},
  journal={Science},
  volume={337},
  number={6099},
  pages={1190--1195},
  year={2012},
  publisher={American Association for the Advancement of Science}
}

@article{nasser2021genome,
  title={Genome-wide enhancer maps link risk variants to disease genes},
  author={Nasser, Joseph and Bergman, Drew T and Fulco, Charles P and Guckelberger, Philine and Doughty, Benjamin R and Patwardhan, Tejal A and Jones, Thouis R and Nguyen, Tung H and Ulirsch, Jacob C and Lekschas, Fritz and others},
  journal={Nature},
  volume={593},
  number={7858},
  pages={238--243},
  year={2021},
  publisher={Nature Publishing Group UK London}
}

@article{chiabrando2012mitochondrial,
  title={The mitochondrial heme exporter FLVCR1b mediates erythroid differentiation},
  author={Chiabrando, Deborah and Marro, Samuele and Mercurio, Sonia and Giorgi, Carlotta and Petrillo, Sara and Vinchi, Francesca and Fiorito, Veronica and Fagoonee, Sharmila and Camporeale, Annalisa and Turco, Emilia and others},
  journal={The Journal of clinical investigation},
  volume={122},
  number={12},
  pages={4569--4579},
  year={2012},
  publisher={American Society for Clinical Investigation}
}

@article{theodoris2023transfer,
  title={Transfer learning enables predictions in network biology},
  author={Theodoris, Christina V and Xiao, Ling and Chopra, Anant and Chaffin, Mark D and Al Sayed, Zeina R and Hill, Matthew C and Mantineo, Helene and Brydon, Elizabeth M and Zeng, Zexian and Liu, X Shirley and others},
  journal={Nature},
  volume={618},
  number={7965},
  pages={616--624},
  year={2023},
  publisher={Nature Publishing Group UK London}
}

@article{fu2023fast,
  title={Fast kernel-based association testing of non-linear genetic effects for biobank-scale data},
  author={Fu, Boyang and Pazokitoroudi, Ali and Sudarshan, Mukund and Liu, Zhengtong and Subramanian, Lakshminarayanan and Sankararaman, Sriram},
  journal={Nature communications},
  volume={14},
  number={1},
  pages={4936},
  year={2023},
  publisher={Nature Publishing Group UK London}
}

@article{cheng2025dnalongbench,
  title={Dnalongbench: a benchmark suite for long-range dna prediction tasks},
  author={Cheng, Wenduo and Song, Zhenqiao and Zhang, Yang and Wang, Shike and Wang, Danqing and Yang, Muyu and Li, Lei and Ma, Jian},
  journal={bioRxiv},
  year={2025}
}

@article{moriguchi2014regulatory,
  title={A regulatory network governing Gata1 and Gata2 gene transcription orchestrates erythroid lineage differentiation},
  author={Moriguchi, Takashi and Yamamoto, Masayuki},
  journal={International journal of hematology},
  volume={100},
  number={5},
  pages={417--424},
  year={2014},
  publisher={Springer}
}

@article{alarcon1991cd3,
  title={The CD3-gamma and CD3-delta subunits of the T cell antigen receptor can be expressed within distinct functional TCR/CD3 complexes.},
  author={Alarcon, B and Ley, SC and Sanchez-Madrid, F and Blumberg, RS and Ju, ST and Fresno, M and Terhorst, C},
  journal={The EMBO Journal},
  volume={10},
  number={4},
  pages={903--912},
  year={1991},
  publisher={Springer}
}

@article{zeng2025gene,
  title={Gene editing without ex vivo culture evades genotoxicity in human hematopoietic stem cells},
  author={Zeng, Jing and Nguyen, My Anh and Liu, Pengpeng and Da Silva, Lucas Ferreira and Levesque, S{\'e}bastien and Lin, Linda Y and Justus, David G and Petri, Karl and Clement, Kendell and Porter, Shaina N and others},
  journal={Cell Stem Cell},
  volume={32},
  number={2},
  pages={191--208},
  year={2025},
  publisher={Elsevier}
}

@article{frangoul2021crispr,
  title={CRISPR-Cas9 gene editing for sickle cell disease and $\beta$-thalassemia},
  author={Frangoul, Haydar and Altshuler, David and Cappellini, M Domenica and Chen, Yi-Shan and Domm, Jennifer and Eustace, Brenda K and Foell, Juergen and de la Fuente, Josu and Grupp, Stephan and Handgretinger, Rupert and others},
  journal={New England Journal of Medicine},
  volume={384},
  number={3},
  pages={252--260},
  year={2021},
  publisher={Mass Medical Soc}
}

@article{squair2021confronting,
  title={Confronting false discoveries in single-cell differential expression},
  author={Squair, Jordan W and Gautier, Matthieu and Kathe, Claudia and Anderson, Mark A and James, Nicholas D and Hutson, Thomas H and Hudelle, R{\'e}mi and Qaiser, Taha and Matson, Kaya JE and Barraud, Quentin and others},
  journal={Nature communications},
  volume={12},
  number={1},
  pages={5692},
  year={2021},
  publisher={Nature Publishing Group UK London}
}

@article{murphy2023avoiding,
  title={Avoiding false discoveries in single-cell RNA-seq by revisiting the first Alzheimer’s disease dataset},
  author={Murphy, Alan E and Fancy, Nurun and Skene, Nathan},
  journal={Elife},
  volume={12},
  pages={RP90214},
  year={2023},
  publisher={eLife Sciences Publications Limited}
}

@inproceedings{kendall2018multi,
  title={Multi-task learning using uncertainty to weigh losses for scene geometry and semantics},
  author={Kendall, Alex and Gal, Yarin and Cipolla, Roberto},
  booktitle={Proceedings of the IEEE conference on computer vision and pattern recognition},
  pages={7482--7491},
  year={2018}
}

@article{klein2025cellflow,
  title={CellFlow enables generative single-cell phenotype modeling with flow matching},
  author={Klein, Dominik and Fleck, Jonas Simon and Bobrovskiy, Daniil and Zimmermann, Lea and Becker, S{\"o}ren and Palma, Alessandro and Dony, Leander and Tejada-Lapuerta, Alejandro and Huguet, Guillaume and Lin, Hsiu-Chuan and others},
  journal={bioRxiv},
  year={2025},
  publisher={Cold Spring Harbor Laboratory}
}

@article{hunt1990altered,
  title={Altered expression of keratin and vimentin in human retinal pigment epithelial cells in vivo and in vitro},
  author={Hunt, Richard C and Davis, Alberta A},
  journal={Journal of cellular physiology},
  volume={145},
  number={2},
  pages={187--199},
  year={1990},
  publisher={Wiley Online Library}
}

@inproceedings{radford2021learning,
  title={Learning transferable visual models from natural language supervision},
  author={Radford, Alec and Kim, Jong Wook and Hallacy, Chris and Ramesh, Aditya and Goh, Gabriel and Agarwal, Sandhini and Sastry, Girish and Askell, Amanda and Mishkin, Pamela and Clark, Jack and others},
  booktitle={International conference on machine learning},
  pages={8748--8763},
  year={2021},
  organization={PmLR}
}

@article{wu2024perturbench,
  title={Perturbench: Benchmarking machine learning models for cellular perturbation analysis},
  author={Wu, Yan and Wershof, Esther and Schmon, Sebastian M and Nassar, Marcel and Osi{\'n}ski, B{\l}a{\.z}ej and Eksi, Ridvan and Yan, Zichao and Stark, Rory and Zhang, Kun and Graepel, Thore},
  journal={arXiv preprint arXiv:2408.10609},
  year={2024}
}

@article{nadig2025transcriptome,
  title={Transcriptome-wide analysis of differential expression in perturbation atlases},
  author={Nadig, Ajay and Replogle, Joseph M and Pogson, Angela N and Murthy, Mukundh and McCarroll, Steven A and Weissman, Jonathan S and Robinson, Elise B and O’Connor, Luke J},
  journal={Nature Genetics},
  volume={57},
  number={5},
  pages={1228--1237},
  year={2025},
  publisher={Nature Publishing Group US New York}
}

@article{gonzalez2025combinatorial,
  title={Combinatorial prediction of therapeutic perturbations using causally inspired neural networks},
  author={Gonzalez, Guadalupe and Lin, Xiang and Herath, Isuru and Veselkov, Kirill and Bronstein, Michael and Zitnik, Marinka},
  journal={Nature Biomedical Engineering},
  pages={1--18},
  year={2025},
  publisher={Nature Publishing Group UK London}
}

@article{lal2024decoding,
  title={Decoding sequence determinants of gene expression in diverse cellular and disease states},
  author={Lal, Avantika and Karollus, Alexander and Gunsalus, Laura and Garfield, David and Nair, Surag and Tseng, Alex M and Gordon, M Grace and Blischak, John and van de Geijn, Bryce and Bhangale, Tushar and others},
  journal={bioRxiv},
  year={2024},
  publisher={Cold Spring Harbor Laboratory}
}

@article{dong2026stack,
  title={Stack: In-Context Learning of Single-Cell Biology},
  author={Dong, Mingze and Adduri, Abhinav and Gautam, Dhruv and Carpenter, Christopher and Shah, Rohan and Ricci-Tam, Chiara and Kluger, Yuval and Burke, Dave P and Roohani, Yusuf Husein},
  journal={bioRxiv},
  year={2026},
  publisher={Cold Spring Harbor Laboratory}
}

@article{avsec2025alphagenome,
  title={Advancing regulatory variant effect prediction with AlphaGenome},
  author={Avsec, {\v{Z}}iga and Latysheva, Natasha and Cheng, Jun and Novati, Guido and Taylor, Kyle R and Ward, Tom and Bycroft, Clare and Nicolaisen, Lauren and Arvaniti, Eirini and Pan, Joshua and others},
  journal={Nature},
  volume={649},
  number={8099},
  pages={1206--1218},
  year={2026},
  publisher={Nature Publishing Group UK London}
}

@article{avsec2021effective,
  title={Effective gene expression prediction from sequence by integrating long-range interactions},
  author={Avsec, {\v{Z}}iga and Agarwal, Vikram and Visentin, Daniel and Ledsam, Joseph R and Grabska-Barwinska, Agnieszka and Taylor, Kyle R and Assael, Yannis and Jumper, John and Kohli, Pushmeet and Kelley, David R},
  journal={Nature methods},
  volume={18},
  number={10},
  pages={1196--1203},
  year={2021},
  publisher={Nature Publishing Group US New York}
}

@article{hingerl2025flashzoi,
  title={Flashzoi: an enhanced Borzoi for accelerated genomic analysis},
  author={Hingerl, Johannes C and Karollus, Alexander and Gagneur, Julien},
  journal={Bioinformatics},
  volume={41},
  number={9},
  pages={btaf467},
  year={2025},
  publisher={Oxford University Press}
}

@article{lorch2026latent,
  title={Latent Causal Diffusions for Single-Cell Perturbation Modeling},
  author={Lorch, Lars and Zhang, Jiaqi and Bunne, Charlotte and Krause, Andreas and Sch{\"o}lkopf, Bernhard and Uhler, Caroline},
  journal={arXiv preprint arXiv:2601.15341},
  year={2026}
}

@article{garau2024multi,
  title={Multi-modal transfer learning between biological foundation models},
  author={Garau-Luis, Juan Jose and Bordes, Patrick and Gonzalez, Liam and Roller, Ma{\v{s}}a and de Almeida, Bernardo and Blum, Christopher and Hexemer, Lorenz and Laurent, Stefan and Lang, Maren and Pierrot, Thomas and others},
  journal={Advances in Neural Information Processing Systems},
  volume={37},
  pages={78431--78450},
  year={2024}
}

@inproceedings{Lipman2022,
  author    = {Lipman, Yaron and Chen, Ricky T. Q. and Ben-Hamu, Heli and Nickel, Maximilian and Le, Matt},
  title     = {Flow Matching for Generative Modeling},
  booktitle = {International Conference on Learning Representations (ICLR)},
  year      = {2023}
}

@inproceedings{tong2023,
  author    = {Tong, Alexander and Malkin, Nikolay and Huguet, Guillaume and Zhang, Yanlei and Rector-Brooks, Jarrid and Fatras, Kilian and Wolf, Guy and Bengio, Yoshua},
  title     = {Improving and Generalizing Flow-Based Generative Models with Minibatch Optimal Transport},
  booktitle = {Transactions on Machine Learning Research (TMLR)},
  year      = {2024}
}

@article{fulco2016systematic,
  title={Systematic mapping of functional enhancer--promoter connections with CRISPR interference},
  author={Fulco, Charles P and Munschauer, Mathias and Anyoha, Rockwell and Munson, Glen and Grossman, Sharon R and Perez, Elizabeth M and Kane, Michael and Cleary, Brian and Lander, Eric S and Engreitz, Jesse M},
  journal={Science},
  volume={354},
  number={6313},
  pages={769--773},
  year={2016},
  publisher={American Association for the Advancement of Science}
}

@article{replogle2022mapping,
  title={Mapping information-rich genotype-phenotype landscapes with genome-scale Perturb-seq},
  author={Replogle, Joseph M and Saunders, Reuben A and Pogson, Angela N and Hussmann, Jeffrey A and Lenail, Alexander and Guna, Alina and Mascibroda, Lauren and Wagner, Eric J and Adelman, Karen and Lithwick-Yanai, Gila and others},
  journal={Cell},
  volume={185},
  number={14},
  pages={2559--2575},
  year={2022},
  publisher={Elsevier}
}

\newpage

\appendix
\onecolumn

\appendix

\setcounter{figure}{0}
\renewcommand{\thefigure}{S\arabic{figure}}
\renewcommand{\theHfigure}{S\arabic{figure}}

\setcounter{table}{0}
\renewcommand{\thetable}{S\arabic{table}}

\setcounter{equation}{0}
\renewcommand{\theequation}{S\arabic{equation}}

\section{Optimal Transport Coupling Details}
\label{sec:ot_details}

\method{} can flexibly adapt different OT pairing algorithms. By default, we employ Hierarchical Refinement Optimal Transport (HiRef)~\cite{halmos2025hierarchical} to compute bijective couplings between control and perturbed cell populations. Below, we detail the complete OT protocol, including the cost function, solver, hyperparameters, and preprocessing steps.

\subsection{Embedding and Latent Space Representation}

\xhdr{Cell Embeddings}
We represent each cell using SCimilarity embeddings~\cite{heimberg2025cell}, a pre-trained foundation model that maps gene expression profiles to a 128-dimensional latent space. Let $\mathbf{e}_i \in \mathbb{R}^{128}$ denote the embedding for cell $i$.

\xhdr{Dimensionality Reduction}
For each perturbation-control matching task, we apply Principal Component Analysis (PCA) to the combined embeddings of control and perturbed cells. Let $\mathcal{C} = \{\mathbf{e}^c_1, \ldots, \mathbf{e}^c_n\}$ and $\mathcal{P} = \{\mathbf{e}^p_1, \ldots, \mathbf{e}^p_m\}$ denote the control and perturbed cell embeddings, respectively. We compute:
\begin{equation}
    \mathbf{Z}_{\text{combined}} = [\mathcal{C}; \mathcal{P}] \in \mathbb{R}^{(n+m) \times 128}
\end{equation}
and apply PCA with $d = \min(50, n+m-1)$ components to obtain the reduced representations $\mathbf{X}^c \in \mathbb{R}^{n \times d}$ and $\mathbf{X}^p \in \mathbb{R}^{m \times d}$.

\subsection{Cost Function}

The transport cost between control cell embedding $i$ and perturbed cell embedding $j$ is defined as the squared Euclidean distance in the PCA-reduced latent space:
\begin{equation}
    C_{ij} = \|\mathbf{z}_c^i - \mathbf{z}_p^j\|_2^2
\end{equation}
where $\mathbf{z}_c^i$ and $\mathbf{z}_p^j$ are the PCA-transformed embeddings.

\subsection{Optimal Transport Solver: Hierarchical Refinement}

We use the Hierarchical Refinement (HiRef) algorithm~\cite{halmos2025hierarchical}, which computes the exact Monge map (bijective coupling) in log-linear time and linear space.

\xhdr{Algorithm Overview}
HiRef leverages low-rank OT to iteratively partition both source and target distributions through a hierarchy of subproblems. At each level, the algorithm:
\begin{enumerate}
    \item Solves a low-rank OT problem to co-cluster source-target pairs
    \item Partitions the data according to the cluster assignments
    \item Recursively refines each partition until reaching singleton matches
\end{enumerate}

\xhdr{Rank Annealing Schedule}
We employ the default optimal rank schedule:
\begin{equation}
    \text{rank\_schedule} = \texttt{optimal\_rank\_schedule}(n, \text{depth}=6, Q_{\max}=\max(2^{15}, n), r_{\max}=2^{10})
\end{equation}
where $n$ is the number of cells to match, $\text{depth}=6$ controls the hierarchy depth, $Q_{\max}$ is the maximum subproblem size, and $r_{\max}=1024$ is the maximum rank at any level.

\subsection{Sample Size Handling and Bootstrapping}

We employ a ``floor without ceiling'' strategy for sample size balancing:

\xhdr{Minimum Cell Threshold}
We set $n_{\min} = 30$ as the minimum number of cells per perturbation group.

\xhdr{Bootstrapping Protocol}
For each perturbation group with $n_{\text{pert}}$ cells:
\begin{itemize}
    \item If $n_{\text{pert}} < n_{\min}$: Bootstrap (sample with replacement) to exactly $n_{\min}$ cells
    \item If $n_{\text{pert}} \geq n_{\min}$: Use all available cells without downsampling
\end{itemize}

\xhdr{Control Sampling}
Controls are sampled (with replacement if necessary) to match the perturbation group size, ensuring 1:1 matching:
\begin{equation}
    |\mathcal{C}_{\text{matched}}| = |\mathcal{P}_{\text{matched}}| = \max(n_{\min}, n_{\text{pert}})
\end{equation}

\xhdr{Cell-Line Aware Matching}
When cell-line annotations are available, we enforce strict within-cell-line matching:
\begin{equation}
    \mathcal{C}_{\text{candidates}} = \{c \in \mathcal{C} : \text{cell\_line}(c) = \text{cell\_line}(\mathcal{P})\}
\end{equation}

\subsection{Hyperparameter Summary}

\begin{table}[h]
\centering
\caption{OT matching hyperparameters}
\label{tab:ot_hyperparams}
\begin{tabular}{lll}
\toprule
\textbf{Parameter} & \textbf{Value} & \textbf{Description} \\
\midrule
PCA components & $\min(50, n+m-1)$ & Latent space dimensionality \\
Cost function & Squared Euclidean & Distance metric in PCA space \\
Hierarchy depth & 6 & Number of refinement levels \\
Max rank ($r_{\max}$) & $2^{10} = 1024$ & Maximum rank per level \\
Max subproblem ($Q_{\max}$) & $\max(2^{15}, n)$ & Largest subproblem size \\
Base rank ($r_0$) & 1 & Binary partitioning \\
Min cells ($n_{\min}$) & 30 & Bootstrap threshold \\
\bottomrule
\end{tabular}
\end{table}

\subsection{Computational Complexity}

HiRef achieves $O(n \log n)$ time complexity and $O(n)$ space complexity, compared to $O(n^2)$ time and space for Sinkhorn-based methods. This enables scaling to datasets with millions of cells.

\subsection{Sensitivity Analysis}

\xhdr{PCA Components}
The choice of 50 PCA components captures $>95\%$ of the variance in SCimilarity embeddings while providing sufficient dimensionality reduction for stable OT computation. We empirically verified that results are robust to choices in the range $[30, 100]$ components.

\xhdr{Bootstrap Threshold}
The minimum cell threshold $n_{\min} = 30$ balances statistical power with representation of rare perturbations.

\xhdr{Regularization}
Unlike Sinkhorn-based OT which requires tuning the entropic regularization parameter $\varepsilon$, HiRef directly computes the unregularized Monge map, eliminating a source of hyperparameter sensitivity.

\subsection{Implementation Details}

All OT computations were performed using the HiRef implementation\footnote{\url{https://github.com/raphael-group/HiRef}} with PyTorch on NVIDIA GPUs. The complete pipeline code is available at \url{https://github.com/mims-harvard/STRAND}.

\subsection{Robustness of OT Pairing}
\label{sec:ot_robustness}
We assess the robustness of OT pairing to the choice of solver and feature space. The pseudo-bulk perturbation displacement is preserved across solver swap (HiRef $\to$ Hungarian) and PCA dimension (PCA-50 vs.\ raw 128-d SCimilarity), with cosine similarity to the default $\approx 1.00$ in all cases. Random pairing inflates transport cost by $\sim 4.8\times$, confirming the structured-pairing signal is non-trivial.

\begin{table}[h]
\centering
\fontsize{9pt}{11pt}\selectfont
\caption{\textbf{OT pairing robustness.} Cosine similarity of pseudo-bulk perturbation displacement vs.\ default (HiRef + PCA-50), and relative transport cost.}
\label{stab:ot_robustness}
\begin{tabular}{llcc}
\toprule
\textbf{Solver} & \textbf{Feature space} & \textbf{Cos.\ to default} & \textbf{Rel.\ transport cost} \\
\midrule
HiRef (default) & PCA-50      & 1.000 & 1.00 \\
HiRef           & Raw 128-d   & 0.998 & 1.04 \\
Hungarian       & PCA-50      & 0.997 & 1.02 \\
Hungarian       & Raw 128-d   & 0.995 & 1.05 \\
Random pairing  & ---         & 0.31  & 4.78 \\
\bottomrule
\end{tabular}
\end{table}

\subsection{PCA Variance Diagnostics}
\label{sec:supp_pca_variance}
Because OT pairing is performed within perturbation-specific matching groups, we validate PCA locally per OT problem.  At $k=50$, K562, Jurkat, and RPE1 all explain an average over $\geq 99.9\%$ of variance across groups. This confirms that the 50 PCs retain at least $99.9\%$ of the variance of SCimilarity latent embedding on average.

\begin{figure}[h]
  \centering
\includegraphics[width=0.5\textwidth]{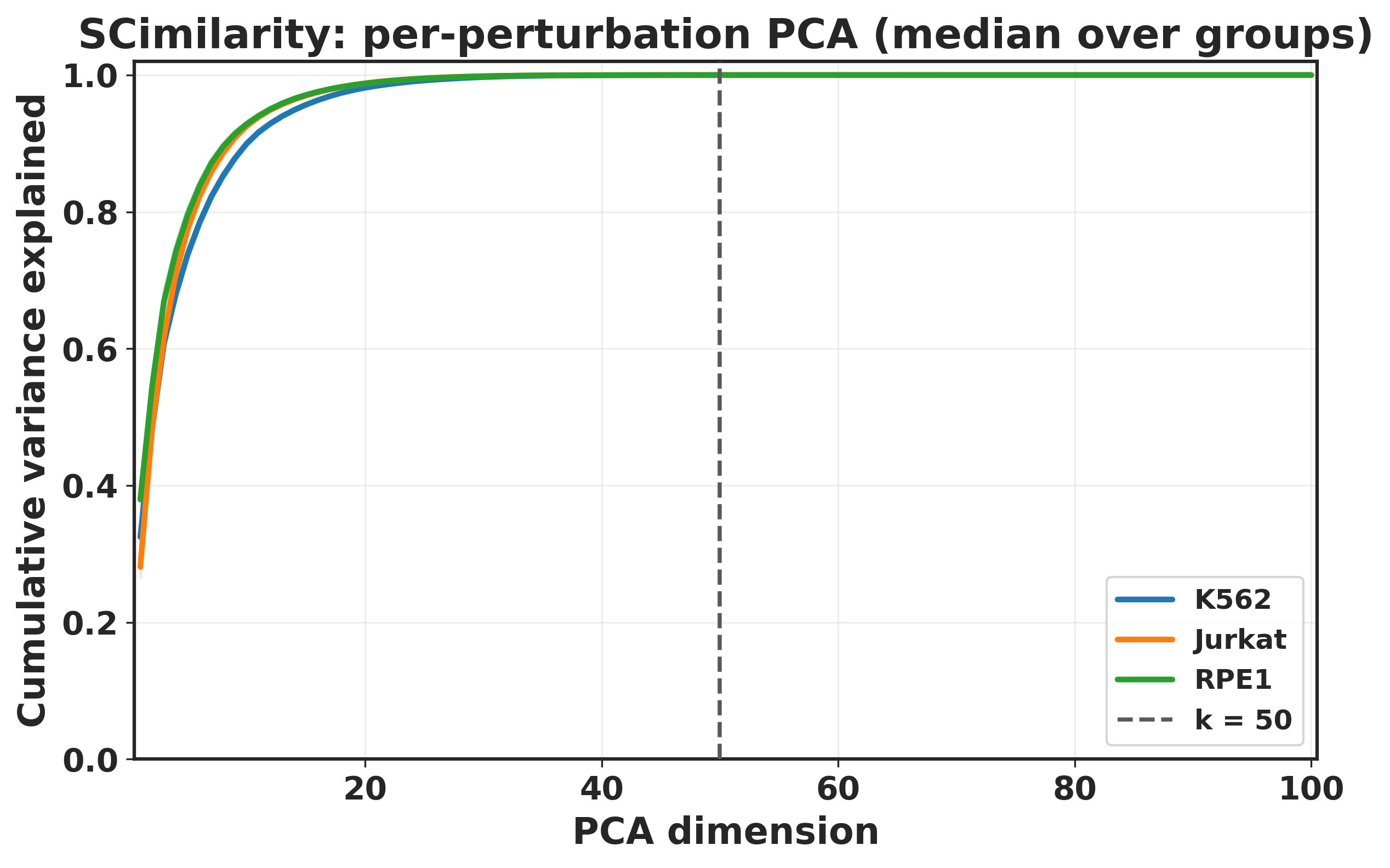}
  \caption{\textbf{Per-perturbation-group PCA variance explained for SCimilarity embeddings.} For each cell line, PCA is fit separately on the concatenated control and perturbed embeddings within each perturbation group, matching the grouping used for OT matching. Lines show the mean cumulative variance explained across groups. The dashed line marks $k=50$, the PCA dimension used for OT matching.}
  \label{fig:supp_pca_variance}
\end{figure}

\section{DNA Perturbation Module Details}
\label{sec:supp_perturb_module}
The cross-attention layer uses 4 heads with $d_k = 64$ per head; projection matrices $W_Q, W_K, W_V \in \mathbb{R}^{d \times d_k}$ are learned. The positional bias $b_\rho$ is parameterized as a learned function of $|j - c_p|$, encoding distance from the perturbation site. The two-layer projection MLP has hidden dimension $2d$, LayerNorm, and GELU activation.

\section{Derivations for I2SB}
\label{sec:supp_theory}

\subsection{Bridge and Forward Process (I2SB)}
We model perturbation effects in a shared latent space.
Let $ \mathbf{z}_0  = \zb_p\in \mathbb{R}^D$ denote the \emph{perturbed} latent (clean endpoint) and
$\mathbf{z}_T = \zb_c  \in \mathbb{R}^D$ the \emph{control} latent (degraded endpoint).
Following I2SB, we define a Gaussian diffusion \emph{bridge} between $(\mathbf{z}_0,\mathbf{z}_T)$.
For discrete steps $t \in \{0,\dots,T\}$, we define the cumulative noise as:
\[
\sigma_t^2 \doteq \sum_{n=0}^{t-1}\beta_n,\qquad
\bar{\sigma}_t^2 \doteq \sigma_T^2 - \sigma_t^2 .
\]
Then the analytic bridge marginal is:
\begin{equation}
q(\mathbf{z}_t \mid \mathbf{z}_0,\mathbf{z}_T)
= \mathcal{N}\!\Big(
\mathbf{z}_t;\;
w_0(t)\,\mathbf{z}_0 + w_1(t)\,\mathbf{z}_T,\;
\sigma^2(t)\mathbf{I}
\Big),
\label{eq:i2sb_bridge}
\end{equation}
with weights and variance:
\[
w_0(t)=\frac{\bar{\sigma}_t^2}{\bar{\sigma}_t^2+\sigma_t^2},\qquad
w_1(t)=\frac{\sigma_t^2}{\bar{\sigma}_t^2+\sigma_t^2},\qquad
\sigma^2(t)=\frac{\sigma_t^2\bar{\sigma}_t^2}{\bar{\sigma}_t^2+\sigma_t^2}.
\]
Equivalently, we sample $\mathbf{z}_t = w_0(t)\mathbf{z}_0 + w_1(t)\mathbf{z}_T + \sigma(t)\boldsymbol{\epsilon}$,
$\boldsymbol{\epsilon}\sim\mathcal{N}(\mathbf{0},\mathbf{I})$.

\xhdr{Schedule}
We use the symmetric I2SB noise schedule (increase then decrease), i.e.,
$\{\beta_n\}$ is an ascending segment concatenated with its reverse, so that the bridge variance $\sigma^2(t)$
peaks near $t\approx T/2$ and vanishes at both endpoints. No condition-dependent
schedule scaling is used.

\subsection{Denoising Model and Training Target}

Although $\mathbf{z}_t$ is sampled from the bridge marginal
$q(\mathbf{z}_t \mid \mathbf{z}_0,\mathbf{z}_T)$,
following I2SB we train the network to predict the
{forward-process noise with respect to the clean endpoint $\mathbf{z}_0$}.
Specifically, we define the training target as:
\[
\boldsymbol{\varepsilon}^{\text{I2SB}}
\;\doteq\;
\frac{\mathbf{z}_t - \mathbf{z}_0}{\sigma_t}.
\]

The denoising objective is:
\[
\mathcal L_{\text{bridge}}
=
\mathbb E_{\mathbf{z}_0,\mathbf{z}_T,t}
\Big[
\big\|
e_\theta(\mathbf{z}_t,t,\mathbf{u}_p,\mathbf{z}_c)
-
\boldsymbol{\varepsilon}^{\text{I2SB}}
\big\|_2^2
\Big],
\]
which is equivalent to the I2SB loss formulation.

\subsection{Sampling (DDPM-style reverse bridge)}
At test time, we start from the control endpoint:
\[
\mathbf{z}_T =  \mathbf{z}_c,
\]
and run DDPM-style reverse steps.
Given current $\mathbf{z}_t$ and $\hat{\boldsymbol{\nu}}=e_\theta(\mathbf{z}_t,t,\mathbf{u}_p,\mathbf{z}_c)$,
recover the predicted clean endpoint:
\[
\hat{\mathbf{z}}_0 = \mathbf{z}_t - \sigma_t \hat{\boldsymbol{\nu}}.
\]
Then sample $\mathbf{z}_{t-1}$ from the Gaussian posterior $p(\mathbf{z}_{t-1}\mid \hat{\mathbf{z}}_0,\mathbf{z}_t)$
(using the same DDPM recursion as in I2SB), iterating $t=T,\dots,1$.

\subsection{Default Inference Configuration}
\label{sec:supp_default_config}

\begin{itemize}[leftmargin=8pt, itemsep=1pt, topsep=2pt]
  \item Samples per perturbation: 256
  \item Classifier-free guidance scale: 1
  \item DDPM sampling parameter $\eta$: 1
\end{itemize}
All in silico perturbation screening case studies use a fixed 10-step diffusion sampling to balance computational efficiency and performance.

For benchmark evaluations, the number of diffusion steps is treated as a tunable hyperparameter. We perform a grid search over $\{6, 10, 14, 20, 50\}$ and select the value that minimizes the Energy Distance between predicted and ground-truth perturbation effects on the validation set.

\section{Sequence-Expression Alignment}
\label{sec:supp_align}

\subsection{Objectives}
We align a sequence-derived perturbation embedding with the corresponding transcriptional effect using a prototypical (perturbation-level) contrastive objective.

\subsection{Setup and Notation}
In a minibatch, let $\mathcal{P}=\{1,\ldots,P\}$ be the set of unique perturbations.
For each perturbation $k\in\mathcal{P}$, let $\mathcal{B}_k$ be the set of matched control-perturbed cell pairs $(\zb_{c,i},\zb_{p,i})$ with perturbation label $k$.
Let $\ub_{k,i}$ denote the sequence-derived perturbation embedding for pair $i$.

\subsection{Perturbation Prototypes (Pseudo-bulk)}
We aggregate single-cell pairs into perturbation-level prototypes:
\begin{equation}
\Delta \bar{\zb}_k
=
\frac{1}{|\mathcal{B}_k|}
\sum_{i\in\mathcal{B}_k}
(\zb_{p,i}-\zb_{c,i}),
\qquad
\bar{\ub}_k
=
\frac{1}{|\mathcal{B}_k|}
\sum_{i\in\mathcal{B}_k}
\ub_{k,i}.
\end{equation}
\xhdr{Optional Weighted Prototypes}
With per-pair weights $w_i$,
\begin{equation}
\Delta \bar{\zb}_k
=
\frac{\sum_{i\in\mathcal{B}_k} w_i(\zb_{p,i}-\zb_{c,i})}{\sum_{i\in\mathcal{B}_k} w_i},
\qquad
\bar{\ub}_k
=
\frac{\sum_{i\in\mathcal{B}_k} w_i\ub_{k,i}}{\sum_{i\in\mathcal{B}_k} w_i}.
\end{equation}

\subsection{Asymmetric Projection and Normalization}
We project only the DNA prototype to the RNA latent dimension to avoid shifting the pretrained RNA space.
Let $\mathrm{Proj}_{\text{DNA}}:\mathbb{R}^{d_{\text{DNA}}}\to\mathbb{R}^{d_{\text{RNA}}}$ be a 2-layer MLP with GELU:
\begin{equation}
\mathrm{Proj}_{\text{DNA}}(\ub)
=
W_2\,\mathrm{GELU}(W_1\ub+b_1)+b_2,
\end{equation}
where $W_1\in\mathbb{R}^{d_h\times d_{\text{DNA}}}$, $W_2\in\mathbb{R}^{d_{\text{RNA}}\times d_h}$ and $d_h{=}128$.
We then $\ell_2$-normalize:
\begin{equation}
\hat{\ub}_k
=
\frac{\mathrm{Proj}_{\text{DNA}}(\bar{\ub}_k)}{\|\mathrm{Proj}_{\text{DNA}}(\bar{\ub}_k)\|_2},
\qquad
\hat{\zb}_k
=
\frac{\Delta\bar{\zb}_k}{\|\Delta\bar{\zb}_k\|_2}.
\end{equation}

\subsection{Soft Targets from RNA Similarity}
To reduce false negatives among biologically similar perturbations, we form soft targets from RNA-RNA prototype similarity.
Define:
\begin{equation}
S_{ij}=\hat{\zb}_i^\top \hat{\zb}_j,\qquad i,j\in\mathcal{P},
\end{equation}
(set $S_{ii}=1$), and a temperature-softened target distribution:
\begin{equation}
T_{ij}
=
\frac{\exp(S_{ij}/\tau_{\text{soft}})}{\sum_{j'\in\mathcal{P}} \exp(S_{ij'}/\tau_{\text{soft}})},
\qquad
\tau_{\text{soft}}=0.1.
\end{equation}

\subsection{Bidirectional Soft-InfoNCE Loss}
We compute cross-modal logits $L_{ij}=\tau\,\hat{\ub}_i^\top \hat{\zb}_j$, where $\tau=\exp(\theta)$ is learnable (initialized from $1/\tau_0$ with $\tau_0{=}0.07$) and clamped to $[5,33]$.
We define row- and column-normalized predicted probabilities and soft targets:
\begin{equation}
\begin{aligned}
p^{\rightarrow}_{ij}
&=
\frac{\exp(L_{ij})}{\sum_{j'}\exp(L_{ij'})},
&\qquad
p^{\leftarrow}_{ij}
&=
\frac{\exp(L_{ij})}{\sum_{i'}\exp(L_{i'j})},
\\
T^{\rightarrow}_{ij}
&=
\frac{\exp(S_{ij}/\tau_{\mathrm{soft}})}
{\sum_{j'}\exp(S_{ij'}/\tau_{\mathrm{soft}})},
&\qquad
T^{\leftarrow}_{ij}
&=
\frac{\exp(S_{ij}/\tau_{\mathrm{soft}})}
{\sum_{i'}\exp(S_{i'j}/\tau_{\mathrm{soft}})}.
\end{aligned}
\end{equation}
Here $T^{\rightarrow}_{ij}=T_{ij}$; because $S$ is symmetric, $T^{\leftarrow}_{ij}=T_{ji}$.
The alignment loss is the average cross-entropy to the soft targets in both directions:
\begin{equation}
\mathcal{L}_{\text{align}}
=
-\frac{1}{2P}\sum_{i\in\mathcal{P}}\sum_{j\in\mathcal{P}}
\big(T^{\rightarrow}_{ij}\log p^{\rightarrow}_{ij}+T^{\leftarrow}_{ij}\log p^{\leftarrow}_{ij}\big).
\end{equation}

\subsection{EMA Stabilization and Gradient Flow}
To reduce minibatch noise, we maintain an EMA bank of RNA prototypes per global perturbation id $\mathrm{gid}_k$:
\begin{equation}
\Delta\bar{\zb}^{\text{EMA}}_k
\leftarrow
\beta\,\Delta\bar{\zb}^{\text{EMA}}_k+(1-\beta)\,\Delta\bar{\zb}_k,
\qquad
\beta=0.9.
\end{equation}
When enabled, we use $\Delta\bar{\zb}^{\text{EMA}}_k$ in place of $\Delta\bar{\zb}_k$ when forming $\hat{\zb}_k$ and $T_{ij}$.
 By default, we detach RNA targets (\texttt{detach\_rna=True}) so gradients update only the DNA adaptor/projection and not the (frozen) RNA encoder.

\subsection{Hyperparameters}
Table~\ref{tab:clip_hyperparams} summarizes hyperparameters.

\begin{table}[h]
\centering
\caption{Hyperparameters for sequence-expression alignment.}
\label{tab:clip_hyperparams}
\begin{tabular}{lc}
\toprule
\textbf{Hyperparameter} & \textbf{Value} \\
\midrule
DNA projection hidden dim ($d_h$) & 128 \\
Projection output dim ($d_{\text{RNA}}$) & varies by RNA encoder \\
Initial temperature ($\tau_0$) & 0.07 \\
Temperature clamp range & $[5.0, 33.0]$ \\
Soft target temperature ($\tau_{\text{soft}}$) & 0.1 \\
EMA momentum ($\beta$) & 0.9 \\
Loss weight ($\lambda_{\text{align}}$) & 1.0 \\
Detach RNA gradients & True \\
\bottomrule
\end{tabular}
\end{table}

\section{Hurdle Truncated-Gaussian Residual Likelihood}
\label{eq:guassian-loss}
We model nonnegative gene expression using a \emph{hurdle} distribution composed of a point mass at zero and a continuous density for strictly positive values. Let $y_{ij}\ge 0$ denote the observed (log-normalized) expression of gene $j\in\{1,\dots,G\}$ in cell $i$, and let $x^{\mathrm{ref}}_{ij}\ge 0$ be the decoded control/reference baseline. We define the residual:
\begin{equation}
r_{ij} \;\triangleq\; y_{ij}-x^{\mathrm{ref}}_{ij},
\end{equation}
and enforce nonnegativity without explicit clamping by modeling $r_{ij}$ with a Normal distribution truncated below at:
\begin{equation}
a_{ij} \;\triangleq\; -x^{\mathrm{ref}}_{ij},
\end{equation}
so that $r_{ij}\in[a_{ij},\infty)$ and equivalently $y_{ij}=x^{\mathrm{ref}}_{ij}+r_{ij}\ge 0$.

\xhdr{Hurdle Gate}
The decoder outputs a Bernoulli logit $\ell_{ij}$ determining whether a gene is nonzero:
\begin{equation}
p_{ij}^{\mathrm{nz}} \;\triangleq\; \Pr(y_{ij}>0)=\sigma(\ell_{ij}),\qquad
p_{ij}^{0}=1-p_{ij}^{\mathrm{nz}}.
\end{equation}

\xhdr{Truncated Normal Residual Model}
Conditioned on $y_{ij}>0$, the residual follows a lower-truncated Normal distribution,
\begin{equation}
r_{ij}\mid (y_{ij}>0)\;\sim\; \mathrm{TN}\!\left(\mu_{ij},\sigma_{ij}^2;\ a_{ij},\infty\right),
\end{equation}
with parameters $(\mu_{ij},\sigma_{ij})$ predicted by the decoder. Let $\phi(\cdot)$ and $\Phi(\cdot)$ denote the standard Normal pdf and cdf. The truncated density for $r\ge a$ is:
\begin{equation}
p_{\mathrm{TN}}(r;\mu,\sigma,a)
=
\frac{1}{\sigma}\,
\frac{\phi\!\left(\frac{r-\mu}{\sigma}\right)}
{1-\Phi\!\left(\frac{a-\mu}{\sigma}\right)}.
\end{equation}
Defining:
\begin{equation}
z_{ij}=\frac{r_{ij}-\mu_{ij}}{\sigma_{ij}},\qquad
\alpha_{ij}=\frac{a_{ij}-\mu_{ij}}{\sigma_{ij}},
\end{equation}
the corresponding log-density is:
\begin{equation}
\log p_{\mathrm{TN}}(r_{ij})
=
\log \phi(z_{ij})-\log\sigma_{ij}-\log\!\bigl(1-\Phi(\alpha_{ij})\bigr).
\end{equation}

\xhdr{Log-likelihood}
Combining the hurdle gate and truncated residual model yields:
\begin{equation}
\log p(y_{ij}\mid x^{\mathrm{ref}}_{ij})
=
\mathbb{I}[y_{ij}=0]\log p_{ij}^{0}
+
\mathbb{I}[y_{ij}>0]\left(
\log p_{ij}^{\mathrm{nz}}+\log p_{\mathrm{TN}}(r_{ij})
\right).
\end{equation}

\xhdr{Reconstruction Objective}
We minimize the negative mean log-likelihood over genes and cells:
\begin{equation}
\mathcal{L}_{\mathrm{rec}}
=
-\frac{1}{|\mathcal{B}|\,G}
\sum_{i\in\mathcal{B}}
\sum_{j=1}^{G}
\log p(y_{ij}\mid x^{\mathrm{ref}}_{ij}),
\end{equation}
with an optional weighted variant when per-cell weights are provided. A mild variance regularizer $\lambda_{\sigma}\,\mathbb{E}[\sigma_{ij}^2]$ is optionally added.

\xhdr{Conditional Mean and Nonnegativity}
For the continuous component, the conditional mean of a lower-truncated Normal satisfies:
\begin{equation}
\mathbb{E}[r\mid r\ge a]
=
\mu+\sigma\,\lambda(\alpha),\qquad
\lambda(\alpha)=\frac{\phi(\alpha)}{1-\Phi(\alpha)},
\end{equation}
where $\lambda(\alpha)$ is the Normal Mills ratio. Consequently,
\begin{equation}
\widehat{y}_{ij}^{\,\mathrm{nz}}
=
x^{\mathrm{ref}}_{ij}+\mathbb{E}[r_{ij}\mid r_{ij}\ge a_{ij}]
\;\ge\;0,
\end{equation}
ensuring nonnegative predictions without ad hoc constraints.

\section{Model Architecture and Training Details}
\xhdr{Model Components}
Our architecture integrates four primary components:
(i) a DNA encoder/adaptor utilizing Borzoi-derived embeddings;
(ii) a conditional diffusion backbone  trained with an I2SB/DDPM objective;
(iii) a supervised RNA predictor head; and
(iv) auxiliary objectives, including CLIP alignment and optional consistency matching.
\xhdr{Notation and Conditioning}
Let $x_c, x_p \in \mathbb{R}^{B \times G}$ denote the control and perturbed RNA profiles, respectively, with corresponding latent representations $z_c, z_p \in \mathbb{R}^{B \times d}$ obtained from the pretrained RNA encoder $e_\psi$.
Let $u_p = g_\phi(E_s, m) \in \mathbb{R}^{B \times d_0}$ denote the perturbation embedding produced by the DNA Perturbation Module.
We construct the projected DNA condition via a linear transformation followed by Layer Normalization:
\begin{align}
    \Delta c_{\mathrm{diff}} &= \operatorname{LN}(W_{\mathrm{diff}} \, u_p) \in \mathbb{R}^{B \times d_c}.
\end{align}
We also maintain a learned null (cell-line specific) embedding $n_{\mathrm{diff}}(u) \in \mathbb{R}^{d_c}$:
\begin{equation}
    c^{\mathrm{null}}_{\mathrm{diff}} = \operatorname{LN}(n_{\mathrm{diff}}(u)) \in \mathbb{R}^{B \times d_c}.
\end{equation}
\xhdr{Training Step Regimes}
To enable classifier-free guidance and robust conditioning, each training iteration samples one of three conditioning modes for the diffusion condition $c_{\mathrm{diff}}$:
\begin{enumerate}
    \item \textbf{Unconditional Mode (\texttt{unconditional}):}
    Uses only null embeddings to serve as the unconditional anchor for guidance. To maintain Distributed Data Parallel (DDP) graph connectivity, the DNA projections are retained with a zero multiplier:
    \begin{equation}
        c_{\mathrm{diff}} = c^{\mathrm{null}}_{\mathrm{diff}}.
    \end{equation}
   \item \textbf{Source-Conditional Mode (\texttt{p\_to\_p}):}
    Utilizes sequence conditioning with source recentering based on Exponential Moving Average (EMA) statistics. We define the shift $\delta z = \hat{\mu} - \hat{\mu}_{c}$, where $\hat{\mu}$ denotes the debiased EMA statistics of the respective latents. The source latent is defined as $z_{\mathrm{src}} = z_{c} + \delta z$.
    In this mode, only DNA projections are active:
    \begin{equation}
        c_{\mathrm{diff}} = \Delta c_{\mathrm{diff}}.
    \end{equation}
    This branch aligns the projected perturbation embedding with the batch-corrected expression level change using CLIP loss
    $ \mathcal{L}_{\mathrm{clip}} = \operatorname{CLIP}(\Delta c_{\mathrm{diff}}, z_{p}, \ell, z_{\mathrm{src}}). $
    \item \textbf{Regular Mode (\texttt{regular}):}
    The standard conditional mode combining both sequence-specific and null information:
    \begin{equation}
        c_{\mathrm{diff}} = \Delta c_{\mathrm{diff}} + c^{\mathrm{null}}_{\mathrm{diff}}.
    \end{equation}
\end{enumerate}

\xhdr{Optimization Objectives}
Given a timestep $t \in \{0, \dots, T\}$, the primary losses follow Section~\ref{sec:method}. The bridge matching loss trains the score network $e_\theta$ to transport control latents to perturbed latents conditioned on the perturbation embedding:
\begin{equation}
    \mathcal{L}_{\mathrm{bridge}} = \mathbb{E}_{z_p, z_c, t}\!\Big[\big\|e_\theta(z_t,\, t,\, u_p,\, z_c) - \epsilon^{\mathrm{I2SB}}\big\|_2^2\Big],
\end{equation}
where $z_t$ is the bridge interpolant (Eq.~1) and $u_p$ is conditioned through $c_{\mathrm{diff}}$ as described above. The reconstruction loss $\mathcal{L}_{\mathrm{rec}}$ evaluates the decoded prediction $\hat{x}_p$ against the observed perturbed expression $x_p$ using the hurdle truncated-Gaussian residual likelihood (App.~\ref{eq:guassian-loss}). The alignment loss $\mathcal{L}_{\mathrm{align}}$ is computed via the CLIP objective (App.~\ref{sec:supp_align}) during the source-conditional training mode.

\xhdr{Total Loss}
The model is trained end-to-end by minimizing a weighted combination of losses (Section~\ref{sec:overall-loss}):
$\mathcal{L} = \lambda_1 \mathcal{L}_{\mathrm{bridge}} + \lambda_2 \mathcal{L}_{\mathrm{align}} + \lambda_3 \mathcal{L}_{\mathrm{rec}}$,
where we utilize uncertainty weighting~\citep{kendall2018multi} with learnable parameters $\sigma_i$ to dynamically adjust each $\lambda_i$:
\begin{equation}
    \mathcal{L}_{\mathrm{total}} = \sum_{i} \frac{1}{2} \left( \frac{\mathcal{L}_i}{\sigma_i^2} + \log \sigma_i^2 \right).
\end{equation}

\section{Data and Evaluation Protocols}
\label{sec:supp_eval}

\subsection{Dataset Statistics}
\label{sec:supp_data_stats}

We adopt the gene-level train/test splits from PerturbQA~\cite{wu2025contextualizing}, which partitions perturbations such that test perturbation genes are entirely absent from training. This setup evaluates a model's ability to generalize to novel genetic perturbations rather than interpolating among seen ones.

\xhdr{Few-shot Setting}
To simulate realistic data scarcity—where only limited observations exist for newly studied perturbations—we modify the zero-shot splits by transferring a small number of cells from the test set into training. Specifically, for each test perturbation gene, we randomly sample $k=5$ cells and move them to the training set, providing minimal supervision while preserving the challenge of limited data.

This procedure yields the following train/test splits:

\begin{table}[ht]
\centering
\footnotesize
\setlength{\tabcolsep}{4pt}
\begin{tabular}{lrrr}
\toprule
\textbf{Dataset} &
\textbf{Genes} &
\textbf{Samples} \\
& &
\textbf{(Train / Test)} \\
\midrule
K562   & 4{,}136 & 318{,}443 /  61{,}233 \\
Jurkat & 6{,}842 & 155{,}393 /  36{,}073 \\
RPE1   & 4{,}760 & 163{,}453 /  46{,}053 \\
\midrule
Combined
       & 2{,}000 & 637{,}289 / 143{,}359 \\
\bottomrule
\end{tabular}
\caption{\textbf{Few-shot dataset statistics}. Train and test counts are reported as paired values.}
\label{stab:low-sample-data-summary}
\end{table}

\xhdr{Zero-Shot Setting}
In the zero-shot setting, we use the original PerturbQA splits without modification. Models are trained exclusively on designated training perturbations and evaluated on held-out test perturbations without any access to test labels or test expression profiles during training.

\begin{table}[ht]
\centering
\footnotesize
\setlength{\tabcolsep}{4pt}
\begin{tabular}{lrrrr}
\toprule
\textbf{Dataset} &
\textbf{Perturb.} &
\textbf{Genes} &
\textbf{Samples} \\
&
\textbf{(Train / Test)} &
&
\textbf{(Train / Test)} \\
\midrule
K562   & 1{,}564 /   267 & 4{,}136 & 317{,}103 /  62{,}573 \\
Jurkat & 1{,}227 /   313 & 6{,}842 & 153{,}823 /  37{,}643 \\
RPE1   & 1{,}596 /   406 & 4{,}760 & 161{,}418 /  48{,}088 \\
\midrule
Combined
       & 2{,}499 /   886 & 2{,}000 & 632{,}339 / 148{,}309 \\
\bottomrule
\end{tabular}
\caption{\textbf{Zero-shot dataset statistics}. Train and test counts are reported as paired values.}
\label{stab:zero-shot-data-summary}
\end{table}

\xhdr{Control Cell Partitioning}
Control (unperturbed) cells serve as the reference distribution for computing perturbation effects. To prevent information leakage between training and evaluation, we partition control cells into two disjoint subsets:
\begin{itemize}
    \item \textbf{Training controls (80\%)}: Used during model training to compute perturbation effects for training genes and as input conditions for generative models.
    \item \textbf{Test controls (20\%)}: Reserved for evaluation; used as input conditions when predicting responses to test perturbations.
\end{itemize}
This partitioning is performed randomly with a fixed seed for reproducibility. The same control partition is used across all methods to ensure fair comparison.

\subsection{Metric Definitions}
\label{sec:supp_metrics}

\begin{itemize}[leftmargin=5pt]
    \item \emph{Mean Squared Error} (MSE), measuring pointwise reconstruction error across all genes. For $N$ perturbations in the test set, let $\pb_i \in \mathbb{R}^G$ denote the ground-truth pseudobulk expression and $\hat{\pb}_i \in \mathbb{R}^G$ denote the predicted pseudobulk expression. We compute:
    \begin{equation}
    \mathrm{MSE}=\frac{1}{NG}\sum_{i=1}^{N}\left\lVert \hat{\pb}_i-\pb_i \right\rVert_2^2.
    \end{equation}
    \item \emph{Direction Match} (Dir.\ Match), evaluating whether the predicted direction of change (up- or down-regulation) for differentially expressed (DE) genes is correct. For each perturbation $t$, $G_t^{\cap}$ is the intersection of the predicted and true DE gene sets using an adjusted $p$-value threshold of $0.05$. Given true log-fold changes $\Delta_{t,g}$ and predicted log-fold changes $\hat{\Delta}_{t,g}$, we compute:
    \begin{equation}
    \mathrm{DirMatch}_t
    =\frac{\left|\left\{ g \in G_t^{\cap} : \mathrm{sgn}\!\left(\hat{\Delta}_{t,g}\right)=\mathrm{sgn}\!\left(\Delta_{t,g}\right)\right\}\right|}
    {\left|G_t^{\cap}\right|}.
    \end{equation}
    \item \emph{Pearson correlation of expression change} ($\Delta$Pearson), measuring correlation between predicted and observed gene-wise perturbation effects. For each perturbation $t$, let the expression change $\Delta_t$ be the element-wise difference between the mean expression of the perturbed cells $\bar{\pb}_t$ and the control cells $\bar{\pmb{c}}_t$: $\Delta_t = \bar{\pb}_t - \bar{\pmb{c}}_t$. Given the predicted expression change $\Delta^{\mathrm{pred}}$ and true expression change $\Delta^{\mathrm{true}}$, we compute the Pearson correlation of expression change as:
    \begin{equation}
    \Delta\mathrm{Pearson} = \mathrm{corr}(\Delta^{\mathrm{pred}}, \Delta^{\mathrm{true}}).
    \end{equation}
\item \emph{Precision@N} (Prec.@N) and \emph{Overlap@N} (Ov.@N).

For each perturbation $t$, let:
\[
S_t^{\mathrm{pred}}=\{g:q_{t,g}^{\mathrm{pred}}<0.05\},
\qquad
S_t^{\mathrm{real}}=\{g:q_{t,g}^{\mathrm{real}}<0.05\}
\]
denote the predicted and real FDR-significant DE gene sets. Let
$\operatorname{Top}_m(S)$ denote the top $m$ genes in set $S$, ranked by
absolute log-fold change.

Precision@N is computed using:
\[
N_{\mathrm{prec}}=|S_t^{\mathrm{pred}}|,
\]
\[
\mathrm{Precision}_{t,N}
=
\frac{
\left|
\operatorname{Top}_{k_{\mathrm{prec}}}(S_t^{\mathrm{pred}})
\cap
\operatorname{Top}_{k_{\mathrm{prec}}}(S_t^{\mathrm{real}})
\right|
}{
N_{\mathrm{prec}}
}.
\]

Overlap@N is computed using:
\[
N_{\mathrm{ov}}=|S_t^{\mathrm{real}}|,
\]
\[
\mathrm{Overlap}_{t,N}
=
\frac{
\left|
\operatorname{Top}_{k_{\mathrm{ov}}}(S_t^{\mathrm{pred}})
\cap
\operatorname{Top}_{k_{\mathrm{ov}}}(S_t^{\mathrm{real}})
\right|
}{
N_{\mathrm{ov}}
}.
\]
    \item \emph{Discrimination Score (Cosine)} (Discr.\ Cos), a rank-based metric that assesses whether each predicted perturbation effect is most similar to its corresponding ground-truth effect (via cosine similarity) relative to all other perturbations~\cite{wu2024perturbench}. A higher score in this metric signifies a strong capacity to prevent mode collapse. Let $\mathcal{P}=\{1,\dots,T\}$ be the set of perturbations. For any $t, p \in \mathcal{P}$, let $\hat{\mathbf{y}}_t$ denote the predicted effect for perturbation $t$, and $\mathbf{y}_p$ denote the ground-truth effect for perturbation mask $\mb$. We define the cosine distance as:
    \begin{equation}
    d_{\cos}(\hat{\mathbf{y}}_t, \mathbf{y}_p) = 1 - \frac{\hat{\mathbf{y}}_t^\top \mathbf{y}_p}{\|\hat{\mathbf{y}}_t\|_2 \|\mathbf{y}_p\|_2}
    \end{equation}
    For each perturbation $t$, we count how many incorrect perturbations are closer to $\hat{\yb}_t$ than the correct one:
    \begin{equation}
    r_t = \sum_{p\neq t} \mathbf{1} \{ d_{\cos}(\hat{\mathbf{y}}_t, \mathbf{y}_p) < d_{\cos}(\hat{\mathbf{y}}_t, \mathbf{y}_t) \}
    \end{equation}
    The final perturbation discrimination score is the normalized average rank across $T$ perturbations:
    \begin{equation}
    \text{Discrimination Score} = 1 - \frac{1}{T}\sum_{t=1}^T \frac{r_t}{T-1}
    \end{equation}
    \item \emph{Energy Distance}:
To compare the distribution of real and predicted single-cell profiles for a given perturbation, we use the (unsquared) energy distance.
For distributions $P,Q$ over $\mathbb{R}^d$,
\begin{equation}
\mathcal{E}(P,Q)
= 2\,\mathbb{E}\|\Xb-\Yb\|_2
- \mathbb{E}\|\Xb-\Xb'\|_2
- \mathbb{E}\|\Yb-\Yb'\|_2,
\end{equation}
For each dataset, we fit a PCA map $T(\cdot)$ (50 components) on the training split and compute a per-perturbation score
$\widehat{\mathcal{E}}\!\left(T(\Xb^{\mathrm{real}}_g),T(\Xb^{\mathrm{pred}}_g)\right)$
using cells matching perturbation $g$. Energy distance is computed per-perturbation.

\item  \emph{Pseudo-bulk mean $R^2$.}
Let $G$ denote the number of genes.
For a perturbation $g$, let $\mathbf{X}^{\mathrm{real}}_g \in \mathbb{R}^{n_g \times G}$ and
$\mathbf{X}^{\mathrm{pred}}_g \in \mathbb{R}^{m_g \times G}$ denote the real and predicted gene expression matrices,
with rows corresponding to cells and columns to genes.
We form pseudo-bulk mean expression vectors by averaging across cells,
\[
\boldsymbol{\mu}^{\mathrm{real}}_g = \frac{1}{n_g}\sum_{i=1}^{n_g} \mathbf{X}^{\mathrm{real}}_{g,i:},
\qquad
\boldsymbol{\mu}^{\mathrm{pred}}_g = \frac{1}{m_g}\sum_{j=1}^{m_g} \mathbf{X}^{\mathrm{pred}}_{g,j:},
\quad
\boldsymbol{\mu}^{\mathrm{real}}_g, \boldsymbol{\mu}^{\mathrm{pred}}_g \in \mathbb{R}^G.
\]
We compute the coefficient of determination across genes,
\[
R^2_g
=
1 - \frac{\lVert \boldsymbol{\mu}^{\mathrm{real}}_g - \boldsymbol{\mu}^{\mathrm{pred}}_g \rVert_2^2}
{\lVert \boldsymbol{\mu}^{\mathrm{real}}_g - \bar{\mu}^{\mathrm{real}}_g \mathbf{1} \rVert_2^2},
\qquad
\bar{\mu}^{\mathrm{real}}_g = \frac{1}{G}\sum_{k=1}^G \mu^{\mathrm{real}}_{g,k},
\]
where $\mathbf{1}\in\mathbb{R}^G$ is the all-ones vector.
We compute $R^2_g$ for each perturbation and summarize results across perturbations for each method.

\end{itemize}

\section{Detailed Performance Benchmarks}
\label{sec:detailed_results}

We provide a comprehensive breakdown of performance metrics across all datasets and settings.

\begin{table*}[h]
\centering
\scriptsize
\setlength{\tabcolsep}{4pt}
\caption{
\textbf{Few shot gene perturbation performance comparison}
Values are reported as mean $\pm$ standard error.
For readability, Ov.@N, Prec.@N, Dir.Match, and Disc.Cos are \emph{scaled by} $\times 10$,
and MSE is \emph{scaled by} $\times 100$.
 Avg. Rank measures the overall rank averaged across all metrics, computed using unrounded raw metric values. Best Avg. Rank is shown in \textbf{bold}.
}
\resizebox{\textwidth}{!}{%
\label{tab:fewshot_results_scaled}
\begin{tabular}{llccccc cc|c}
\toprule
 &  &
 \multicolumn{2}{c}{\textbf{Distributional Fidelity}} &
 \multicolumn{3}{c}{\textbf{DE Signal}} &
 \multicolumn{2}{c}{\textbf{Global Fit}} &
 \textbf{Summary} \\
\cmidrule(lr){3-4}
\cmidrule(lr){5-7}
\cmidrule(lr){8-9}
\cmidrule(lr){10-10}
\textbf{Dataset} & \textbf{Method} &
\textbf{Discr. Cos $\uparrow$} &
\textbf{E-Dist $\downarrow$} &
\textbf{Dir. Match $\uparrow$} &
\textbf{Prec.@N $\uparrow$} &
\textbf{Ov.@N $\uparrow$} &
\textbf{$\Delta$Pearson $\uparrow$} &
\textbf{MSE $\downarrow$} &
\textbf{Avg. Rank $\downarrow$} \\
\midrule
\multirow{6}{*}{\textbf{K562}}
 & BioLord & 5.02 $\pm$ 0.18 & 7.72 $\pm$ 0.11 & 5.13 $\pm$ 0.04 & 0.66 $\pm$ 0.03 & 0.31 $\pm$ 0.02 & 0.02 $\pm$ 0.00 & 0.95 $\pm$ 0.05 & 4.86 \\
 & GEARS & 5.12 $\pm$ 0.17 & 0.96 $\pm$ 0.06 & 6.63 $\pm$ 0.06 & 0.63 $\pm$ 0.03 & 0.51 $\pm$ 0.03 & 0.22 $\pm$ 0.01 & 0.92 $\pm$ 0.04 & 3.14 \\
 & Linear & 5.04 $\pm$ 0.18 & 7.56 $\pm$ 0.10 & 3.58 $\pm$ 0.05 & 0.18 $\pm$ 0.01 & 0.51 $\pm$ 0.02 & 0.11 $\pm$ 0.00 & 0.81 $\pm$ 0.04 & 4.14 \\
 & PerturbMean & 5.02 $\pm$ 0.18 & 7.56 $\pm$ 0.10 & 8.50 $\pm$ 0.05 & 0.15 $\pm$ 0.01 & 0.64 $\pm$ 0.03 & 0.55 $\pm$ 0.01 & 0.78 $\pm$ 0.04 & 2.86 \\
 & STATE & 5.38 $\pm$ 0.17 & 12.73 $\pm$ 0.12 & 5.23 $\pm$ 0.12 & 0.66 $\pm$ 0.03 & 0.32 $\pm$ 0.02 & 0.05 $\pm$ 0.01 & 5.36 $\pm$ 0.27 & 4.43 \\
\rowcolor{gray!10}  & STRAND & 7.10 $\pm$ 0.17 & 1.53 $\pm$ 0.05 & 7.47 $\pm$ 0.10 & 0.67 $\pm$ 0.03 & 1.05 $\pm$ 0.06 & 0.33 $\pm$ 0.01 & 0.80 $\pm$ 0.04 & \textbf{1.57} \\
\midrule
\multirow{6}{*}{\textbf{Jurkat}}
 & BioLord & 5.07 $\pm$ 0.16 & 8.48 $\pm$ 0.13 & 6.13 $\pm$ 0.06 & 0.44 $\pm$ 0.03 & 0.18 $\pm$ 0.02 & 0.14 $\pm$ 0.00 & 0.95 $\pm$ 0.05 & 4.43 \\
 & GEARS & 5.64 $\pm$ 0.17 & 1.27 $\pm$ 0.07 & 6.71 $\pm$ 0.06 & 0.40 $\pm$ 0.03 & 0.23 $\pm$ 0.02 & 0.24 $\pm$ 0.01 & 0.86 $\pm$ 0.05 & 3.14 \\
 & Linear & 5.03 $\pm$ 0.16 & 8.31 $\pm$ 0.11 & 3.66 $\pm$ 0.04 & 0.18 $\pm$ 0.03 & 0.56 $\pm$ 0.02 & 0.09 $\pm$ 0.00 & 0.78 $\pm$ 0.05 & 4.29 \\
 & PerturbMean & 5.02 $\pm$ 0.16 & 8.31 $\pm$ 0.11 & 8.01 $\pm$ 0.05 & 0.07 $\pm$ 0.01 & 0.25 $\pm$ 0.02 & 0.45 $\pm$ 0.00 & 0.70 $\pm$ 0.03 & 3.00 \\
 & STATE & 6.52 $\pm$ 0.17 & 13.57 $\pm$ 0.20 & 5.84 $\pm$ 0.11 & 0.45 $\pm$ 0.03 & 0.20 $\pm$ 0.02 & 0.13 $\pm$ 0.01 & 4.54 $\pm$ 0.15 & 4.29 \\
\rowcolor{gray!10}  & STRAND & 7.19 $\pm$ 0.16 & 1.82 $\pm$ 0.07 & 7.58 $\pm$ 0.09 & 0.44 $\pm$ 0.03 & 0.64 $\pm$ 0.04 & 0.34 $\pm$ 0.01 & 0.78 $\pm$ 0.04 & \textbf{1.86} \\
\midrule
\multirow{6}{*}{\textbf{RPE1}}
 & BioLord & 5.03 $\pm$ 0.14 & 11.24 $\pm$ 0.21 & 6.72 $\pm$ 0.05 & 0.99 $\pm$ 0.04 & 0.63 $\pm$ 0.03 & 0.26 $\pm$ 0.01 & 1.99 $\pm$ 0.08 & 4.86 \\
 & GEARS & 5.41 $\pm$ 0.15 & 2.36 $\pm$ 0.11 & 7.59 $\pm$ 0.08 & 1.00 $\pm$ 0.04 & 1.36 $\pm$ 0.06 & 0.44 $\pm$ 0.01 & 1.62 $\pm$ 0.06 & 3.00 \\
 & Linear & 5.05 $\pm$ 0.14 & 10.25 $\pm$ 0.15 & 4.31 $\pm$ 0.05 & 2.17 $\pm$ 0.07 & 1.73 $\pm$ 0.05 & 0.33 $\pm$ 0.01 & 1.55 $\pm$ 0.06 & 3.29 \\
 & PerturbMean & 5.01 $\pm$ 0.14 & 10.26 $\pm$ 0.15 & 8.69 $\pm$ 0.05 & 0.37 $\pm$ 0.02 & 0.80 $\pm$ 0.02 & 0.57 $\pm$ 0.01 & 1.46 $\pm$ 0.06 & 3.43 \\
 & STATE & 6.97 $\pm$ 0.14 & 11.42 $\pm$ 0.19 & 6.03 $\pm$ 0.08 & 1.02 $\pm$ 0.04 & 0.57 $\pm$ 0.03 & 0.19 $\pm$ 0.01 & 5.27 $\pm$ 0.07 & 4.71 \\
\rowcolor{gray!10}  & STRAND & 7.04 $\pm$ 0.14 & 2.62 $\pm$ 0.10 & 7.93 $\pm$ 0.08 & 1.01 $\pm$ 0.04 & 1.76 $\pm$ 0.07 & 0.47 $\pm$ 0.01 & 1.39 $\pm$ 0.05 & \textbf{1.71} \\
\bottomrule
\end{tabular}
}
\label{stab:low-sample}
\end{table*}

\begin{table*}[h]
\centering
\scriptsize
\setlength{\tabcolsep}{4pt}
\renewcommand{\arraystretch}{1.15}
\caption{
\textbf{Zero-shot gene perturbation performance comparison}
Values are reported as mean $\pm$ standard error.
For readability, Ov.@N, Prec.@N, Dir.Match, and Disc.Cos are \emph{scaled by} $\times 10$,
and MSE is \emph{scaled by} $\times 100$. (N/A) represents that the method does not apply to the task and is therefore not evaluated.  Avg. Rank measures the overall rank averaged across all metrics, computed using unrounded raw metric values. Best Avg. Rank is shown in \textbf{bold}.
}
\resizebox{\textwidth}{!}{%
\label{tab:zeroshot_gene_results_scaled}
\begin{tabular}{llccccc cc|c}
\toprule
 &  &
 \multicolumn{2}{c}{\textbf{Distributional Fidelity}} &
 \multicolumn{3}{c}{\textbf{DE Signal}} &
 \multicolumn{2}{c}{\textbf{Global Fit}} &
 \textbf{Summary} \\
\cmidrule(lr){3-4}
\cmidrule(lr){5-7}
\cmidrule(lr){8-9}
\cmidrule(lr){10-10}
\textbf{Dataset} & \textbf{Method} &
\textbf{Discr. Cos $\uparrow$} &
\textbf{E-Dist $\downarrow$} &
\textbf{Dir. Match $\uparrow$} &
\textbf{Prec.@N $\uparrow$} &
\textbf{Ov.@N $\uparrow$} &
\textbf{$\Delta$Pearson $\uparrow$} &
\textbf{MSE $\downarrow$} &
\textbf{Avg. Rank $\downarrow$} \\
\midrule
\multirow{5}{*}{\textbf{K562}}
 & BioLord & 5.07 $\pm$ 0.18 & 7.70 $\pm$ 0.11 & 5.31 $\pm$ 0.04 & 0.68 $\pm$ 0.03 & 0.31 $\pm$ 0.02 & 0.04 $\pm$ 0.00 & 0.92 $\pm$ 0.05 & 3.29 \\
 & Linear & 5.06 $\pm$ 0.18 & 7.56 $\pm$ 0.10 & 3.46 $\pm$ 0.05 & 0.18 $\pm$ 0.01 & 0.49 $\pm$ 0.02 & 0.11 $\pm$ 0.00 & 0.78 $\pm$ 0.04 & 3.00 \\
 & PerturbMean & 5.02 $\pm$ 0.18 & 7.56 $\pm$ 0.10 & 8.50 $\pm$ 0.05 & 0.15 $\pm$ 0.01 & 0.64 $\pm$ 0.03 & 0.55 $\pm$ 0.01 & 0.78 $\pm$ 0.04 & 2.14 \\
 & STATE & N/A & N/A & N/A & N/A & N/A & N/A & N/A & N/A \\
\rowcolor{gray!10}  & STRAND & 5.14 $\pm$ 0.18 & 1.74 $\pm$ 0.08 & 6.58 $\pm$ 0.10 & 0.69 $\pm$ 0.03 & 0.85 $\pm$ 0.04 & 0.21 $\pm$ 0.01 & 0.85 $\pm$ 0.04 & \textbf{1.57} \\
\midrule
\multirow{5}{*}{\textbf{Jurkat}}
 & BioLord & 5.05 $\pm$ 0.16 & 8.51 $\pm$ 0.13 & 5.88 $\pm$ 0.04 & 0.46 $\pm$ 0.03 & 0.20 $\pm$ 0.02 & 0.10 $\pm$ 0.00 & 0.84 $\pm$ 0.04 & 3.00 \\
 & Linear & 5.02 $\pm$ 0.16 & 8.31 $\pm$ 0.11 & 3.56 $\pm$ 0.04 & 0.17 $\pm$ 0.03 & 0.56 $\pm$ 0.02 & 0.09 $\pm$ 0.00 & 0.72 $\pm$ 0.03 & 2.86 \\
 & PerturbMean & 5.02 $\pm$ 0.16 & 8.31 $\pm$ 0.11 & 8.01 $\pm$ 0.05 & 0.07 $\pm$ 0.01 & 0.25 $\pm$ 0.02 & 0.45 $\pm$ 0.00 & 0.70 $\pm$ 0.03 & 2.29 \\
 & STATE & N/A & N/A & N/A & N/A & N/A & N/A & N/A & N/A \\
\rowcolor{gray!10}  & STRAND & 5.26 $\pm$ 0.16 & 1.97 $\pm$ 0.09 & 7.04 $\pm$ 0.09 & 0.46 $\pm$ 0.03 & 0.52 $\pm$ 0.04 & 0.29 $\pm$ 0.01 & 0.76 $\pm$ 0.03 & \textbf{1.86} \\
\midrule
\multirow{5}{*}{\textbf{RPE1}}
 & BioLord & 5.04 $\pm$ 0.14 & 11.29 $\pm$ 0.21 & 6.90 $\pm$ 0.06 & 1.08 $\pm$ 0.04 & 0.74 $\pm$ 0.03 & 0.26 $\pm$ 0.01 & 1.92 $\pm$ 0.08 & 3.43 \\
 & Linear & 5.02 $\pm$ 0.14 & 10.26 $\pm$ 0.15 & 4.27 $\pm$ 0.05 & 2.19 $\pm$ 0.07 & 1.73 $\pm$ 0.05 & 0.33 $\pm$ 0.01 & 1.47 $\pm$ 0.06 & 2.43 \\
 & PerturbMean & 5.01 $\pm$ 0.14 & 10.28 $\pm$ 0.15 & 8.69 $\pm$ 0.05 & 0.37 $\pm$ 0.02 & 0.80 $\pm$ 0.02 & 0.57 $\pm$ 0.01 & 1.46 $\pm$ 0.06 & 2.43 \\
 & STATE & N/A & N/A & N/A & N/A & N/A & N/A & N/A & N/A \\
\rowcolor{gray!10}  & STRAND & 5.13 $\pm$ 0.14 & 3.13 $\pm$ 0.14 & 7.86 $\pm$ 0.08 & 1.09 $\pm$ 0.04 & 1.84 $\pm$ 0.07 & 0.48 $\pm$ 0.01 & 1.53 $\pm$ 0.06 & \textbf{1.71} \\
\bottomrule
\end{tabular}
}
\label{stab:zero-shot}
\end{table*}

\begin{table}[h]
\scriptsize
    \centering
\setlength{\tabcolsep}{2pt}
\begin{tabular}{llcccccccc}
\toprule
\textbf{Dataset} & \textbf{Method} & \textbf{Discr. Cos $\uparrow$} & \textbf{E-Dist $\downarrow$} & \textbf{Dir. Match $\uparrow$} & \textbf{Prec.@N $\uparrow$} & \textbf{Ov.@N $\uparrow$} & \textbf{$\Delta$Pearson $\uparrow$} & \textbf{MSE $\downarrow$} & \textbf{Avg. Rank $\downarrow$} \\
\midrule
\multirow{6}{*}{\textbf{K562}}
 & BioLord & 5.06 $\pm$ 0.22 & 7.72 $\pm$ 0.13 & 5.28 $\pm$ 0.04 & 0.70 $\pm$ 0.04 & 0.32 $\pm$ 0.03 & 0.04 $\pm$ 0.00 & 0.93 $\pm$ 0.06 & 4.14 \\
 & GEARS & 4.82 $\pm$ 0.21 & 1.01 $\pm$ 0.08 & 6.57 $\pm$ 0.06 & 0.69 $\pm$ 0.04 & 0.56 $\pm$ 0.03 & 0.21 $\pm$ 0.01 & 0.90 $\pm$ 0.05 & 2.86 \\
 & Linear & 5.08 $\pm$ 0.21 & 7.58 $\pm$ 0.12 & 3.48 $\pm$ 0.05 & 0.20 $\pm$ 0.02 & 0.50 $\pm$ 0.02 & 0.11 $\pm$ 0.00 & 0.80 $\pm$ 0.05 & 3.57 \\
 & PerturbMean & 4.80 $\pm$ 0.21 & 7.58 $\pm$ 0.12 & 8.43 $\pm$ 0.05 & 0.15 $\pm$ 0.01 & 0.66 $\pm$ 0.03 & 0.54 $\pm$ 0.01 & 0.79 $\pm$ 0.05 & 2.57 \\
 & STATE & N/A & N/A & N/A & N/A & N/A & N/A & N/A & N/A \\
\rowcolor{gray!10}  & STRAND & 5.15 $\pm$ 0.22 & 1.70 $\pm$ 0.09 & 6.53 $\pm$ 0.12 & 0.71 $\pm$ 0.04 & 0.87 $\pm$ 0.05 & 0.21 $\pm$ 0.02 & 0.86 $\pm$ 0.05 & \textbf{1.86} \\
\midrule
\multirow{6}{*}{\textbf{Jurkat}}
 & BioLord & 5.18 $\pm$ 0.19 & 8.63 $\pm$ 0.16 & 5.90 $\pm$ 0.04 & 0.48 $\pm$ 0.04 & 0.21 $\pm$ 0.03 & 0.10 $\pm$ 0.00 & 0.87 $\pm$ 0.05 & 3.86 \\
 & GEARS & 5.77 $\pm$ 0.20 & 1.40 $\pm$ 0.09 & 6.84 $\pm$ 0.07 & 0.46 $\pm$ 0.04 & 0.26 $\pm$ 0.03 & 0.26 $\pm$ 0.01 & 0.83 $\pm$ 0.04 & 2.57 \\
 & Linear & 5.13 $\pm$ 0.19 & 8.42 $\pm$ 0.13 & 3.57 $\pm$ 0.05 & 0.18 $\pm$ 0.03 & 0.56 $\pm$ 0.02 & 0.10 $\pm$ 0.01 & 0.75 $\pm$ 0.04 & 3.57 \\
 & PerturbMean & 4.97 $\pm$ 0.19 & 8.41 $\pm$ 0.13 & 7.99 $\pm$ 0.06 & 0.08 $\pm$ 0.01 & 0.24 $\pm$ 0.02 & 0.45 $\pm$ 0.01 & 0.74 $\pm$ 0.04 & 2.86 \\
 & STATE & N/A & N/A & N/A & N/A & N/A & N/A & N/A & N/A \\
\rowcolor{gray!10}  & STRAND & 5.35 $\pm$ 0.19 & 1.92 $\pm$ 0.10 & 7.12 $\pm$ 0.09 & 0.48 $\pm$ 0.04 & 0.53 $\pm$ 0.05 & 0.29 $\pm$ 0.01 & 0.79 $\pm$ 0.04 & \textbf{2.14} \\
\midrule
\multirow{6}{*}{\textbf{RPE1}}
 & BioLord & 5.13 $\pm$ 0.17 & 11.58 $\pm$ 0.25 & 6.93 $\pm$ 0.06 & 1.15 $\pm$ 0.05 & 0.79 $\pm$ 0.04 & 0.27 $\pm$ 0.01 & 2.00 $\pm$ 0.09 & 3.86 \\
 & GEARS & 5.08 $\pm$ 0.18 & 4.12 $\pm$ 0.22 & 5.10 $\pm$ 0.04 & 1.00 $\pm$ 0.05 & 0.72 $\pm$ 0.04 & 0.01 $\pm$ 0.01 & 2.29 $\pm$ 0.10 & 4.29 \\
 & Linear & 5.25 $\pm$ 0.17 & 10.40 $\pm$ 0.18 & 4.34 $\pm$ 0.06 & 2.31 $\pm$ 0.08 & 1.80 $\pm$ 0.06 & 0.35 $\pm$ 0.01 & 1.51 $\pm$ 0.07 & 2.57 \\
 & PerturbMean & 5.21 $\pm$ 0.17 & 10.42 $\pm$ 0.18 & 8.67 $\pm$ 0.07 & 0.39 $\pm$ 0.02 & 0.84 $\pm$ 0.03 & 0.58 $\pm$ 0.01 & 1.51 $\pm$ 0.07 & 2.57 \\
 & STATE & N/A & N/A & N/A & N/A & N/A & N/A & N/A & N/A \\
\rowcolor{gray!10}  & STRAND & 5.30 $\pm$ 0.18 & 3.06 $\pm$ 0.16 & 7.95 $\pm$ 0.10 & 1.17 $\pm$ 0.05 & 1.99 $\pm$ 0.09 & 0.50 $\pm$ 0.02 & 1.55 $\pm$ 0.07 & \textbf{1.71} \\
\bottomrule
\end{tabular}
  \caption{\textbf{Zero-shot performance comparison with GEARS}. We selected the subset of genes for which GEARS can make predictions. 189 genes for K562, 236 genes for Jurkat, and 269 genes for RPE1. }
    \label{stab:zero-shot-subset}
\end{table}

\begin{table}[h]
\scriptsize
    \centering
\setlength{\tabcolsep}{4pt}
 \caption{\textbf{Ablation on CLIP loss}}
\resizebox{\textwidth}{!}{%
\begin{tabular}{llccccccc|c}
\toprule
\textbf{Dataset} & \textbf{Method} & \textbf{Discr. Cos $\uparrow$} & \textbf{E-Dist $\downarrow$} & \textbf{Dir. Match $\uparrow$} & \textbf{Prec.@N } & \textbf{Ov.@N $\uparrow$} & \textbf{$\Delta$Pearson $\uparrow$} & \textbf{MSE $\downarrow$} & \textbf{Avg. Rank $\downarrow$} \\
\midrule
\multirow{2}{*}{\textbf{K562}}
 & STRAND (w/o CLIP) & 6.81 $\pm$ 0.17 & 1.42 $\pm$ 0.05 & 7.34 $\pm$ 0.11 & 0.67 $\pm$ 0.03 & 1.01 $\pm$ 0.05 & 0.34 $\pm$ 0.01 & 0.76 $\pm$ 0.04 & 1.57 \\
  & STRAND & 7.10 $\pm$ 0.17 & 1.53 $\pm$ 0.05 & 7.47 $\pm$ 0.10 & 0.67 $\pm$ 0.03 & 1.05 $\pm$ 0.06 & 0.33 $\pm$ 0.01 & 0.80 $\pm$ 0.04 & \textbf{1.43} \\
\midrule
\multirow{2}{*}{\textbf{Jurkat}}
 & STRAND (w/o CLIP) & 6.73 $\pm$ 0.16 & 1.64 $\pm$ 0.07 & 7.50 $\pm$ 0.09 & 0.44 $\pm$ 0.03 & 0.61 $\pm$ 0.05 & 0.34 $\pm$ 0.01 & 0.76 $\pm$ 0.04 & 1.57 \\
  & STRAND & 7.19 $\pm$ 0.16 & 1.82 $\pm$ 0.07 & 7.58 $\pm$ 0.09 & 0.44 $\pm$ 0.03 & 0.64 $\pm$ 0.04 & 0.34 $\pm$ 0.01 & 0.78 $\pm$ 0.04 & \textbf{1.43} \\
\midrule
\multirow{2}{*}{\textbf{RPE1}}
 & STRAND (w/o CLIP) & 6.23 $\pm$ 0.15 & 2.84 $\pm$ 0.12 & 7.78 $\pm$ 0.09 & 0.99 $\pm$ 0.04 & 1.75 $\pm$ 0.07 & 0.46 $\pm$ 0.02 & 1.53 $\pm$ 0.06 & 2.00 \\
  & STRAND & 7.04 $\pm$ 0.14 & 2.62 $\pm$ 0.10 & 7.93 $\pm$ 0.08 & 1.01 $\pm$ 0.04 & 1.76 $\pm$ 0.07 & 0.47 $\pm$ 0.01 & 1.39 $\pm$ 0.05 & \textbf{1.00} \\
\bottomrule
\end{tabular}
}

    \label{stab:clip-ablation}
\end{table}

\begin{table}[t]
\scriptsize
\centering
\setlength{\tabcolsep}{4pt}
\caption{\textbf{Ablation of the perturbation module}.}
\resizebox{\textwidth}{!}{%
\begin{tabular}{llccccccc|c}
\toprule
Dataset & Method & Discr. Cos $\uparrow$ & E-Dist $\downarrow$ & Dir. Match $\uparrow$ & Prec.@N $\uparrow$ & Ov.@N $\uparrow$ & $\Delta$Pearson $\uparrow$ & MSE $\downarrow$ & Avg. Rank $\downarrow$ \\
\midrule
\multirow{3}{*}{\textbf{K562}}
& STRAND
& 7.10 $\pm$ 0.17 & 1.53 $\pm$ 0.05 & 7.47 $\pm$ 0.10 & 0.67 $\pm$ 0.03 & 1.05 $\pm$ 0.06 & 0.33 $\pm$ 0.01 & 0.80 $\pm$ 0.04 & \textbf{1.29} \\
& \makecell[l]{STRAND\\(w/o Perturbation Module)}
& 5.12 $\pm$ 0.18 & 1.80 $\pm$ 0.08 & 6.82 $\pm$ 0.08 & 0.67 $\pm$ 0.03 & 0.75 $\pm$ 0.03 & 0.25 $\pm$ 0.01 & 0.85 $\pm$ 0.04 & 2.57 \\
& \makecell[l]{STRAND\\(w/o Latent Transport Module)}
& 5.00 $\pm$ 0.18 & 2.45 $\pm$ 0.08 & 7.13 $\pm$ 0.08 & 0.70 $\pm$ 0.04 & 0.96 $\pm$ 0.05 & 0.33 $\pm$ 0.01 & 0.86 $\pm$ 0.04 & 2.14 \\
\midrule
\multirow{3}{*}{\textbf{Jurkat}}
& STRAND
& 7.19 $\pm$ 0.16 & 1.82 $\pm$ 0.07 & 7.58 $\pm$ 0.09 & 0.44 $\pm$ 0.03 & 0.64 $\pm$ 0.04 & 0.34 $\pm$ 0.01 & 0.78 $\pm$ 0.04 & \textbf{1.00} \\
& \makecell[l]{STRAND\\(w/o Perturbation Module)}
& 5.03 $\pm$ 0.17 & 2.02 $\pm$ 0.12 & 7.04 $\pm$ 0.07 & 0.42 $\pm$ 0.03 & 0.42 $\pm$ 0.03 & 0.29 $\pm$ 0.01 & 0.81 $\pm$ 0.05 & 2.43 \\
& \makecell[l]{STRAND\\(w/o Latent Transport Module)}
& 4.98 $\pm$ 0.16 & 2.61 $\pm$ 0.11 & 7.22 $\pm$ 0.07 & 0.43 $\pm$ 0.03 & 0.31 $\pm$ 0.02 & 0.32 $\pm$ 0.01 & 0.84 $\pm$ 0.04 & 2.57 \\
\midrule
\multirow{3}{*}{\textbf{RPE1}}
& STRAND
& 7.04 $\pm$ 0.14 & 2.62 $\pm$ 0.10 & 7.93 $\pm$ 0.08 & 1.01 $\pm$ 0.04 & 1.76 $\pm$ 0.07 & 0.47 $\pm$ 0.01 & 1.39 $\pm$ 0.05 & \textbf{1.29} \\
& \makecell[l]{STRAND\\(w/o Perturbation Module)}
& 5.04 $\pm$ 0.14 & 3.20 $\pm$ 0.14 & 8.00 $\pm$ 0.08 & 0.98 $\pm$ 0.04 & 1.71 $\pm$ 0.06 & 0.50 $\pm$ 0.01 & 1.62 $\pm$ 0.07 & 1.71 \\
& \makecell[l]{STRAND\\(w/o Latent Transport Module)}
& 5.00 $\pm$ 0.14 & 4.37 $\pm$ 0.18 & 7.57 $\pm$ 0.06 & 0.95 $\pm$ 0.04 & 1.17 $\pm$ 0.04 & 0.42 $\pm$ 0.01 & 1.85 $\pm$ 0.08 & 3.00 \\
\bottomrule
\end{tabular}
}
\label{tab:perturbation_latent_diffusion_ablation_se}
\end{table}

\section{Overview of Existing Models}
\label{sec:supp_baselines}

We compare our method against the following models:
\textbf{GEARS}~\cite{Roohani2023}, \textbf{BioLord}~\cite{Piran2024}, \textbf{STATE}~\cite{Adduri2025},
and two baseline models:
\textbf{Linear} model on Borzoi embeddings~\cite{linder2025predicting}, and \textbf{PerturbMean}~\cite{vinas2025systema}.\textbf{STACK}~\cite{dong2026stack} is included in leave-one-cell-line-out transfer task,  where its in-context formulation is most
directly aligned with the task.

\xhdr{GEARS}~\cite{Roohani2023}
GEARS is a graph neural network that integrates gene--gene coexpression and GO-derived perturbation similarity graphs to learn perturbation embeddings and predict post-perturbation gene expression. It represents the dominant paradigm of incorporating \emph{gene-level biological priors}, allowing us to test whether sequence-resolved DNA conditioning provides advantages beyond relational gene graphs.

We used the default GO-based perturbation similarity graph from the official GEARS implementation, with $k=20$ neighbors for perturbation embedding learning. GEARS was trained for $20$ epochs using a hidden size of $64$ and a learning rate of $10^{-3}$. Because perturbation embeddings are defined over the GO graph, GEARS is limited to perturbations whose target genes are present in the training graph.

\xhdr{BioLord}~\cite{Piran2024}
BioLord is a disentanglement-based variational autoencoder that factorizes gene expression into cellular context and perturbation components, enabling counterfactual prediction of perturbation effects.

To introduce functional context, we incorporate GO-derived perturbation neighborhood features from the GEARS GO graph. For each perturbation, we select the top $20$ GO neighbors ranked by edge importance and use these weighted neighborhood features as ordered attributes for conditioning. BioLord was trained for $200$ epochs using AdamW for each dataset.

\xhdr{STATE}~\cite{Adduri2025}
STATE is a transformer-based foundation model trained on over 100M perturbed cells, representing the frontier of scale in expression-based perturbation modeling.

We extracted basal cell embeddings using the pretrained 600M-parameter STATE embedding model (epoch-16 checkpoint from the original work). A STATE transition model was then trained from scratch using default hyperparameters to predict perturbed expression from basal embeddings, preventing data leakage from overlapping Replogle datasets. Evaluation of STATE was restricted to low-sample settings, as the original work does not benchmark zero-shot prediction of unseen perturbations within a cell line.

\xhdr{STACK}~\cite{dong2026stack}
STACK is an in-context learning model for single-cell gene expression prediction that
processes batches of cells jointly using dual-axis tabular attention over cells and gene
groups. Its in-context formulation allows source-cell examples to provide perturbation
context for imputing query-cell states, making it a task-aligned baseline for
leave-one-cell-line-out transfer.

We evaluated STACK in the cross-cell-line setting using the official implementation and
default configuration. For each held-out cell line, STACK was trained on the two source
cell lines and used to predict perturbation responses in the held-out cell line without
target-cell-line fine-tuning. At inference, each batch contains a visible source-cell
perturbation context together with query control cells from the held-out cell line. STACK
then performs iterative masked generation, progressively imputing the query-cell expression
profiles; the final generated profiles are used as the predicted perturbed population. We
evaluate these predictions using the same train/test splits and metrics as \method{}.

\xhdr{Linear}
We include a linear regression model trained on Borzoi~\citep{linder2025predicting} DNA embeddings (7\,kb TSS-centered context, mean-pooled) to isolate the contribution of sequence information under minimal architectural assumptions. Recent benchmarks indicate that such linear models can be competitive in perturbation prediction~\citep{ahlmann2025deep}. Implementation details and a 524\,kb context-width control are provided in App.~\ref{sec:supp_linear}.

\xhdr{PerturbMean}~\cite{vinas2025systema}
PerturbMean predicts the average perturbation effect observed during training and serves as a strong non-parametric baseline. Systema demonstrated that this baseline often matches or exceeds GEARS, scGPT, and CPA, making it a useful reference for assessing whether models capture perturbation-specific effects beyond systematic variation.

Following the evaluation framework of Viñas Torné et al.~(2025), we employ the perturbed mean as a nonparametric baseline. Let $\mathcal{P}_{\text{train}}$ denote the set of all perturbed cells in the training data, and let $\mathbf{x}_i \in \mathbb{R}^G$ denote the gene expression profile of cell $i$. The perturbed mean is defined as:
\begin{equation}
\boldsymbol{\mu}_{\text{pert}}
\;=\;
\frac{1}{\lvert \mathcal{P}_{\text{train}} \rvert}
\sum_{i \in \mathcal{P}_{\text{train}}}
\mathbf{x}_i \, .
\end{equation}

This baseline uses $\boldsymbol{\mu}_{\text{pert}}$ as the predicted post-perturbation expression profile for all test perturbations, regardless of the specific gene targeted. The predicted perturbation effect is then computed as:
\begin{equation}
\widehat{\Delta \mathbf{x}}
\;=\;
\boldsymbol{\mu}_{\text{pert}} - \boldsymbol{\mu}_{\text{ctrl}} \, ,
\end{equation}
where $\boldsymbol{\mu}_{\text{ctrl}}$ is the mean expression of control cells. This prediction is compared against the ground-truth perturbation effect $\Delta \mathbf{x}$ for each held-out perturbation $X$.

\subsection{RNA Representations}
\label{sec:rna_rep_pipeline}

Both STRAND and STATE leverage a pretrained RNA foundation model to obtain cell embeddings. To preserve maximal transcriptome information at the representation stage, we compute RNA embeddings from the \emph{full} gene expression matrix (no HVG selection or gene filtering beyond dataset-provided gene definitions).
For downstream perturbation prediction and evaluation (e.g., pseudobulk shifts and metrics), we operate on the Standardized expression matrix described below.

\xhdr{Standardized Expression Matrices}
For each cell line (K562, RPE1, Jurkat), we build an AnnData object with a consistent gene axis defined by a canonical gene list (either cell-line-specific or the across-cell-line union; see \texttt{reformat\_adata\_genes}). All preprocessing starts from the raw count matrix $X$.

\xhdr{Per-Cell Library-Size Normalization}
For each dataset we normalize (the per-cell-line files and the final combined HVG matrix), we apply Scanpy’s library-size normalization with \texttt{sc.pp.normalize\_total(adata, target\_sum=1e4)}, which rescales each cell’s total UMI count to $10{,}000$. For cell $i$ with total count $c_i$, each gene count $x_{ij}$ is transformed as:
\[
x'_{ij} = x_{ij}\cdot \frac{10{,}000}{c_i}.
\]

\xhdr{Logarithmic Transformation}
After library-size normalization, we apply a log transform using \texttt{sc.pp.log1p(adata)}:
\[
x''_{ij} = \log(1 + x'_{ij}).
\]
This stabilizes variance and limits the influence of extreme counts.

\xhdr{HVG Selection for Cross-Cell-Line (Combined) Dataset}
\label{sec:combined-data-description}
To construct the final combined dataset, we concatenate union-gene raw count matrices across cell lines and select highly variable genes using only control cells. We run \texttt{sc.pp.highly\_variable\_genes(..., flavor="seurat\_v3", n\_top\_genes=2000)} on the raw counts to avoid selecting genes whose variance is driven by perturbations. We then restrict the full concatenated matrix to these 2{,}000 HVGs. Only after HVG selection do we apply library-size normalization to $10{,}000$ counts per cell followed by \texttt{log1p}, yielding the expression matrix used for downstream analyses.

For the combined cross–cell-line dataset, HVG selection is performed prior to normalization and log transformation. For per–cell-line datasets, normalization is performed directly without HVG restriction.

\section{Other Baselines}

\subsection{Linear Baseline}
\label{sec:supp_linear}

The linear baseline model is based on the assumption that the effect of a perturbation can be approximated as a linear transformation of the perturbed gene's characteristics and the cell's covariates, learned as an additive shift to the cell's baseline state.

\xhdr{Model Formulation}
Let $x_{i} \in \mathbb{R}^G$ denote the gene expression profile of cell $i$ under a perturbation $g$ (where $G$ is the number of genes), and let $x_{i, \text{base}} \in \mathbb{R}^G$ denote the corresponding baseline expression profile. In our experimental setup, $x_{i, \text{base}}$ corresponds to the expression of a control cell (no perturbation) matched to the sample $i$, or the mean expression of control cells if direct matching is unavailable.

We formalize the prediction task as learning the perturbation effect $\delta_i \in \mathbb{R}^G$, defined as the residual between the perturbed and baseline states:
\begin{equation}
    \delta_i = x_{i} - x_{i, \text{base}}
\end{equation}

The model constructs a feature vector $z_i$ for each sample. For single-dataset experiments, $z_i$ consists solely of the DNA embedding of the perturbed gene $e_g \in \mathbb{R}^D$, obtained by mean-pooling frozen Borzoi embeddings over a 7\,kb TSS-centered window. For the case of the combined dataset that consists of multiple cell lines, we augment the feature vector with the cell line identity as a one-hot encoded covariate $c_i$ to account for cell-type specific baseline shifts:
\begin{equation}
    z_i = \begin{cases}
        e_g & \text{single dataset} \\
        [e_g; c_i] & \text{combined dataset}
    \end{cases}
\end{equation}
In cases where the sample is a control (no perturbation), we utilize a zero vector for the gene embedding component $e_g$.

The predicted perturbation shift $\hat{\delta}_i \in \mathbb{R}^G$ is computed via a learned linear projection:
\begin{equation}
    \hat{\delta}_i = W z_i + b
\end{equation}
where $W$ is the weight matrix mapping the feature space to the gene expression space, and $b \in \mathbb{R}^G$ is the bias vector.

The final predicted expression profile $\hat{x}_i$ is obtained by applying the predicted shift to the baseline:
\begin{equation}
    \hat{x}_i = x_{i, \text{base}} + \hat{\delta}_i
\end{equation}

\xhdr{Training and Implementation}
The parameters $W$ and $b$ are estimated using Lasso regression (Linear least squares with $L_1$ regularization) to minimize the mean squared error between the predicted shifts $\hat{\delta}$ and the observed shifts $\delta$, while enforcing sparsity in the weight matrix. The objective function minimizes:
\begin{equation}
    \mathcal{L} = \sum_{i} || \delta_i - (W z_i + b) ||_2^2 + \alpha ||W||_1
\end{equation}
where $\alpha$ is the regularization strength (defaulting to $\alpha=0.05$). Prior to training, both the input features $z$ and the target residuals $\delta$ are standardized using Z-score normalization.

\xhdr{Context-Width Control}
To test whether the 7\,kb context disadvantages the Linear baseline relative to \method{}'s 524\,kb context, we reran Linear with \method{}'s full 524\,kb Borzoi context, changing only the pooling window (identical frozen encoder, splits, training, and evaluation). This control uses 2,000 HVGs. Widening Linear from 7\,kb to 524\,kb yielded nearly identical performance across the evaluated metrics (Table~\ref{stab:linear_context}); Linear's Discrimination Cosine stayed at chance level in every cell line at both widths (chance $=5.0$), while \method{} reached 7.11--7.23.

\begin{table}[h]
\centering
\caption{\textbf{Linear context-width control.} Mean over K562, Jurkat, and RPE1 (2,000 HVGs). DiscCos, DirMatch, P@N, and O@N are scaled by $\times$10; MSE by $\times$100.}
\label{stab:linear_context}
\small
\setlength{\tabcolsep}{4pt}
\begin{tabular}{llccccccc}
\toprule
Setting & Model & DiscCos $\uparrow$ & DirMatch $\uparrow$ & P@N $\uparrow$ & O@N $\uparrow$ & $\Delta$Pear $\uparrow$ & MSE $\downarrow$ & Rank $\downarrow$ \\
\midrule
\multirow{3}{*}{Few-shot ($k=5$)} & Linear 7\,kb & 5.060 & 6.774 & 1.728 & 1.983 & 0.293 & 1.754 & 1.91 \\
 & Linear 524\,kb & 5.046 & 6.767 & 1.720 & 1.970 & 0.291 & 1.764 & 2.86 \\
 & \method{} & \textbf{7.159} & \textbf{8.014} & \textbf{2.427} & \textbf{2.626} & \textbf{0.460} & 1.754 & \textbf{1.22} \\
\midrule
\multirow{3}{*}{Zero-shot} & Linear 7\,kb & 5.025 & 7.057 & 1.801 & 1.950 & 0.328 & \textbf{1.720} & 2.00 \\
 & Linear 524\,kb & 5.021 & 7.058 & 1.796 & 1.942 & 0.327 & 1.726 & 2.67 \\
 & \method{} & \textbf{5.389} & \textbf{7.981} & \textbf{2.158} & \textbf{2.397} & \textbf{0.457} & 1.750 & \textbf{1.33} \\
\bottomrule
\end{tabular}
\end{table}

\subsection{Concatenation-Based Conditioning Baseline}
\label{sec:supp_concat}

To assess whether the performance gains of \method{} arise from conditioning the diffusion dynamics rather than increased model capacity, we implement a concatenation-based baseline that replaces transport-based conditioning with feature fusion.
This baseline mirrors common conditioning strategies used in prior perturbation prediction methods.

\subsection{Input Representations}

Let $z_c \in \mathbb{R}^d$ denote the latent embedding of a control cell, obtained from the same pretrained RNA encoder used in \method{}.
Let $E_s \in \mathbb{R}^{L \times d_0}$ denote the token-level DNA embeddings produced by Borzoi for the genomic region associated with the perturbation.
As in \method{}, a learnable perturbation adaptor $f_\phi$ maps $E_s$ to a fixed-dimensional perturbation embedding:
\[
z_{\text{DNA}} = f_\phi(E_s) \in \mathbb{R}^d.
\]

At inference time, the baseline observes only $(z_c, z_{\text{DNA}})$.

\subsection{Concatenation Predictor Architecture}

The concatenation baseline predicts the perturbed cell embedding $\hat{z}_p$ via a deterministic feedforward network:
\[
\hat{z}_p = g_\psi\big([z_c \, ; \, z_{\text{DNA}}]\big),
\]
where $[\cdot ; \cdot]$ denotes vector concatenation and
$g_\psi : \mathbb{R}^{2d} \rightarrow \mathbb{R}^d$ is a multilayer perceptron.

The predictor $g_\psi$ consists of:
\begin{itemize}
    \item An input layer of dimension $2d$
    \item Two hidden layers with ReLU activations
    \item An output layer of dimension $d$
\end{itemize}

The hidden layer widths are matched to the diffusion score network used in \method{} to control for parameter count.

\subsection{Gene-Level Expression Reconstruction}

The predicted perturbed embedding $\hat{z}_p$ is passed to the same gene-level decoder used in \method{}, which reconstructs gene expression values via the hurdle truncated Gaussian residual likelihood.
This ensures that differences in performance arise solely from the conditioning mechanism, not from downstream decoding.

\subsection{Training Objective}

The concatenation baseline is trained without any diffusion or bridge-matching objective.
The reconstruction loss $\mathcal{L}_{\text{rec}}$ is the same hurdle truncated Gaussian likelihood used in \method{}. The same contrastive loss between $z_{\text{DNA}}$ and perturbed embeddings is included to ensure comparable supervision.

We apply the same uncertainty weighting as \method{} to balance the active objectives:
\[
\mathcal{L}_{\text{total}} = \sum_{i \in \{\text{rec},\, \text{CLIP}\}} \frac{1}{2}\left(\frac{\mathcal{L}_i}{\sigma_i^2} + \log \sigma_i^2\right),
\]
where $\sigma_i$ are learnable uncertainty parameters.
Note that, unlike \method{}, the diffusion losses are absent; only the reconstruction and contrastive terms participate in the weighting.

\section{Additional Results and Case Studies}
\label{sec:supp_results}

\subsection{Unseen Cell-Line Perturbation Prediction}
\label{sec:supp_cross_cellline}
Top-$K$ curves rank genes by the absolute observed expression change of each perturbation and report, over the top $K$ genes, the Pearson correlation between predicted and observed changes and the fraction of genes whose predicted change has the correct sign. Changes are computed against the same control mean for all methods, and no DE significance threshold is applied, unlike Dir.\ Match in App.~\ref{sec:supp_metrics}.

\begin{figure}[t]
  \centering
  \includegraphics[height=0.79\textheight]{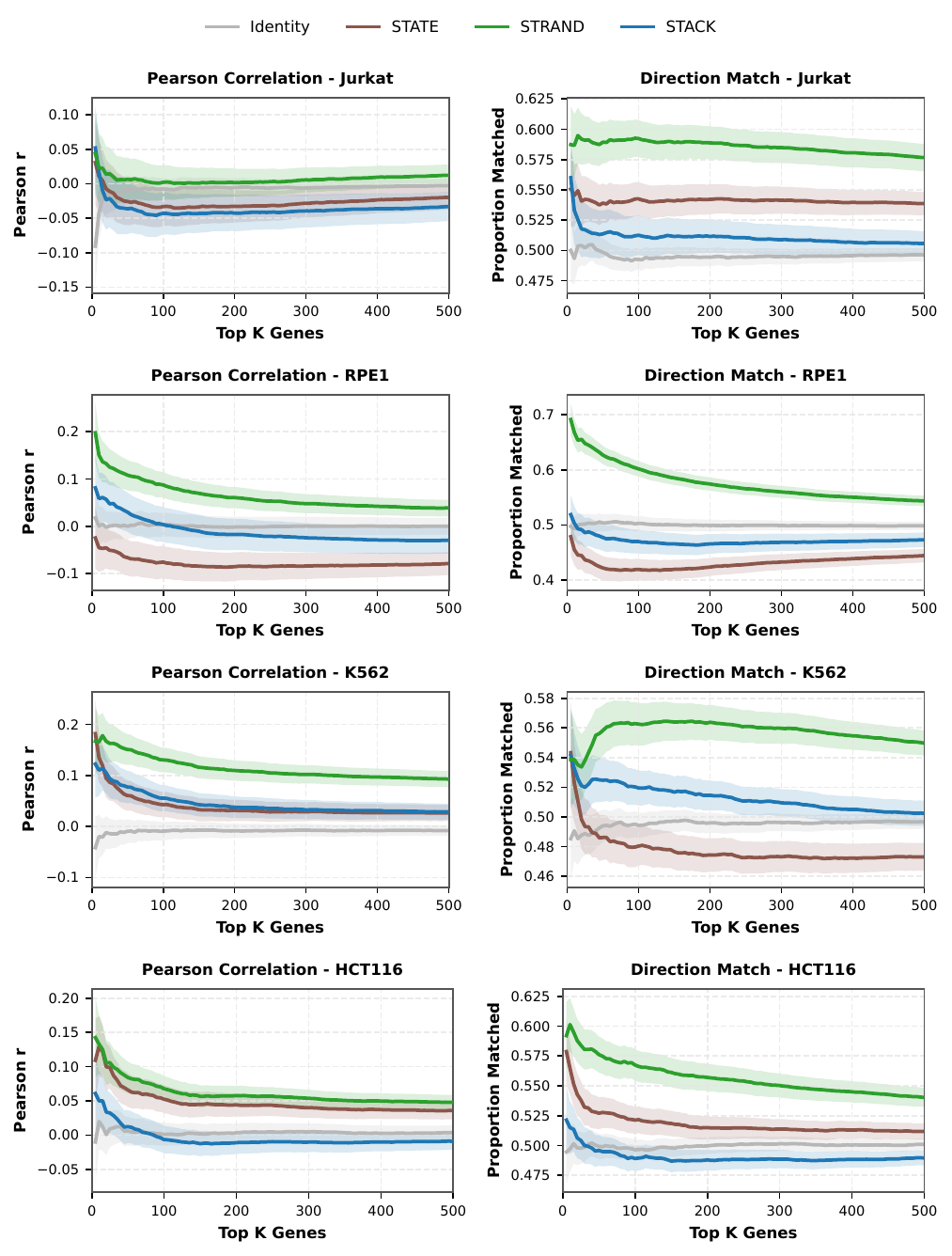}
  \caption{\textbf{Unseen cell-line perturbation prediction.}
  \method{} (green) consistently outperforms STACK (blue), STATE (brown) and the Identity baseline (gray).}
  \label{fig:supp_cross_cellline}
\end{figure}

\paragraph{External-study transfer to HCT116.}
\label{sec:supp_hct116}
The bottom row of Fig.~\ref{fig:supp_cross_cellline} applies a checkpoint trained on K562, RPE1, and Jurkat to the genome-wide dual-sgRNA CRISPRi screen of HCT116 from X-Atlas/Orion~\cite{huang2025x}. No HCT116 perturbation data are used, and all methods receive the same HCT116 control cells at inference. We pool all 109 batches, keep perturbations whose target gene is also perturbed in training, and take the 300 with the most cells, sampling 200 cells from each; we also sample 10{,}000 control cells (seed 7). Expression changes are computed against the control mean on the 2{,}000-gene training panel, of which 1{,}903 genes are measured in HCT116 (the rest are set to zero), with library-size normalization over the panel genes (App.~\ref{sec:rna_rep_pipeline}). Averaged over $K$ from 5 to 100, \method{} reaches a direction match of $0.578$ (95\% bootstrap CI $[0.564, 0.593]$), versus $0.534$ for STATE and $0.499$ for STACK and Identity, a paired gain of $0.045$ $[0.029, 0.059]$ over STATE. Its paired Pearson gain is $0.074$ $[0.050, 0.099]$ over STACK and $0.015$ $[-0.004, 0.033]$ over STATE.

\subsection{Additional Case Studies}
\label{sec:supp_case_study}

\begin{table}[t]
\centering
\caption{ \textbf{Sequence-level perturbation impact across FLVCR1 TSSs in K562.}
Among five candidate TSSs, the alternative FLVCR1b promoter (Transcript 202) exhibits the largest perturbation impact, measured by Energy Distance.
}
{
\begin{tabular}{lcccl}
\toprule
 \textbf{Transcript} & \multicolumn{2}{c}{\textbf{Transcript Annotation}}
& \multicolumn{1}{c}{\textbf{Impact}}
& \\
\cmidrule(lr){2-3} \cmidrule(lr){4-4}
\textbf{ID}
& \textbf{TSS (chr1)}
& \textbf{Type}
& \textbf{E-dist. $\uparrow$}
& \\
\midrule
201 & 212,858,275 & Canonical (FLVCR1a)   & 0.564          \\
202 & 212,858,916 & Alternative (FLVCR1b) & \textbf{0.683} \\
203 & 212,864,412 & Unknown               & 0.450          \\
204 & 212,883,409 & Unknown               & 0.580          \\
205 & 212,863,631 & Unknown               & 0.470          \\
\bottomrule
\end{tabular}
\label{stab:TSS-screen}
}
\end{table}

\begin{figure}[t]
\centering
\begin{minipage}[t]{0.48\linewidth}
\vspace{0pt}
\small
\paragraph{CD3D in Jurkat cells.}
CD3D encodes a component of the T-cell receptor complex and is specifically expressed in T lymphocytes. Jurkat, a T-cell leukemia line, shows dramatically higher perturbation sensitivity across the CD3D locus compared to K562 and RPE1 (Figure~\ref{sfig:jurkat-cd3d}). The elevated baseline and multiple peaks throughout the gene body suggest that Jurkat cells maintain active regulatory architecture at this locus, consistent with CD3D's essential role in T-cell signaling \cite{alarcon1991cd3}. The peaks observed at approximately 1kb and 5–6kb downstream of the TSS may correspond to intronic enhancers or other cis-regulatory elements active specifically in the T-cell context.
\end{minipage}\hfill
\begin{minipage}[t]{0.48\linewidth}
\vspace{0pt}
\centering
\includegraphics[width=\linewidth]{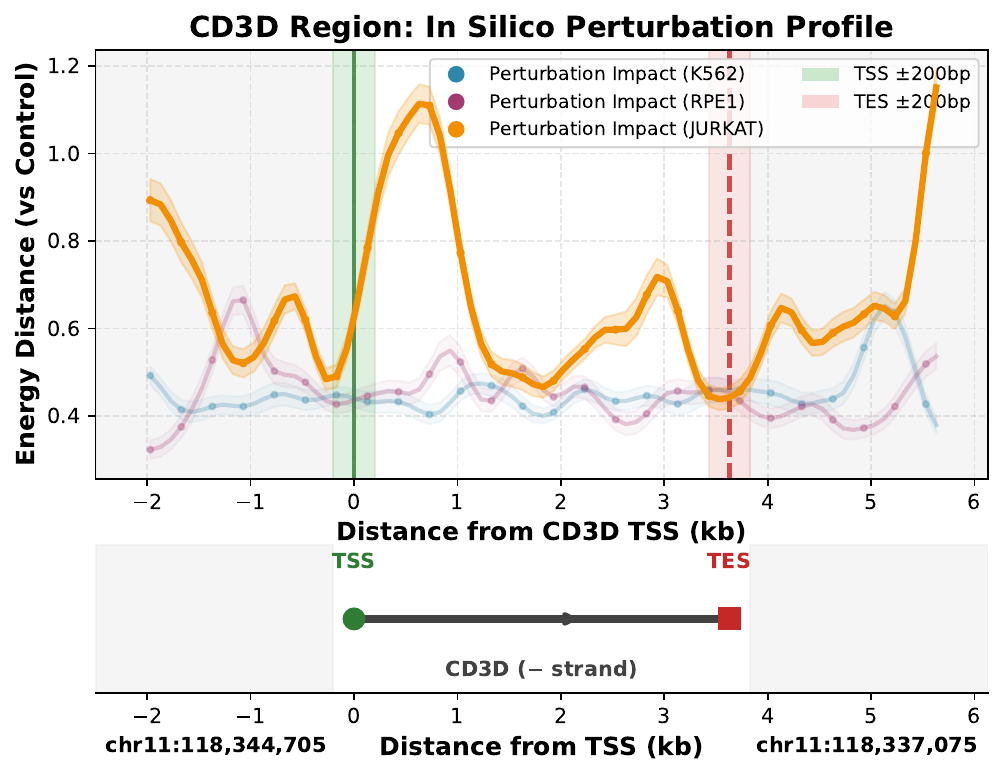}
\caption{\textbf{Jurkat - CD3D.}}
\label{sfig:jurkat-cd3d}
\end{minipage}
\end{figure}

\begin{figure}[t]
\centering
\begin{minipage}[t]{0.48\linewidth}
\vspace{0pt}
\small
\paragraph{KRT18 in RPE1 cells.}
KRT18 (Keratin 18) is an intermediate filament protein characteristic of simple epithelia. RPE1, a retinal pigment epithelial cell line, shows the strongest perturbation response at this locus, with prominent peaks near the TSS and in the downstream region around 5kb (Figure~\ref{sfig:rpe-krt18}). In contrast, the hematopoietic lines K562 and Jurkat show substantially lower and flatter profiles. This pattern recapitulates KRT18's known epithelial-specific expression pattern \cite{hunt1990altered}. Interestingly, the downstream peak in RPE1 cells (approximately 5kb from TSS) shows even higher sensitivity than the TSS itself, suggesting the presence of a strong distal regulatory element that may be critical for KRT18 expression in epithelial cells.
\end{minipage}\hfill
\begin{minipage}[t]{0.48\linewidth}
\vspace{0pt}
\centering
\includegraphics[width=\linewidth]{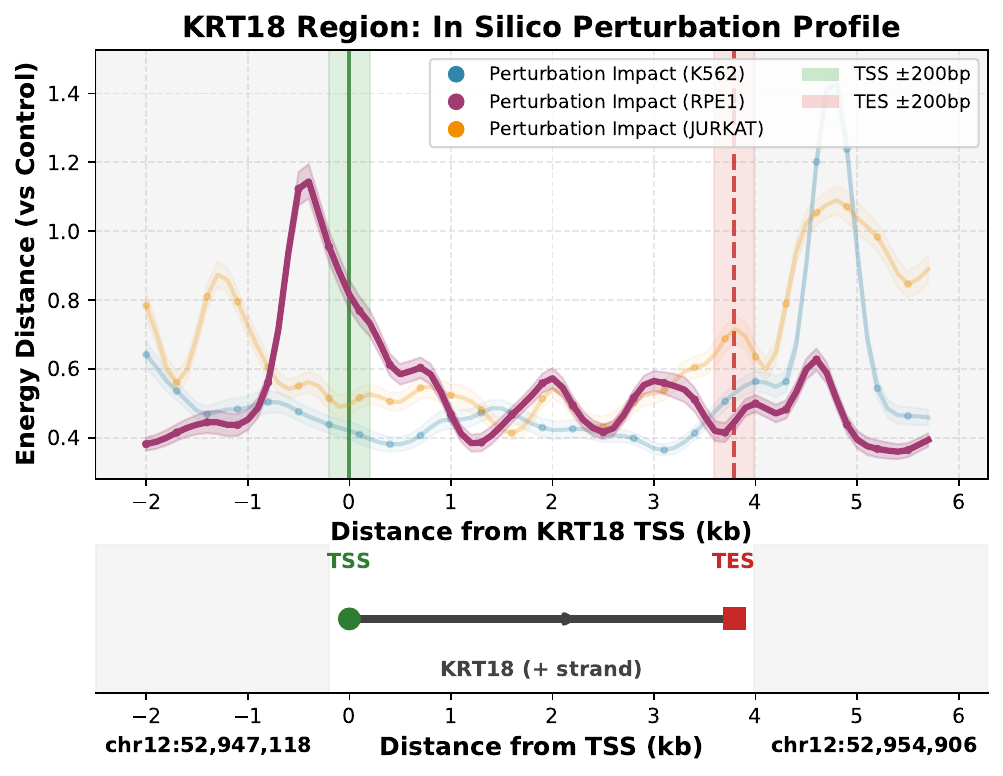}
\caption{\textbf{RPE1 - KRT18.}}
\label{sfig:rpe-krt18}
\end{minipage}
\end{figure}

These profiles are consistent with known cell-type-specific regulatory landscapes, and \method{} obtains them without explicit supervision on chromatin accessibility or transcription factor binding data. The approach recovers known regulatory patterns at canonical sites (e.g., TSS regions) and highlights putative novel regulatory elements (particularly the intragenic and downstream peaks) that could guide experimental prioritization for functional validation studies.

\section{Bridge Matching vs.\ Conditional Flow Matching}
\label{sec:supp_bridge_ablation}
We isolate the choice of generative transport backbone by holding the OT coupling, DNA perturbation module, decoder, and data splits fixed and swapping the I2SB bridge objective for a deterministic OT-CFM~\cite{Lipman2022,tong2023} objective. Across 21 metric–cell-line comparisons (7 metrics × 3 cell lines),  I2SB outperforms OT-CFM on $18/21$ paired comparisons; mean MSE improves from $0.0145$ to $0.0099$ ($-31.7\%$).

\begin{table}[h]
\scriptsize
\centering
\setlength{\tabcolsep}{4pt}
\caption{\textbf{Flow matching vs. diffusion.} Bridge matching (I2SB) vs. flow-matching paired ablation.}
\resizebox{\textwidth}{!}{%
\begin{tabular}{llccccccc|c}
\toprule
Dataset & Method & Discr. Cos $\uparrow$ & E-Dist $\downarrow$ & Dir. Match $\uparrow$ & Prec.@N $\uparrow$ & Ov.@N $\uparrow$ & $\Delta$Pearson $\uparrow$ & MSE $\downarrow$ & Avg. Rank $\downarrow$ \\
\midrule
\multirow{2}{*}{\textbf{K562}} & STRAND (FM) & 5.05 $\pm$ 0.18 & 2.39 $\pm$ 0.10 & 4.31 $\pm$ 0.13 & 0.68 $\pm$ 0.04 & 0.42 $\pm$ 0.02 & -0.08 $\pm$ 0.01 & 1.10 $\pm$ 0.06 & 1.86 \\
 & STRAND (I2SB) & 7.10 $\pm$ 0.17 & 1.53 $\pm$ 0.05 & 7.47 $\pm$ 0.10 & 0.67 $\pm$ 0.03 & 1.05 $\pm$ 0.06 & 0.33 $\pm$ 0.01 & 0.80 $\pm$ 0.04 & \textbf{1.14} \\
\midrule
\multirow{2}{*}{\textbf{Jurkat}} & STRAND (FM) & 4.97 $\pm$ 0.16 & 2.72 $\pm$ 0.12 & 4.56 $\pm$ 0.10 & 0.45 $\pm$ 0.03 & 0.28 $\pm$ 0.02 & -0.10 $\pm$ 0.01 & 0.93 $\pm$ 0.04 & 1.86 \\
 & STRAND (I2SB) & 7.19 $\pm$ 0.16 & 1.82 $\pm$ 0.07 & 7.58 $\pm$ 0.09 & 0.44 $\pm$ 0.03 & 0.64 $\pm$ 0.04 & 0.34 $\pm$ 0.01 & 0.78 $\pm$ 0.04 & \textbf{1.14} \\
\midrule
\multirow{2}{*}{\textbf{RPE1}} & STRAND (FM) & 4.93 $\pm$ 0.14 & 5.13 $\pm$ 0.22 & 4.50 $\pm$ 0.07 & 1.13 $\pm$ 0.04 & 0.97 $\pm$ 0.04 & -0.19 $\pm$ 0.01 & 2.33 $\pm$ 0.10 & 1.86 \\
 & STRAND (I2SB) & 7.04 $\pm$ 0.14 & 2.62 $\pm$ 0.10 & 7.93 $\pm$ 0.08 & 1.01 $\pm$ 0.04 & 1.76 $\pm$ 0.07 & 0.47 $\pm$ 0.01 & 1.39 $\pm$ 0.05 & \textbf{1.14} \\
\bottomrule
\end{tabular}
}
\label{tab:flow_matching_vs_diffusion_se}
\end{table}

\section{Perturbation Coordinate Construction and Sensitivity}
\label{sec:supp_tss_protocol}
Because most public perturbation datasets retain gene-level labels rather than exact gRNA coordinates, we provide a deterministic coordinate-recovery pipeline to make \method{} applicable to such datasets in a reproducible manner. Gene symbols are normalized to current HGNC-approved symbols, mapped to stable Ensembl gene IDs (GRCh38, Ensembl 110), and resolved to canonical TSSs in GENCODE v44 \texttt{basic} annotations using the priority MANE\_Select $>$ Ensembl\_canonical $>$ APPRIS, with strand-aware coordinate handling. The result is a TSS-centered proxy when exact perturbation coordinates are unavailable; we extract a $\pm 200$\,bp CRISPRi-accessible window ($\mb$) and a $w=524{,}288$\,bp sequence context ($s$) centered on the TSS.

\xhdr{Exact-gRNA Sensitivity Analysis}
For CRISPRi perturbations with available guide-coordinate annotations, we repeated the few-shot benchmark using exact gRNA-centered perturbation regions, changing only the coordinate definition. Results are comparable to the default TSS-centered proxy across K562, Jurkat, and RPE1 (Table~\ref{stab:coordinate-sensitivity}). Exact gRNA coordinates slightly improve average rank in Jurkat and RPE1, while the TSS proxy slightly outperforms in K562. This indicates that the main conclusions are robust to the operational coordinate choice in the current benchmark.

One likely reason for this stability is that more than 70\% of annotated guides in the analyzed datasets fall within the default TSS-proximal window. We therefore use the TSS-centered proxy as the default coordinate definition for consistency across datasets and to maximize coverage when exact guide coordinates are unavailable. Modeling guide-dependent efficacy, motif-distance effects, and distal enhancer perturbations remains an important direction for datasets with systematic locus-resolved perturbation labels.

\begin{table}[h]
\scriptsize
\centering
\setlength{\tabcolsep}{4pt}
\caption{\textbf{Sensitivity to perturbation-coordinate definition.}
Few-shot benchmark results using the default $\pm 200$\,bp TSS-centered proxy versus exact gRNA-centered perturbation regions when guide coordinates are available. Only the coordinate definition is changed; data splits, model architecture, training protocol, and evaluation metrics are fixed. Dir. Match, Prec.@N, Ov.@N, and Discr. Cos are scaled by $\times 10$; MSE is scaled by $\times 100$.
}
\resizebox{\textwidth}{!}{%
\begin{tabular}{llccccccc|c}
\toprule
\textbf{Dataset} & \textbf{Method} & \textbf{Discr. Cos $\uparrow$} & \textbf{E-Dist $\downarrow$} & \textbf{Dir. Match $\uparrow$} & \textbf{Prec.@N $\uparrow$} & \textbf{Ov.@N $\uparrow$} & \textbf{$\Delta$Pearson $\uparrow$} & \textbf{MSE $\downarrow$} & \textbf{Avg. Rank $\downarrow$} \\
\midrule
\multirow{2}{*}{\textbf{K562}}
 & $\pm 200$ TSS & 5.16 $\pm$ 0.18 & \textbf{1.70 $\pm$ 0.08} & \textbf{6.54 $\pm$ 0.10} & \textbf{0.72 $\pm$ 0.04} & \textbf{0.87 $\pm$ 0.04} & \textbf{0.22 $\pm$ 0.01} & \textbf{0.87 $\pm$ 0.05} & \textbf{1.14} \\
 & gRNA & \textbf{5.24 $\pm$ 0.18} & 1.75 $\pm$ 0.08 & 6.39 $\pm$ 0.10 & 0.71 $\pm$ 0.04 & 0.83 $\pm$ 0.04 & 0.20 $\pm$ 0.01 & 0.89 $\pm$ 0.05 & 1.86 \\
\midrule
\multirow{2}{*}{\textbf{Jurkat}}
 & $\pm 200$ TSS & 5.26 $\pm$ 0.16 & 1.95 $\pm$ 0.09 & 6.84 $\pm$ 0.09 & \textbf{0.50 $\pm$ 0.03} & 0.54 $\pm$ 0.04 & 0.26 $\pm$ 0.01 & 0.80 $\pm$ 0.04 & 2.00 \\
 & gRNA & \textbf{5.36 $\pm$ 0.17} & \textbf{1.81 $\pm$ 0.09} & \textbf{7.04 $\pm$ 0.08} & \textbf{0.50 $\pm$ 0.03} & \textbf{0.55 $\pm$ 0.04} & \textbf{0.29 $\pm$ 0.01} & \textbf{0.77 $\pm$ 0.04} & \textbf{1.00} \\
\midrule
\multirow{2}{*}{\textbf{RPE1}}
 & $\pm 200$ TSS & 5.16 $\pm$ 0.14 & 3.31 $\pm$ 0.14 & 7.78 $\pm$ 0.08 & 1.25 $\pm$ 0.05 & 1.98 $\pm$ 0.07 & 0.46 $\pm$ 0.01 & 1.64 $\pm$ 0.07 & 2.00 \\
 & gRNA & \textbf{5.22 $\pm$ 0.15} & \textbf{3.06 $\pm$ 0.14} & \textbf{7.83 $\pm$ 0.08} & \textbf{1.26 $\pm$ 0.05} & \textbf{2.04 $\pm$ 0.08} & \textbf{0.47 $\pm$ 0.01} & \textbf{1.57 $\pm$ 0.07} & \textbf{1.00} \\
\bottomrule
\end{tabular}
}
\label{stab:coordinate-sensitivity}
\end{table}

\section{HDAC6 Multi-Peak Case Study}
\label{sec:supp_hdac6}
Sliding-window perturbation profiles around \textit{HDAC6} reveal a secondary high-impact peak at approximately $+13.65$\,kb downstream of the GATA1 TSS, with E-distance $0.811$ ($\sim 2\times$ background) and within $400$\,bp of the validated e-HDAC6 enhancer center reported in CRISPRi screens by Fulco et al.; CRISPRi silencing of e-HDAC6 in K562 reduces \textit{GATA1} expression by $\sim 33\%$ and \textit{HDAC6} expression by $\sim 22\%$ in that work, providing independent functional evidence that this distal peak is biologically meaningful.

\section{Computational Efficiency}
\label{app:compute}

We profile computational cost on a single NVIDIA H100
(80 GB) GPU with eight data-loading workers.
Table~\ref{stab:compute} summarizes preprocessing and
training time across five matched benchmark configurations.
We use the benchmark training budgets: 20 epochs for
GEARS and \method{}, and 40,000 steps for STATE.
The DNA and RNA encoders in \method{} remain frozen.

\begin{table}[t]
\centering
\caption{\textbf{Training runtime.}
Single-GPU wall-clock hours aggregated over five
matched configurations (Jurkat and RPE1 with both
gene panels, and K562 with 2k HVGs).
Preprocessing includes method-specific data preparation.}
\label{stab:compute}
\small
\begin{tabular}{lrrr}
\toprule
& \multicolumn{3}{c}{Runtime (hours)} \\
\cmidrule(lr){2-4}
\textbf{Method} &
\textbf{Preprocessing} &
\textbf{Training} &
\textbf{Total} \\
\midrule
GEARS     & 22.3 &  6.2 & 28.5 \\
STATE     &  2.8 & 19.4 & 22.2 \\
\method{} &  9.4 & 15.3 & 24.7 \\
\bottomrule
\end{tabular}
\end{table}

\xhdr{Coordinate-Level Inference}
For each genomic coordinate, \method{} processes a
524-kb sequence window and predicts 256 cells in
2.96\,s. Sequence encoding requires
$0.180 \pm 0.009$\,s, accounting for only 6.1\%
of inference time, while transport and decoding
require $2.785 \pm 0.020$\,s.
Sequence encoding is performed once per coordinate
and shared across the predicted cells.
Scanning 81--118 coordinates at 100-bp intervals
takes 3.5--7.6 minutes without retraining.
With 48 predicted cells per perturbation in Jurkat,
\method{} requires 0.28\,s, compared with 5.5\,s
for STACK under matched output size and excluding startup.

Beyond the five shared configurations in Table~\ref{stab:compute},
the full six-configuration benchmark required approximately
29.5 GPU-hours for \method{} and 29.1 GPU-hours for STATE;
GEARS did not complete within 40 GPU-hours. Nine ablation runs
(three variants across three cell lines) required at most
29.4 GPU-hours in total, reusing the SCimilarity embeddings and
OT couplings. A separate inference-only sweep over diffusion steps
$T\in\{2,6,10,20,50,100\}$, distinct from the validation-time
selection in App.~\ref{sec:supp_default_config}, required
3.6 GPU-hours without retraining.

\xhdr{Scaling with Dataset Size}
We also assess runtime scaling by increasing the Jurkat dataset
from 19,183 to 191,466 cells while holding the perturbations,
task, and training budgets fixed. Over this approximately tenfold
increase, measured computational cost grows by 1.7$\times$ for
\method{}, 2.1$\times$ for STATE, and 9.4$\times$ for GEARS.

\section{Limitations and Future Work}
\label{sec:limitations}

\xhdr{Limitations}
The genome-coverage figure reported in
Table~\ref{tab:perturbation_resolution} refers to architectural input
support: \method{} can condition on sequence-defined loci across a much
larger fraction of the genome than gene-identifier methods. Empirical
supervision, however, is currently limited by available CRISPRi perturbation
datasets. Most benchmark annotations are gene-centered, and we therefore use
the same perturbation-coordinate protocol as the main benchmark, with
additional guide-coordinate analyses showing that \method{} can condition on
more precise perturbation loci when such annotations are available. Within this supervision regime,
position-resolved predictions at TSS-proximal loci, including FLVCR1,
GATA1, CD3D, and KRT18, are evaluated against independently established
cell-type-specific regulatory biology. Predictions at distal or otherwise
unsupervised loci are extrapolative from the learned sequence-conditioned
perturbation adaptor; the e-HDAC6 distal peak provides targeted supporting
evidence for this capability. Systematic distal-enhancer validation will
require perturbation datasets with dense, locus-resolved regulatory
annotations.

\method{} operates at 128\,bp model resolution, set by Flashzoi/Borzoi output
binning, rather than at single-nucleotide resolution. Our
\emph{in silico} sequence perturbation experiments do not account for
practical constraints such as sgRNA design feasibility or target
druggability, which may limit experimental realizability. The regulatory
context window supported by \method{} is capped by the input capacity of the
upstream DNA foundation model. Finally, the current model is trained on
CRISPRi perturbation data; evaluating other perturbation modalities, such as
CRISPRa or base editing, is an important direction for future work.

\xhdr{Future Work}

Several promising directions remain. Integrating alternative DNA and RNA foundation models may improve representation quality and generalization. Incorporating long-range regulatory interactions, such as distal enhancers and chromatin contacts, would enable more comprehensive modeling of sequence-to-expression relationships. Extending the perturbation module to multiple target loci, for example by encoding each locus separately and combining the resulting perturbation embeddings, would enable combinatorial and trans-chromosomal perturbation prediction. Cross-context generalization also remains limited by data scale; we anticipate improved performance as larger, more diverse perturbation datasets across additional cell types become available.

\section*{Broader Impact}
This work advances machine learning methods for modeling genetic perturbation effects by introducing sequence-conditioned generative transport for single-cell transcriptomics. By enabling locus-specific prediction of perturbation responses at 128\,bp model resolution, the proposed framework may support hypothesis generation in functional genomics, guide experimental design, and improve interpretation of non-coding genetic variation relevant to human disease.

The model is intended for in silico analysis and does not directly enable genome editing or clinical intervention. However, as with many computational methods in genomics, downstream applications could influence experimental prioritization in biomedical research. Misinterpretation of model predictions without experimental validation could lead to incorrect biological conclusions. \method{} is intended as a predictive and exploratory tool to inform experimental validation, not replace it.

The datasets used are publicly available and de-identified, and the work does not introduce new privacy risks. We do not foresee immediate negative societal impacts beyond those already associated with the broader use of machine learning in biological research.


\end{document}